\documentclass{ametsocV6.1}

\usepackage[english]{babel}
\usepackage[utf8x]{inputenc}
\usepackage[T1]{fontenc}
\usepackage{textcomp}
\usepackage{subfigure}
\usepackage{amsmath}
\usepackage{amssymb}

\usepackage{amsthm}
\usepackage{bm}
\usepackage{xcolor}
\usepackage{natbib}
\usepackage{csquotes}
\usepackage{cancel}
\usepackage{siunitx}

\newtheorem{theorem}{Theorem}

\usepackage{physics}
\usepackage{url}
\usepackage[hidelinks]{hyperref}

\newcommand{\lb}{\left<}
\newcommand{\rb}{\right>}
\newcommand{\lp}{\left(}
\newcommand{\rp}{\right)}
\newcommand{\ti}{\textit}
\newcommand{\pt}{\partial_t}
\newcommand{\px}{\partial_x}
\newcommand{\py}{\partial_y}
\newcommand{\pz}{\partial_z}
\newcommand{\bs}{\boldsymbol}

\newcommand{\oxy}[1]{{\lb {#1} \rb}}

\newcommand{\B}{\bs{B}}
\newcommand{\beq}{\begin{equation}}
\newcommand{\eeq}{\end{equation}}
\newcommand{\bu}{\boldsymbol{u}}

\newcommand{\ba}{\boldsymbol{a}}
\newcommand{\bx}{\boldsymbol{x}}

\newcommand{\bsig}{\bs{\sigma}}

\newcommand{\bnabla}{\boldsymbol{\nabla}}

\usepackage{scalerel,stackengine}
\newcommand\reallywidehat[1]{%
\savestack{\tmpbox}{\stretchto{%
  \scaleto{%
    \scalerel*[\widthof{\ensuremath{#1}}]{\kern-.6pt\bigwedge\kern-.6pt}%
    {\rule[-\textheight/2]{1ex}{\textheight}}
  }{\textheight}%
}{0.5ex}}%
\stackon[1pt]{#1}{\tmpbox}%
}
\AtBeginDocument{\RenewCommandCopy\qty\SI}
 
\title{Stochastic compressible Navier--Stokes equations and their approximations for ocean modelling}

\author{Gilles \textsc{Tissot}\aff{a}\correspondingauthor{Gilles Tissot, gilles.tissot@inria.fr},
Quentin \textsc{Jamet}\aff{a,b,c},
\'Etienne \textsc{M\'emin}\aff{a}
}

\affiliation{\aff{a}INRIA Centre de l'Universit\'e de Rennes, IRMAR -- UMR CNRS 6625,
             av. G\'en\'eral Leclerc, 35042, Rennes, France
             \\\aff{b}LOPS, Ifremer, ZI Pointe du Diable -- CS 10070, Plouzan\'e, 29280, France
             \\\aff{c} SHOM, 13 rue du Chatellier, 29200 Brest, France
             }

\begin{document}
\newcommand{\bcdot}{\boldsymbol{\cdot}}
\newcommand{\mathsfbi}[1]{\bm{\mathsf{#1}}}
\newcommand{\dt}{\,\mathrm{d}t}
\newcommand{\dx}{\,\mathrm{d}\bm{x}}
\newcommand{\ddt}[2]{\frac{\displaystyle \partial{#1}}{\displaystyle \partial{#2}}}
\newcommand{\dddt}[2]{\frac{\displaystyle \partial^{2}{#1}}{\displaystyle \partial{#2}^{2}}}
\newcommand{\ddddt}[2]{\frac{\displaystyle \partial^{3}{#1}}{\displaystyle \partial{#2}^{3}}}
\newcommand{\dddddt}[2]{\frac{\displaystyle \partial^{4}{#1}}{\displaystyle \partial{#2}^{4}}}
\newcommand{\dBt}{\mathrm{d}\bm{B}_t}
\newcommand{\Bt}{\bm{B}_t}
\newcommand{\dBs}{\mathrm{d}\bm{B}_s}
\newcommand{\dBsp}{\mathrm{d}\bm{B}_{s'}}
\newcommand{\sdBt}{\boldsymbol{\sigma}_t \mathrm{d}\bm{B}_t}
\newcommand{\sdBh}{\hat{\boldsymbol{\sigma}}_t \mathrm{d}\bm{B}_t}
\newcommand{\sdBz}{\hat{\boldsymbol{\sigma}}_0 \mathrm{d}\bm{B}_t}
\newcommand{\sdBs}{\boldsymbol{\sigma}_s \mathrm{d}\bm{B}_s}
\newcommand{\sdBsp}{\boldsymbol{\sigma}_{s'} \mathrm{d}\bm{B}_{s'}}
\newcommand{\rouge}[1]{\textcolor{red}{#1}}
\newcommand{\bleu}[1]{\textcolor{blue}{#1}}
\newcommand{\monvert}[1]{\textcolor{forestgreen}{#1}}
\newcommand{\here}{\textcolor{red}{\hrule}\begin{center} \textcolor{red}{Stopped here!}\end{center} \textcolor{red}{\hrule}\clearpage}
\newcommand{\Div}{\nabla \bcdot}
\newcommand{\mat}[1]{\mathsfbi{#1}}
\newcommand{\Grad}{\nabla}
\newcommand{\Diffst}[1]{\frac{1}{2}\Div (\mathsfbi{a} \Grad #1)}
\newcommand{\Adv}[2]{\left(#1\bcdot\nabla\right)#2}
\newcommand{\adv}[2]{\Bigr(#1\bcdot\nabla\Bigr)#2}
\newcommand{\sadv}[2]{\bigl(#1\bcdot\nabla\bigr)#2}
\newcommand{\udrift}{{\bm{u}^\star}}
\newcommand{\sds}{\boldsymbol{\sigma}_t(\Div\boldsymbol{\sigma}_t)}
\newcommand{\Rem}{\frac{1}{Re}}
\newcommand{\RePrm}{\frac{1}{RePr}}
\newcommand{\gmg}{\frac{\gamma-1}{\gamma}}
\newcommand{\covariation}[2]{\dd_t\left\langle \int_0^{\bcdot} #1,\int_0^{\bcdot} #2 \right\rangle_t}
\newcommand{\molec}[1]{} 
\newcommand{\delete}[1]{}
\newcommand{\strat}[2]{#1\circ #2}
\newcommand{\mgt}[1]{\textcolor{teal}{#1}}
\newcommand{\EM}[1]{\textcolor{red}{#1}}
\newcommand{\cgt}[1]{\textcolor{blue}{\textbf{#1}}}
\newcommand{\mr}{\mathrm}
\newcommand{\mb}{\mathbb}  
\newcommand{\dif}{\mr{d}}
\newcommand{\Exp}{\mb{E}} 
\newcommand{\ie}{\textit{i.e.}~}
\newcommand{\cf}{\textit{c.f.}~}
\newcommand{\eg}{\textit{e.g.}~}

\abstract{
This paper presents a joint theoretical and numerical study of a stochastic version of the compressible Navier--Stokes equations within the location uncertainty (LU) framework, applied to problems related to upper ocean vertical mixing.
This approach builds on an extended stochastic form of the Reynolds transport theorem, incorporating stochastic source terms.
As in the deterministic case, this conservation theorem is applied to mass, mass of species (such as salinity), momentum, and total energy, leading to transport equations for the primitive variables: density, mass fraction of species, velocity, and temperature.
We subsequently apply the Boussinesq approximations to this general system, and recover existing formulations of the incompressible stochastic Navier--Stokes and stochastic Boussinesq equations.
We employ this new framework in a Boussinesq large-eddy simulation of temperature-driven free convection event, and highlight the potential of stochastic transport to reproduce penetrative convection effects at the base of the mixed layer under the Boussinesq approximation. 
Compression effects identified in our stochastic Boussinesq hydrostatic model are  found to be negligible in the temperature equation when expressed in terms of internal energy, in agreement with Boussinesq approximation. 
However, when expressed in terms of potential energy, compression effects become significant, and reveal interesting properties of the stochastic pressure terms within the mixed layer. 
We believe this later results open new physical modelling perspective enabling to represent oceanic dynamics within a stochastic framework that more fully accounts for physical uncertainties and approximations, while also providing a basis for improving the energetic consistency of subgrid-scale vertical models used in ocean general circulation models.
}

\maketitle


\section{Introduction}
Stochastic representations of fluid flow dynamics are increasingly employed across a range of  applications to account for uncertainty in system evolution and to retain, at coarse resolution, essential features of high-resolution dynamics such as bifurcations and spontaneous stochasticity associated with singularities and the nonuniqueness of classical solutions.
This transition from an intractable, fully deterministic, high-resolution description to a tractable, large-scale stochastic representation is critical across a wide range of domains, from engineering to climate science, where uncertainties arise from turbulence, physical modelling approximations, or numerical discretisation.

We consider the location uncertainty (LU) framework, which provides a physically principled stochastic formulation for compressible fluid flows. It relies on decomposing the displacement of fluid particles into a time-differentiable velocity field and a highly fluctuating component represented as a Brownian motion.

This Lagrangian decomposition leads to a stochastic formulation of the Reynolds transport theorem, which in turn enables a direct Eulerian stochastic representation of the flow through the specification of stochastic physical invariants.

This approach was first proposed in \citet{memin2014} for incompressible fluid flows, resulting in a stochastic version of the Navier--Stokes equations.
The ability of this model to represent fine-scale deterministic Navier--Stokes dynamics at a coarse scale has been recently analyzed in \citet{Debusshe-Hug-Memin-2023}. 
Beyond establishing the existence of weak (probabilistic) solutions in three dimensions, with uniqueness in two dimensions, this stochastic representation exhibits a well-behaved vanishing-noise limit and converges to the corresponding deterministic equations.
This property provides strong physical consistency for the LU framework and can be exploited in practice to enable efficient exploration of the associated attractors \citep{chapron2018}.
Various derivations within this framework have been developed and numerically assessed for a range of ocean models, including Boussinesq \citep{resseguier2017e}, quasi-geostrophic (QG) \citep{bauer2020a,Li-Memin-Tissot-2023}, surface quasi-geostrophic (SQG) \citep{resseguier2017e}, shallow water \citep{brecht2021} models or 1D Stokes-Ekmann models \citep{li2025}.

In the ocean, variations in density, temperature, and salinity, as well as compressibility effects, play a key role in driving flow dynamics.
However, the stochastic models cited previously have generally relied on the Boussinesq approximation, which assumes small compressibility from the outset.
While this is a reasonable approximation in many cases, it can become limiting,
particularly  when accurate energy budget are considered \citep{young2010}.
In recent years, intensive research efforts have been devoted to incorporate compressibility effects into ocean flow models \citep{tailleux2012, eden2015, dewar2015}.
It has been shown \citep{mcdougall2003, graham2013} that, even under the Boussinesq approximation assumptions, the work of compression and dilatation is not negligible due to density variations.
In such cases, potential enthalpy is a conserved quantity, leading to the introduction of a variable known as conservative temperature.
This conservation holds exactly in the absence of molecular or eddy dissipation.
However, the topic becomes more controversial when conservation laws are expressed in an averaged sense for large-scale modelling, particularly when energy is transferred to unresolved (small-scale) motions through mixing processes \citep{mcdougall2021}.
For example, \citet{tailleux2015} explored the differences between the non-conservation of potential temperature and conservative temperature. 

In the numerical proof of concept proposed in the present paper, we want to evaluate how such effects might contribute in a stochastic formulation of sub-grid scale vertical turbulence schemes, with a focus on oceanic convection.
This choice is motivated by the significant impacts of ocean convection on climate, where incorrect representations may lead to significant biases in the Atlantic meridional overturning circulation (AMOC) simulated by climate models \citep{heuze2020}, and by renew interests in its parameterisation in coarse resolution models \cite[e.g.][]{giordani2020,perrot2025}.
These phenomenological schemes based on multi-fluid averaging procedure have shown to provide accurate plume vertical advection parameterisation, but leaving open the question of how compression effects contribute.

Other popular approaches are based on Turbulent Kinetic Energy (TKE)
\citep[also termed generic length scale (GLS), e.g.][]{umlauf2003} sub-grid scale models, which instead rely on Reynolds averaged Navier--Stokes (RANS) equations.
These approaches are also derived for Boussinesq fluids \citep{mellor1982,stull1988,sander1998}, obscuring the energetic consistency between temperature and momentum fluxes used in the derivation of such models.
Our aim here is to take a step forward in clarifying this energy consistency by avoiding a Boussinesq assumption from the start, a turbulent closure in the transport equations of turbulent variables, or phenomenological assumptions in the definition of average operators.

The stochastic LU framework, which provides a conservative representation, is a natural setting for formulating conservation of total energy—the sum of kinetic, potential, and internal energy—while explicitly separating resolved slow scales from unresolved fast scales.

This formulation allows us to keep track of the appropriate orders of magnitude of the different terms in the temperature equation depending on which energy reservoir they are associated with.
By a slight abuse of language, inherited from the large eddy simulation (LES) community, we will often refer to resolved scales as large scales, and to unresolved processes, modelled here as temporally decorrelated stochastic terms, as small-scale processes.
The latter are modelled as temporally decorrelated stochastic terms that are nonetheless smooth in space, meaning they are uncorrelated in time but spatially correlated.
In clear, our closure relies on this decorrelation assumption, and on the definition of the small scale velocity noise covariance, which we believe to be easier to characterise in a physically consistent manner than Reynolds stresses.

While the assumption of temporal decorrelation is a mathematical idealisation, it allows for a rigorous treatment using stochastic calculus.
As we will show, this formalism allows for a systematic derivation of (correlated in time) noise-induced terms, analogous to Stokes drift and diffusion, that are often introduced empirically in evolution equations.
Within this framework, it becomes possible to derive consistent evolution equations for non-conserved quantities, such as temperature, in a compressible setting.
It can be noted that turbulence modelling for the evolution equations of the thermodynamic variables usually also requires a Boussinesq-like assumption \citep[see for instance][]{shyy1997}, while in our proposed model the small scale interactions terms are deduced from the stochastic small scale velocity.

Another key aspect is that Boussinesq models cannot sustain acoustic waves, which has significant implications for two major applications.
First, in ocean acoustics, where accurate wave propagation is essential -- for example, in tsunami detection \citep{dubois2023}.
Second, in numerical simulations of non-Boussinesq models, where pseudo-compressibility methods \citep{chorin1967} are used to compute pressure dynamically and avoid solving a full three-dimensional Poisson equation \citep{auclair2018} at each time step.
To develop numerical oceanic stochastic LU models that include non-hydrostatic pressure corrections, a full stochastic derivation from the compressible equations appears necessary.
Relevant approximations can then be introduced to design physically consistent schemes.

A compressible stochastic system cannot simply be derived from its incompressible counterpart, as the latter represents a limiting case of the former.
In this work, we start from the classical physical conservation laws to derive a general stochastic compressible Navier--Stokes system.
We then verify that this system remains consistent with previously proposed incompressible stochastic models.
By applying the Boussinesq approximation, we also obtain a stochastic Boussinesq model that incorporates thermodynamic effects.

The present paper builds upon the workshop proceedings \cite{tissotSTUOD2023}, which were purely theoretical and derived the governing equations used here. In the present work, these derivations have been revised and clarified, and are presented in a more complete and self-consistent form.
In addition, that previous work relied on a relaxation of the Boussinesq approximation through the inclusion of a pseudo-compressibility term in the spirit of \citet{auclair2018}, which is not retained in the present study.
The main focus of this paper is the development and application of diagnostics within numerical simulations of oceanic convection. A key new contribution is the explicit inclusion of species mass fractions $Y_i$, which, while not directly used in the diagnostics, constitute an essential component of realistic ocean modelling and are therefore retained for generality.
Finally, all derivations of the stochastic governing equations are provided in detail in the appendices for completeness.

In this paper, all derivations are carried out without accounting for Coriolis effects, although their inclusion would be straightforward.
Section~\ref{sec:SRTT} briefly reviews the LU formalism and introduces a convenient form of the stochastic Reynolds transport theorem for conserved quantities balanced by external sources or stochastic fluxes.
In Section~\ref{sec:compressible}, we derive the stochastic compressible Navier--Stokes equations.
Section~\ref{sec:Boussinesq} applies the Boussinesq approximations, and we verify consistency with previously proposed stochastic isochoric models \citep{resseguier2017e}.
The resulting stochastic Boussinesq model is extended to incorporate thermodynamic effects, revisiting concepts introduced in \cite{tissotSTUOD2023} and complemented here with additional physical interpretation and a broader discussion of the literature.
Diagnostics of the stochastic terms in the temperature transport equation are performed in Section~\ref{sec:numerical}, based on a large-eddy simulation of convective plumes, to assess the physical relevance of stochastic compressible effects under the Boussinesq and hydrostatic approximations.
Finally, Section~\ref{sec:conclusion} provides concluding remarks and directions for future research.
Technical details regarding stochastic calculus, derivation of the equations, details relevant to energy budget diagnostics, and additional numerical considerations are provided in the appendices.

\section{Stochastic Reynolds transport theorem}
\label{sec:SRTT}
The transport of conserved quantities under stochastic displacement of fluid parcels is described by the stochastic Reynolds transport theorem (SRTT), introduced in \cite{memin2014}.
When stochastic source terms are included in the budget, additional correction terms associated with stochastic fluctuations (quadratic variation) must be taken into account \citep{resseguier2017e}.
In this section, we briefly present the modelling framework under location uncertainty and rewrite the SRTT with stochastic source terms in a form suitable for subsequent developments.

In the modelling framework under location uncertainty \citep{memin2014}, the infinitesimal displacement of a particle located at $\bm{X}_t$ at time $t$ is expressed in differential form as
\begin{equation}
    \mathrm{d}\bm{X}_t=\bm{u}(\bm{X}_t,t)\mathrm{d}t+ (\boldsymbol{\sigma}_t \mathrm{d}\bm{B}_t)(\bm{X_t}),
    \label{eq:stdisplacement}
\end{equation}
where $\bm{u} = (u,v,w)^T$ is a time-differentiable (Lagrangian) velocity component, and $\mathrm{d}\bm{B}_t$ denotes the increment of a (cylindrical) functional Wiener process \citep{DaPrato}, which models unresolved, time-decorrelated velocity contributions.
The operator $\boldsymbol{\sigma}_t$ is a correlation operator defined as an integral operator involving a spatial convolution over the domain $\Omega$, using a user-defined symmetric matrix-valued correlation kernel\footnote{Note that for a non-symmetric kernel, a noise with the same (Gaussian) law can be defined using a symmetric kernel associated with the symmetric square root of the covariance operator.} $\boldsymbol{\check{\sigma}}$, such that
\begin{equation}
\left(\boldsymbol{\sigma}_t \mathrm{d}\bm{B}t\right)^{i}(\bm{x}) = \int_\Omega \boldsymbol{\check{\sigma}}^{ij}(\bm{x}, \bm{x}', t)\mathrm{d}\bm{B}_t^j(\bm{x}')\mathrm{d}\bm{x}',
\end{equation}
where the Einstein summation convention over repeated indices is used.
For a kernel that is bounded in both time and space, the kernel integral operator $\boldsymbol{\sigma}_t$ is Hilbert-Schmidt and the noise defined above is well defined.
Associated with $\boldsymbol{\sigma}_t$, we define the (matrix-valued) variance tensor $\mathsfbi{a}$ as
\begin{equation}
\mathsfbi{a}_{ij}(\bm{x}) = \int_\Omega\boldsymbol{\check{\sigma}}^{ik}(\bm{x},\bm{x}',t) \boldsymbol{\check{\sigma}}^{kj}(\bm{x}',\bm{x},t) \mathrm{d}\bm{x}'.
\end{equation}
In the general case, this quantity corresponds to the quadratic variation of the noise:
\begin{equation}
\mathsfbi{a}_{ij} \mathrm{d} t= \covariation{\sdBs^{i}}{\sdBs^{j}}.
\end{equation}
The quadratic variation (briefly recalled in Appendix~\ref{sec:bracket}) is a stochastic  process of finite variation. In the case of a deterministic correlation operator, its infinitesimal increment can be interpreted as a one-point covariance tensor:
\begin{equation}
    \mathsfbi{a}_{ij}(\bm{x})\dd t=\mathbb{E}\left(\left(\boldsymbol{\sigma}_t \mathrm{d}\bm{B}_t\right)^{i}(\bm{x})\left(\boldsymbol{\sigma}_t \mathrm{d}\bm{B}_{t}\right)^{j}(\bm{x})\right),
    \label{eq:a}
\end{equation}
and we therefore refer to it, by a slight abuse of language, as the variance tensor.

In the following, and for the sake of simplicity, we assume that the correlation operator $\boldsymbol{\sigma}_t$ is deterministic.
However, straightforward extensions can be made to the case of a random correlation operator (see, for instance, \citet{Li-Memin-Tissot-2023}, where both $\mathsfbi{a}$ and $\boldsymbol{\sigma}_t$ are random and derived from a Dynamic Mode Decomposition -- DMD).
It is worth noting that, since these operators are  Hilbert-Schmidt, a convenient spectral representation of the noise can be obtained using the eigenfunction basis of the noise correlation operator:
\begin{equation}
   \begin{split}
      &\left(\boldsymbol{\sigma}_t \mathrm{d}\bm{B}_t\right)(\bm{x})= 
       \sum_{i=1}^{\infty} \lambda_{i}(t)^{1/2} \boldsymbol{\varphi}_i (\bm{x},t) \mathrm{d}\beta_{i,t}
      \\&\text{ and }
      \mathsfbi{a}_{ij}(\bm{x},t) =
      \sum_{k=1}^{\infty} \lambda_{k}(t)  \boldsymbol{\varphi}_{k}^{i} (\bm{x},t)  \boldsymbol{\varphi}_{k}^{j} (\bm{x},t),
   \end{split}
   \label{eq:modal_noise}
\end{equation}
where $\lambda_i$ and $\boldsymbol{\varphi}_i$ denotes respectively the positive (decreasing) eigenvalues and eigenfunctions of the noise correlation operator $\boldsymbol{\sigma}_t$, and $\dd \beta_{i,t}$ are increments of independent scalar Wiener processes.
This is the strategy that we follow in Section~\ref{sec:numerical} to estimate the noise and its variance from model outputs, with Fourier modes taken as the eigenfunctions for the spectral decomposition.

Within this framework, the stochastic transport operator of a scalar quantity $q$ is defined by
(\citet{memin2014} -- see also \citet{bauer2020a} for a simplified and comprehensive  exposition dedicated to geophysical fluids)
\begin{equation}
        \mathbb{D}_t q
        \triangleq \,
        \dd_tq
        +\adv{\underbrace{\bigl(\bm{u}-\frac{1}{2}\Div\mathsfbi{a}+\sds\bigr)}_{\udrift}}{q}\dt
        +\Adv{\sdBt}{q}
        -\Diffst{q}\dt,
        \label{eq:stochastic_transport}
\end{equation}
where $\udrift$, referred to as the \emph{drift velocity}, corresponds to the resolved velocity corrected for the effects induced by both the inhomogeneity and the divergence of the noise correlation tensor, respectively.
The drift correction, $\bm{u}^\star-\bm{u}$,
is termed the Itô-Stokes drift, due to its connection with the wave induced Stokes drift \citep{bauer2020a}.
The second term in the drift velocity involves the quantity $\sds$ defined as the matrix kernel product
\begin{equation}
    \bigl(\sds\bigr)^{i}(\bm{x}) = \int_\Omega \boldsymbol{\check{\sigma}}^{ik}(\bm{x},\bm{x}',t) \partial_{x^{j}} \boldsymbol{\check{\sigma}}^{jk}(\bm{x}',\bm{x},t)\mathrm{d}\bm{x}',
\end{equation}
which is associated with the divergence of the noise.
The last term of the transport operator corresponds to a noise-induced diffusion and the noise term takes the form of an advection term often referred to as transport noise in the literature.

It can be verified, by applying the It\^{o} integration by parts formula (see Appendix~\ref{sec:bracket} for a brief recap on covariation and It\^o integration by parts) that under suitable boundary conditions and for divergence-free noise, the $L_2$ norm of $q$ is preserved path-wise (that is, for each realisation of the noise).
In this case, the energy injected by the stochastic transport term (via its covariation) precisely compensates the dissipation induced by the stochastic diffusion term \citep{bauer2020a,resseguier2017e}.
This balance can be interpreted as an immediate instance of a fluctuation-dissipation relationship between the transport noise (whose energy is expressed through its covariation) and the stochastic diffusion.
Note that this balance is seen only in the It\^{o} setting.
For Stratonovich integral it is implicit and reveals itself only when expressing expectation and variance -- which requires to go back to It\^{o} integral.
Physical relevance of the drift velocity and the stochastic diffusion $\Diffst{q}\dt$ has been extensively highlighted in previous studies \citep[\eg][]{bauer2020a,chandramouli2018,pinier2019}.
Mathematically, the emergence of such a term together with a transport noise has been also verified as a suitable limit of  a coupled system of 3D slow/fast Navier--Stokes equation with additive Brownian forcing on the fast component \citep{Debussche-Pappalattera2023,Debussche-Hofmanova2023}.
In the wake of F. Flandoli program, transport noise has been recently the subject of intensive research works due to well posedness properties \citep{Agresti-et-al-2022,Brzezniak-Slavik-2021,Crisan-Flandoli-Holm-2019,Debusshe-Hug-Memin-2023,Flandoli-et-al-10,Flandoli-Galeati-Luo-2021, Flandoli-Luo-2021, Goodair-et-al2022, Lang2023,Mikulevicius-Rosovskii-2005} of several fluid dynamics models as well as due to the emergence of enhanced dissipation and mixing \citep{Flandoli:2022sb,Flandoli-Russo2023}.
Transport noise has been also exploited with variational formulation \citep{Debussche-Memin25,Holm2015,street-crisan2021}.
Compressible stochastic systems have been recently studied in~\citet{breit-et-al-2021} for barotropic fluids and \citet{boadi2025} for isentropic Euler equations, both hypotheses allowing to avoid the derivation of an energy equation.
The systems analysed in these works are comparable to ours.
In this study, exploiting the modelling capabilities of the LU framework, we derive, from physical principles, the thermodynamic evolution equations for active tracers.
These evolution equations are derived from the stochastic Reynolds transport theorem and from evaluations of the associated work terms.
Their derivation is described below.

We now aim to derive evolution equations for conserved quantities subject to forcing terms.
The variation of $q$, integrated over a transported volume $\mathcal{V}(t)$,  can be expressed as the following budget
\begin{equation}
    \dd \int_{\mathcal{V}(t)} q \dx
    =  \int_{\mathcal{V}(t)}\left(Q_t\dt + \bm{Q}_\sigma\bcdot\dBt\right)\dx.
   \label{eq:start}
\end{equation}
The right-hand side (RHS) terms correspond to forcing (respectively, work) when this general expression is associated with the momentum (respectively, energy) equation.
The real-valued source terms $Q_t\dt + \bm{Q}_\sigma\bcdot\dBt$, which may arise from physical sources or fluxes,  represent, respectively, a time-differentiable component and a martingale contribution defined through a vector-valued kernel $\check{\boldsymbol{Q}}$.
The product, $\bm{Q}_\sigma\bcdot\dBt$ is to be interpreted as the inner product between the vector-valued kernel, $\check{\boldsymbol{Q}}$, and the Brownian vector $\dBt$:
\begin{equation}
\label{Q-prod}
(\bm{Q}_\sigma\bcdot\dBt) (\bm{x}) = \int_\Omega   \check{\boldsymbol{Q}}^{j}(\bm{x},\bm{y},t) \dBt^j(\bm{y})  \mathrm{d}\bm{y}.
\end{equation}

Even though the operator $\bm{Q}_\sigma$ is assumed to be deterministic in the applications considered in this paper, it could be extended to stochastic process under the condition that $\dd \bm{Q}_\sigma$ is differentiable in the Lagrangian frame.
Under this assumption, the stochastic Reynolds transport theorem takes the explicit form:
\begin{equation}
    \dd_t q
    + \Div\Bigl(\bigl((\bm{u}-\frac{1}{2}\Div\mathsfbi{a})\dt+\sdBt\bigr) q\Bigr)
    + \Div(\boldsymbol{\sigma}_t\bm{Q}_\sigma)\dt
    -\Diffst{q}\dt
    =
    Q_t\dt + \bm{Q}_\sigma\bcdot\dBt
    .
    \label{eq:SRTT_transport}
\end{equation}
The demonstration is provided in Appendix~\ref{sec:strat-ito}.

The absence of the term $\sds$ in the modified drift is a notable feature of this formulation.
It is not neglected; rather it cancels, when rewriting
$\strat{\boldsymbol{\sigma}_t}{\dBt}$ in It\^o form.
As will be detailed below, this contribution reappears when transforming the conservative form of the equations into the corresponding non-conservative form, \ie when deriving transport equations for the primitive variables.

From a modelling perspective, several approaches to specifying the noise have recently been proposed in the literature. We provide below a brief overview of the current state of the art.

Early models introduced noise based on the assumption of homogeneous, stationary turbulence \citep{resseguier2017e}.
Building on the modal representation of the noise~\eqref{eq:modal_noise}, spectral methods based on Empirical Orthogonal function (EOF) -- also referred to as Proper Orthogonal Decomposition (POD) -- have been developed to account for stationary, inhomogeneous turbulence across various flow regimes \citep{bauer2020b, brecht2021, Cotter-et-al-2020, Gugole-Franzke-2019}.

To introduce temporal evolution in the noise and ensure its interaction with the resolved scales, projection techniques have been proposed to enforce a steering of the small-scale dynamics by the large-scale flow \citep{Li-Memin-Tissot-2023, tucciarone2023}.
Advanced models for wave-dominated flows, leveraging resolvent analysis and spectral proper orthogonal decomposition \citep{towne2018}, have also been derived \citep{tissotJFM2021, Tissot-et-al-2023}.
In the context of the shallow-water wave model, the interaction between large-scale waves (\eg internal waves) and stochastic perturbations has been investigated in \citet{memin2023linear}.

Methods based on dynamic mode decomposition have been proposed in \citet{Li-Memin-Tissot-2023}.
Furthermore, a technique using a Girsanov change of measure has been developed to handle non-centered noise with slow components \citep{Li-Memin-Tissot-2023}, or to guide the dynamics toward available observations \citep{Dufee-Memin-Crisan-2023}.

Efficient noise models assuming self-similarity have been introduced in \citet{brecht2021, kadriharouna2017}.
Comparative studies assessing different noise models using statistical criteria in simplified ocean dynamics frameworks are reported in \citet{brecht2021, Li-Memin-Tissot-2023, resseguier2021}.

Finally, the possibility of representing various unresolved physical processes (such as waves, turbulence, deep convection, \textit{etc.}) through a combination of different noise models remains an open and promising direction for future research.

\paragraph*{}
{\bf Remark on the decorrelation assumption}: The LU framework relies on a decorrelation assumption for the small-scale components.
This mathematical idealisation enables a clean differential calculus for stochastic processes and, more importantly, leads naturally  to the emergence of time-correlated terms associated with covariations between these processes.
These terms -- corresponding to additional fluctuation-induced drifts or diffusion -- are often introduced empirically in large-scale models;
here, they arise directly from calculus.
This decorrelation assumption has been recently relaxed in an incompressible setting using a variational approach \citep{Debussche-Memin25}.
Extending such a framework to compressible flows remains significantly more challenging and will be the subject of future investigations.

\section{Stochastic compressible Navier--Stokes equations}
\label{sec:compressible}
To derive the stochastic compressible Navier--Stokes equations, we first apply the stochastic Reynolds transport theorem (SRTT)~\eqref{eq:SRTT_transport} to the mass, mass of species, momentum, and total energy equations.
This procedure requires a  precise definition of the physical variables involved.

\subsection{Adimensionalisation}\label{subsec:adim}
We consider  time $t$, spatial coordinate $\bm{x}=(x,y,z)^T$ in the domain $\Omega$, and the associated Euclidean canonical basis $(\bm{e}_x,\bm{e}_y,\bm{e}_z)$.
Physical quantities are denoted by a superscript $\bullet^\phi$, while other quantities are dimensionless.
Adimensionalisation is performed using reference values denoted by the subscript: $\bullet_\text{\tiny ref}$.
We introduce the characteristic length $L_\text{\tiny ref}$, velocity $u_\text{\tiny ref}$, density $\rho_\text{\tiny ref}$, sound speed $c_\text{\tiny ref}$ and viscosity $\mu_\text{\tiny ref}$.
This yields the following non-dimensional variables:
\begin{equation}
   \begin{split}
      &\bm{x}=\frac{\bm{x}^\phi}{L_\text{\tiny ref}}
      \,;\quad
      t=\frac{t^\phi{u_\text{\tiny ref}}}{L_\text{\tiny ref}}
      \,;\quad
      \bm{u}=\frac{\bm{u}^\phi}{u_\text{\tiny ref}}
      \,;\quad
      c=\frac{c^\phi}{u_\text{\tiny ref}}
      \,;\quad
      M=\frac{u_\text{\tiny ref}}{c_\text{\tiny ref}}
      \,;\quad
      \rho=\frac{\rho^\phi}{\rho_\text{\tiny ref}}
      \,;\quad
      \mu=\frac{\mu^\phi}{\mu_\text{\tiny ref}}
      \,;\\
      &
      p=\frac{p^\phi}{\rho_\text{\tiny ref} u_\text{\tiny ref}^2}
      \,;\quad
      T=\frac{T^\phi c_p^\phi}{u_\text{\tiny ref}^2}
      \,;\quad
      \gamma = \frac{c_p^\phi}{c_v^\phi}
      \,;\quad
      e=\frac{e^\phi}{u_\text{\tiny ref}^2}=\frac{T}{\gamma}
      \,;\quad
      \bm{g}=\frac{\bm{g}^\phi L_{\text{\tiny ref}}}{u_\text{\tiny ref}^2}
      ,
   \end{split}
   \label{eq:adim}
\end{equation}
where $\bm{u}$ is the velocity vector, $c$ the speed of sound, and $M$ the Mach number (\ie the ratio of typical particle velocity to typical sound speed).
The quantities $\rho$,  $p$, $\mu$ denote the density, the pressure and  dynamic viscosity, respectively.
We denote by $T$ the temperature, $\gamma$ the heat capacity ratio, $(c_p,c_v)$ the heat capacities at constant pressure/volume, $e$ the specific internal energy and $\bm{g}=-g\bm{e}_z$ stands for the acceleration vector due to gravity.
We also consider the mass fraction $Y_i$ of the species~$i$.
Salinity $S$ discussed in Section~\ref{sec:Boussinesq} is a special case of such a mass fraction transported by the flow.
Finally, we introduce the Reynolds, Prandtl ans Schmidt numbers:
\begin{equation}
    Re=\frac{\rho_\text{\tiny ref} u_\text{\tiny ref} L_\text{\tiny ref}}{\mu_\text{\tiny ref}} 
    \quad;\quad
    Pr=\frac{c_p^\phi\mu^\phi}{k_T^\phi} 
    \quad;\quad
    Sc_i=\frac{\mu^\phi}{\rho^\phi D_i^\phi} ,
\end{equation}
where $k_T^\phi$ is the thermal conductivity, and $D_i$ the mass diffusivity of species $i$.

\subsection{Conservative formulation}
Applying the stochastic Reynolds transport theorem (SRTT) equation~\eqref{eq:SRTT_transport} to density ($q=\rho$), mass concentration of a species $i$ ($q=\rho Y_i$), momentum per unit volume ($q=\rho\bm{u}$) and total energy density ($q=\rho E$) leads to the system written in conservative form:
\begin{multline}
    \dd_t \rho
    + \Div\Bigl(\bigl((\bm{u}-\frac{1}{2}\Div\mathsfbi{a})\dt+\sdBt\bigr)\rho\Bigr)
    =
    \Diffst{\rho}\dt.
    \\
    \shoveleft{
    \dd_t (\rho Y_i)
    + \Div\Bigl(\bigl((\bm{u}-\frac{1}{2}\Div\mathsfbi{a})\dt+\sdBt\bigr)\rho Y_i\Bigr)
    + \Div\bigl(\boldsymbol{\sigma}_t(\rho\bm{Q}_{\sigma}^{Y_i})\bigr)\dt
    =
    \Diffst{\rho Y_i}\dt
    }
    \\
    + \frac{1}{ReSc_i}\Div\bigl(\Grad (\rho Y_i)\bigr)\dt
    + \rho Q_t^{Y_i} \dt
    + \rho \bm{Q}_{\sigma}^{Y_i}\bcdot\dBt.
    \\
    \shoveleft{
    \dd_t(\rho u_i)
    + \Div\left(\left((\bm{u}-\frac{1}{2}\Div\mathsfbi{a})\dt+\sdBt\right)\rho u_i\right)
    + \Div(\boldsymbol{\sigma}_t\bm{F}_\sigma^{\rho u_i})\dt
    =
    -\ddt{p}{x_i} \dt
     - \ddt{\dd p_t^\sigma}{x_i}-\rho g \delta_{i3}\dt
     }
    \\
    + \Rem\ddt{\tau_{ij}(\bm{u})}{x_j}\dt + 
     \Rem\ddt{\tau_{ij}(\sdBt)}{x_j}
    +\Diffst{(\rho u_i)}\dt
    \\\shoveleft{
     \dd_t (\rho E)
     + \Div\left(\left((\bm{u}-\frac{1}{2}\Div\mathsfbi{a})\dt+\sdBt\right)\rho E\right)
     + \Div(\boldsymbol{\sigma}_t\bm{F}^{\rho E}_\sigma)\dt
     =
     \Diffst{(\rho E)}\dt
     }
     \\
    +\dd W- \Div(\dd \bm{q}).
    \label{eq:conservative}
\end{multline}
A semi-martingale source $Q_t^{Y_i}\dt+\bm{Q}_{\sigma}^{Y_i}\bcdot\dBt$, with $\bm{Q}_{\sigma}^{Y_i}$ a correlation operator defined via a vector-valued kernel, is considered.
This source may arise from physico-chemical reactions or dilution effects.
For simplicity, we do not consider additional diffusive mass fluxes related to differences in molecular properties between species, such as the Soret effect \citep[which is induced by differences in thermal diffusivity between species under a temperature gradient;][]{mortimer1980}.
However, such effects could be included without conceptual difficulty.

In the above equation we define
\[
\bm{F}_\sigma^{\rho u_i}\bcdot\dBt
= \delete{-\rho u_i \Div(\sdBt)} -\ddt{\dd p_t^\sigma}{x_i} +\Rem\Div(\tau_i(\sdBt))
.\]
Throughout the paper, we adopt the following notational convention: the martingale right-hand-side forcing term of a transport equation of any variable $q$ is written as $\bm{F}_\sigma^q\bcdot\dBt$, where $\bm{F}_\sigma^q$ is a correlation operator expressed through a (possibly time dependent) vector-valued kernel.
In line with previous correlation tensor product expressions, the product ($\boldsymbol{\sigma}_t \bm{F}_{\sigma}^{\rho u_i}$) is understood as a matrix vector kernel product.

The forces involved here are caused by pressure gradient, viscous stresses $\mathsfbi{\tau}$ and gravity ($\delta_{i3}$ being non-null only for the vertical component).
The pressure gradient impulse is decomposed into a time-differentiable part $-\partial p/\partial x_i\dt$ and a martingale component $-\partial \dd p_t^\sigma/\partial x_i$.
The precise form of the viscous stress $\mathsfbi{\tau}$, of the elementary work of the forces $\dd W$ (pressure and viscous) and the heat fluxes $\dd \bm{q}$ (mass and thermal diffusions, and external heat source) are detailed in Appendix~\ref{sec:details_non_conservative}.

The system~\eqref{eq:conservative} is complemented by an equation of state
$p=f(\rho,T,Y_i)$, from which the random pressure $\dd p_t^\sigma$ can be obtained by differentiation (see Appendix~\ref{sec:details_non_conservative} for details).
In the context of oceanic flows, however, the equation of state is more commonly formulated in terms of density rather than pressure.
A formulation adapted to such flows is provided in Section~\ref{sec:Boussinesq}.

The conservative formulation~\eqref{eq:conservative} follows directly from the SRTT and is well suited for numerical discretisations based on conservation principles, such as finite-volume or discontinuous Galerkin methods.
In the deterministic setting, alternative formulations of the energy equation are often used, involving enthalpy, entropy -- derived from the second law of thermodynamics \citep[e.g.][]{chassaingbook,vallisbook} --, pressure (for isentropic flows in aeroacoustics applications), or related variables.
However, the algebra underlying these transformations in the deterministic case does not generally lead to simple or convenient expressions in the stochastic setting, due to the presence of quadratic variation terms.
In particular, transport equations for entropy are sometimes introduced under specific assumptions. In the general case, the entropy transport equation is obtained by combining the energy transport equation with the fundamental thermodynamic relation (the Gibbs relation; see, for instance, \citealt[p.~229]{landaubook}).
We do not derive an entropy transport equation in this work, but it could be obtained by following these classical steps.

\subsection{Definition of the energy}
At this stage, the form of the total energy must be specified, as it is closely tied to the physical mechanisms at play.
In the present study, we consider the total energy density to be
\begin{equation}
\rho E = \rho ( e + \frac{1}{2}\|\bm{u}\|^2 + gz ),
\end{equation}
\ie the product of the density and the sum of the specific internal energy  $e=\frac{T}{\gamma}$, the specific kinetic energy $\frac{1}{2}\|\bm{u}\|^2$ and the gravitational potential energy per unit mass $gz$.
In this definition, the kinetic energy involves only the resolved scales.
The kinetic energy of the small-scale component is excluded, as it is pathwise ill-defined (potentially infinite).
As will be discussed below, we assume that the work associated with the noise is balanced by variations of its kinetic energy, corresponding to rapidly oscillating contributions that appear as higher-order terms.
Nevertheless, the influence of the small-scale velocity on all components of the total energy is accounted for through the stochastic Reynolds transport theorem (SRTT) and through the evaluation of the associated work terms.

\subsection{Definition of the work of forces and heat fluxes}
We now detail the heat fluxes and the works associated with the different forces.

The work due to the time-dif\-fer\-enti\-able pressure represents how pressure acts through the displacement of the control surface.
It is obtained by integrating the pressure force, multiplied by the surface displacement, over a transported control volume and applying Green's formulae.
The resulting expression is consistent with the deterministic formulation, but additionally involves the drift velocity, as detailed in Appendix~\ref{sec:displacement_control_surface}.
The pressure work is given by
\begin{equation}
    \begin{split}
    \int_{\Omega(t)}\dd W_{p}\,\dx=&\int_{\partial \Omega(t)} \left(-p\,\bm{n}\,\dd S\right) \bcdot (\udrift\dd t +\sdBt)
               \\ =& - \int_{\Omega(t)} \Div \bigl(p\, (\udrift\dd t +\sdBt)\bigr)\,\dx.
    \end{split}
\end{equation}
The minus sign arises from the convention of using the outward unit normal vector $\bm{n}$ convention.
We can then identify
\begin{equation}
    \dd W_{p}= - \Div \bigl(p\, (\udrift\dd t +\sdBt)\bigr).
        \label{pressure-work}
\end{equation}
Similarly, the work associated with the viscous stress of the resolved component is given by
\begin{equation}
    \dd W_{\tau} = \Rem\Div \bigl(\mat{\tau}(\bm{u})\left(\udrift\dt+\sdBt\right)\bigr).
        \label{viscous-work}
\end{equation}
Following Section~\ref{sec:calc}\ref{sec:work_random}, we also account for the work of the random pressure,
\begin{equation}
    \dd W_{rp} = -\Div{\left(\udrift\dd p_t^\sigma\right)},
\end{equation}
and the work of the random viscous stress,
\begin{equation}
    \dd W_{r\tau}= \frac{1}{Re}\Div{\left(\mat{\tau}(\sdBt)\udrift\right)}.
            \label{random-viscous-work}
\end{equation}
As discussed rigorously in Section~\ref{sec:calc}\ref{sec:work_random}, we do not include the work of the random forces directly associated with $\sdBt$, as such highly irregular (in time) contributions are assumed to be balanced by variations in the kinetic energy of $\sdBt$.

There is no net work contribution from gravity to the total energy, as the gain in kinetic energy induced by the gravitational force is exactly compensated by a corresponding loss in potential energy.
This reflects the conservative nature of the gravity force.

The sum of all the above work contributions yields the global work:
\begin{equation}
 \dd W= \dd W_{p} +\dd W_{rp} + \dd W_{\tau} +  \dd W_{r\tau}.
\end{equation}

In presence of several species, the energy budget is affected by mass diffusion when the species have different partial heat capacities $c_{v,i}$ (such that $c_v = \sum_i Y_ic_{v,i}=\frac{1}{\gamma}$).
This results in an additional flux term:
\begin{equation}
   \dd \bm{q}_{Y_i} = -\sum_i \frac{1}{Re Sc_i} (\rho Y_i e_i) \Grad Y_i \dd t,
\end{equation}
where the internal energy of the species $i$ is given by  $e_i = c_{v,i} T$.
The Dufour effect \citep[\ie a heat flux driven by gradients in chemical potential gradient;][]{mortimer1980} is neglected in the present study, but could be readily included if needed.

The thermal conductivity is introduced by expressing the heat flux using Fourier's law:
\begin{equation}
 \dd \bm{q}=-\frac{\rho}{RePr}\nabla T\dd t.
 \label{thermal-conductivity}
\end{equation}
Finally, we consider a time-differentiable spatially distributed heat flux term $\dot{Q}\dt$, which accounts for all other possibles heat sources, such as hydrothermal vents in the ocean, latent heat in the atmosphere, or radiative effects, as in Section~\ref{sec:numerical}.
These effects are essential for the energy balance and influences the vertical dynamics, especially during convective events.
This source term is written here as a general spatially distributed forcing, but it can also be confined to the surface boundary, as is the case in the numerical results of Section~\ref{sec:numerical}.

All ingredients are now in place for the derivation of a set of transport equations for the primitive variables, as shown in Section~\ref{subsec:non-conserv}.

\subsection{Non-conservative formulation}\label{subsec:non-conserv}
The conservative formulation has the drawback of requiring nonlinear transformations to recover the primitive variables $(\rho,\bm{u},T)$, which complicates subsequent approximations (Boussinesq, hydrostatic, linearisation, \textit{etc.}) and processing steps.
Many fluid mechanics and ocean models are instead expressed directly in terms of primitive variables. Prominent examples include operational codes implementing the primitive equations, such as NEMO \citep{madec2017} and CROCO (used and presented in Section~\ref{sec:numerical})\footnote{Conservative formulations may nevertheless be reintroduced after approximations for numerical discretisation purposes.}.
Following a similar approach to that used in the deterministic framework (see Appendix~\ref{sec:details_non_conservative}), we derive the non-conservative form of the stochastic compressible Navier--Stokes equations:
\begin{subequations}
\begin{align}
         &
         \dd_t \rho
        + \Div\Bigl(\bigl((\bm{u}-\frac{1}{2}\Div\mathsfbi{a})\dt+\sdBt\bigr)\rho\Bigr)
        =
        \Diffst{\rho}\dt.
        \label{eq:non_conservative_rho}
        \\&
        \mathbb{D}_tY_i
        +\sum_k\covariation{\left(\sdBs\right)^k}{\ddt{}{x_k}\bm{Q}_\sigma^{Y_i}\bcdot\dBs}
        =
        \frac{1}{ReSc_i}\Div(\Grad Y_i)\dt
        + Q_{Y_i} \dt
        + \bm{Q}_{\sigma}^{Y_i}\bcdot\dBt.
        \label{eq:non_conservative_Y}
        \\&
        \rho\,\mathbb{D}_tu_i
        +\rho\sum_k\covariation{\left(\sdBs\right)^k}{\ddt{}{x_k}\bm{F}_\sigma^{u_i}\bcdot\dBs}
        \nonumber\\&\qquad
        =-\ddt{p}{x_i} \dt - \ddt{ \dd p_t^\sigma}{x_i} + \Rem\ddt{\tau_{ij}(\bm{u})}{x_j}\dt + \Rem\ddt{\tau_{ij}(\sdBt)}{x_j}-\rho g \delta_{i3}\dt,
        \label{eq:non_conservative_u}
        \\&
         \frac{\rho}{\gamma} \mathbb{D}_tT
         + \underbrace{\frac{\rho}{\gamma} \sum_k\covariation{{\sdBs}^k }{ \ddt{}{x_k}\left(\bm{F}^T_\sigma\bcdot{\dBs}\right)}}_{A_T}
         +
         \underbrace{
         \frac{\rho}{2}
         \sum_i
        {\covariation{ \bm{F}_\sigma^{u_i}\bcdot{\dBs} }{ \bm{F}_\sigma^{u_i}\bcdot{\dBs} }}}_{A_u}
        \nonumber\\&\qquad
        = -\underbrace{p\Div(\udrift\dd t +\sdBt)}_{P_t}
          -\underbrace{\dd p_t^\sigma\Div\udrift}_{P_\sigma}
          +\underbrace{\Rem\mathsfbi{\tau}(\bm{u}):\Grad\left(\udrift \dt +\sdBt\right)}_{V_t}
          +\underbrace{\Rem\mathsfbi{\tau}(\sdBt):\Grad\udrift}_{V_\sigma}
        \nonumber\\&\qquad
          +\underbrace{{\left((\udrift-\bm{u})\dt+\sdBt\right)\bcdot\left(-\Grad p + \rho\bm{g} + \Rem\Div\mathsfbi{\tau}(\bm{u})\right)} \delete{- \rho g \left(\sdBt\right)_z}}_{D_t}
          \nonumber\\&\qquad
          +\underbrace{{\left(\udrift-\bm{u}\right)\bcdot\left(-\Grad \dd p_t^\sigma + \Rem\Div\mathsfbi{\tau}(\sdBt)\right)
}}_{D_\sigma}
        \nonumber\\&\qquad
        +\sum_i \frac{1}{Re Sc_i} \Div \left((\rho Y_i c_{v,i}T) \Grad Y_i\right) \dd t
        +\frac{\rho}{RePr}\Div(\Grad T)\dt
        + \dot{Q}\dt,
    \label{eq:non_conservative_T}
\end{align}
\label{eq:non_conservative}
\end{subequations}
with the martingale of the right-hand sides
\begin{equation}
    \begin{split}
        \bm{F}_\sigma^{u_i}\bcdot\dBs &= \frac{1}{\rho}\left(-\ddt{\dd p_s^\sigma}{x_i} +\Rem\ddt{\tau_{ij}(\sdBs)}{x_j}\right)
        \\
        \frac{\rho}{\gamma} \bm{F}^T_\sigma\bcdot\dBt
        =&
        -p\Div\sdBt
        -\dd p_t^\sigma\Div\udrift
        +\Rem\mathsfbi{\tau}(\bm{u}):\Grad\sdBt
        +\Rem\mathsfbi{\tau}(\sdBt):\Grad\udrift
        \\&
        +\sdBt\bcdot\left(-\Grad p + \rho\bm{g} + \Rem\Div\mathsfbi{\tau}(\bm{u})\right)
        +\left(\udrift-\bm{u}\right)\bcdot\left(-\Grad \dd p_t^\sigma + \Rem\Div\mathsfbi{\tau}(\sdBt)\right).
    \end{split}
\end{equation}

The set of equations~\eqref{eq:non_conservative}, which involves now the stochastic transport operator \eqref{eq:stochastic_transport}, describes the respective evolution of the density~\eqref{eq:non_conservative_rho}, species mass fractions~\eqref{eq:non_conservative_Y}, momentum~\eqref{eq:non_conservative_u}, and temperature~\eqref{eq:non_conservative_T}. This system is complemented by the equation of state.
The continuity equation has an intuitive form, in which the velocity term gathers the large-scale velocity, the noise-induced component, and the small-scale Brownian contribution, together with a noise-induced diffusion term on the left-hand side.

The species mass fractions satisfy an advection–diffusion equation with semimartingale forcing (\textit{i.e.} combining time-differentiable and Brownian components).
In addition, a covariation term appears between the small-scale velocity and the martingale forcing; this term arises from stochastic calculus rules.

The expression~\eqref{eq:non_conservative_u} closely resembles the momentum equation of the incompressible Navier--Stokes equations \citep[{eq. 41 with incompressibility assumption}]{memin2014}.
The only additional term is the covariation between the (martingale part of the) forcing and the small-scale component~$\sdBt$.
This term, which is usually difficult to evaluate analytically, can be neglected on physical and modeling grounds, as, for instance, in~\citet[I, Appendix~E]{resseguier2017e}.

In equation~\eqref{eq:non_conservative_T}, we recover all terms present in the deterministic framework, now expressed using the stochastic transport operator in place of the classical material derivative.
However, additional covariation terms emerge due to the stochastic formulation.

In particular, the term $A_T$ arises from the work of the martingale components of the fluxes (both mechanical and thermal), while $A_u$ reflects the increase in kinetic energy due to covariations of the martingale forcing in the momentum equation.
On the right-hand side, we remark that the drift velocity contributes to the compression/dilatation work of both the time-differentiable pressure (in $P_t$) and the martingale pressure (in $P_\sigma$), in agreement with the control-surface displacement analysis performed in Appendix~\ref{sec:displacement_control_surface}.
Finally, we note that in the case of isochoric flow, the conditions $\Div \udrift=0$ and $\Div \sdBt = 0$ consistently imply the absence of compression/dilatation work.

The terms $V_t$ and $V_\sigma$ represent the work done by viscous stresses that are, respectively, smooth in time and of martingale type.
Additionally, the terms $D_t$ and $D_\sigma$ correspond to the works resulting from the alignment of the drift and small-scale martingale velocities with the applied forces.
We refer to $D_t$ and $D_\sigma$ collectively as \emph{drift works}.

Focusing on the term $-(\udrift-\bm{u})\bcdot\Grad p$, we interpret this specific drift work as analogous to baropycnal work \citep{aluie2013}, a concept arising in the context of compressible large eddy simulation (LES).
In standard compressible LES, baropycnal work arises from the alignment between the large-scale pressure gradient and the Reynolds stress generated by correlations between small-scale density and velocity fluctuations
 -- specifically, the term $\frac{1}{\overline{\rho}}\Grad \overline{p} \bcdot  \overline{\rho'\bm{u}'}$, where $\overline{\bullet}$ denotes a large-scale filtering and $\bullet'$ the small-scale components.
The quantity $\frac{1}{\overline{\rho}} \overline{\rho'\bm{u}'}$ has the dimensions of a velocity.
In our stochastic framework, the analogous effective displacement induced by small-scale effects is captured directly by the drift velocity over a finite time interval.
This interpretation extends naturally to the other drift work terms, as well as to those associated with viscous stresses and stochastic (martingale) contributions.

In the presence of gravity, the vertical component of the drift work takes the form
\begin{equation}
((w^\star-w)\dt + \left(\sdBt\right)_z )\left(-\ddt{p}{z}-\rho g\right).
\end{equation}
In oceanic applications, it is customary to decompose the pressure into hydrostatic and non-hydrostatic components, $p=p_H + p_{NH}$ and to express the density as $\rho=\rho_0+\rho'$, where $\rho_0$ is a constant reference density and $\rho'$ the deviation.
Under these decompositions, the vertical drift-work component becomes
\begin{equation}
      ((w^\star-w)\dt+\left(\sdBt\right)_z)\left(-\ddt{p_{NH}}{z}-\rho' g\right)
      =\rho_0((w^\star-w)\dt+\left(\sdBt\right)_z)\left(-\frac{1}{\rho_0}\ddt{p_{NH}}{z} +b \right)
   , 
\end{equation}
where $b=-\rho'g/\rho_0$ denotes the buoyancy.
This expression makes explicit the contribution of vertical small-scale mass fluxes interacting with non-hydrostatic pressure gradients and buoyancy effects.
Notably, the hydrostatic pressure component $p_H$ does not appear in the drift work, as it is exactly balanced by the background stratification under hydrostatic equilibrium.
This drift work term is therefore expected to be relatively small under the Boussinesq approximation.
A finer decomposition of pressure and density will be considered in Section~\ref{sec:Boussinesq} in the context of the Boussinesq approximation.

It is also worth noting that, in the special case of divergence-free homogeneous noise -- where the variance tensor is spatially uniform -- the drift term vanishes identically,  as $\udrift-\bm{u} = 0$ cancels, leading to zero drift work.

It has been shown in \cite{tissotSTUOD2023} that performing the low Mach approximation on the system~\eqref{eq:non_conservative}, allows us to recover the incompressible stochastic Navier--Stokes equations originally proposed by \citet{memin2014}.
In the following, we apply the Boussinesq assumption to derive ocean circulation models.

\section{Boussinesq-hydrostatic approximation}
\label{sec:Boussinesq}

In this section, starting from the stochastic compressible Navier--Stokes equations, we derive the Boussinesq approximation by considering small density fluctuations.
These fluctuations are neglected, except when they appear in combination with $\bm{g}$, leading to the classical definition of buoyancy.
We also apply the hydrostatic approximation, using the standard aspect-ratio scaling  $D=H/L_{\text{\tiny ref}}\ll 1$, where $H$ is the characteristic water depth.
At the scales of interest, we neglect molecular diffusion effects in the equations for species, momentum, and temperature.
For simplicity, we do not account for rotational effects; however, the Coriolis pseudo-force could be straightforwardly incorporated following the approach of \citet{tucciarone2023}.
The vertical coordinate $z\in[-H,\eta]$ ranges from the bottom boundary at ($-H$) to the free surface ($\eta$).
The density is expanded asymptotically as
\begin{equation}
    \rho=\rho_0 + \epsilon \rho_1(z)+\epsilon\rho_2(x,y,z,t)+o(\epsilon),
    \label{eq:density_expansion}
\end{equation}
where $\rho_1(z)$ represents the horizontally and time averaged stratification term, and $\epsilon\ll 1$ is a small parameter.
Importantly, we do not require $\rho_1$ to dominate $\rho_2$.
The buoyancy is defined as $b=-\epsilon\,g\rho_2/\rho_0$.

The pressure field is decomposed as follows:
\begin{equation}
    p=\underbrace{p_0(z)}_{\mathcal{O}(1/\epsilon)}+ p_1 + p_2 + \mathcal{O}(\epsilon),
\end{equation}
where $p_0$ and $p_1$ satisfy hydrostatic balance:
\begin{equation}
    \ddt{p_0}{z}=- g\rho_0
    \quad \text{and} \quad
    \ddt{p_1}{z}=-\epsilon g\rho_1.
\end{equation}
The pressure terms $p_2$ and the martingale pressure $\dd p_s^\sigma$  are determined through a generalisation of the hydrostatic balance, which accounts for some of the  non-hydrostatic effects by balancing, in the vertical momentum equation, the vertical pressure gradient with buoyancy, stochastic diffusion, corrective drift, and stochastic advection of the vertical velocity component~$w$.

We consider a regime in which the hydrostatic approximation in the deterministic framework is only marginally valid -- that is, it holds approximately but is close to breaking down.
In this intermediate regime, sufficiently strong noise can invalidate the hydrostatic assumption.
From this perspective, the goal is to model weak non-hydrostatic effects via stochastic modelling.
By scaling analysis (assuming a weak aspect ratio and noise with large amplitude), the terms  $\dd_tw$ and the advective term $\Adv{\bm{u}}{w}$ are neglected, while the vertical velocity advection by the noise and the noise-induced terms are retained.
More precisely, denoting by $L^\sigma$ the characteristic amplitude scale of $\sdBt$, the advection of $\sdBt$ cannot be neglected if\footnote{If the frame rotation is taken into account, the ratio $R_o/Bu$ (Rossby over Burger) is additionally involved, but does not alter the existence of  this intermediate regime.}
\begin{equation*}
L^\sigma/L_{\text{\tiny ref}}\sim 1/(Fr\, D)^2,
\end{equation*} where $Fr=u_{\text{\tiny ref}}/(N H)$ is the Froude number.
Moreover, stochastic diffusion and drift velocity effects become significant when
 \begin{equation*}
 (L^\sigma/L_{\text{\tiny ref}})^2\tau/T_{\text{\tiny ref}}\sim  1/(Fr\, D)^2,
 \end{equation*} where $\tau$ denotes the decorrelation time.
Since $\dd_tw$ is neglected, martingale and time-differentiable terms can be cleanly separated, allowing the remaining pressure terms to be computed by vertical integration similarly to the classical hydrostatic scheme:
\begin{equation}
    \begin{split}
        &p_2=-\rho_0\int_z^\eta  b + \Adv{\frac{1}{2}\Div\mathsfbi{a}}{w}+\Diffst{w} \,\dd z ,
        \\
        &\dd p_t^\sigma=\rho_0 \int_{-H}^{z} \Adv{\sdBt}{w} \,\dd z.
    \end{split}
    \label{eq:dpt_boussinesq}
\end{equation}
It can be noted that, contrary to the classical buoyancy-driven component, which is integrated downward from the surface, the boundary condition for the random pressure $\dd p_t^\sigma$ is defined to be zero at the bottom.
As a consequence, it is integrated upward from the bottom.
Here, we have neglected $\dd_t w$; however, the martingale component of vertical transport could induce fast vertical accelerations, rather than contributing only to martingale pressure fluctuations.
An intermediate assumption is to consider that the time-differentiable part of $\dd_t w$ is negligible -- consistent with the classical hydrostatic balance -- while retaining its martingale part.
This martingale component could be explicitly diagnosed from the vertical velocity increments, thereby providing an additional correction term to $\dd p_t^\sigma$ in equation~\eqref{eq:dpt_boussinesq}.

Considering these approximations in the compressible Navier--Stokes equations~\eqref{eq:non_conservative} leads to the system
\begin{equation}
    \left\{
    \begin{aligned}
       &
       \mathbb{D}_t u_i
       =
       -\frac{1}{\rho_0}\ddt{p_2}{x_i} \dt - \frac{1}{\rho_0}\ddt{ \dd p_t^\sigma}{x_i}
       \quad\text{for}\quad i=\{u,v\}\\&
        w=\frac{1}{2}(\Div\mathsfbi{a})_z-\int_{-H}^z\left(\ddt{u^\star}{x}+\ddt{v^\star}{y}\right)\,\dd z
       \\&
       \Div(\sdBt)=0
       \\&
       p_2=-\rho_0\int_z^\eta b + \Adv{\frac{1}{2}\Div\mathsfbi{a}}{w}+\Diffst{w} \,\dd z
       \\&
       \dd p_t^\sigma = \rho_{0}\int_{-H}^z \Adv{\sdBt}{w}\,\dd z
       \\&
       \frac{\rho_0}{\gamma} \mathbb{D}_tT=
       - g (\eta-z)\mathbb{D}_t\rho_{\textit{\tiny BQ}}
       +{\left(\frac{1}{2}\Div\mathsfbi{a}\dt-\sdBt\right)\bcdot\Grad p_2}
       +{\frac{1}{2}\Div\mathsfbi{a}\bcdot\Grad \dd p_t^\sigma}
       - A_T
       - A_u 
       + \dot{Q}\dt
       \\&
       \mathbb{D}_tS
       =
       Q_t^{S} \dt+ \bm{Q}_{\sigma}^{S}\bcdot\dBt
       \\&
       b=g\left(1+\epsilon\frac{\rho_1}{\rho_0}-\frac{\rho_{\text{\tiny BQ}}}{\rho_0}\right)
       .
    \end{aligned}
    \right.
    \label{eq:boussinesq_EOS}
\end{equation}
with $S$ the salinity ($Y_i=S$, with $Q_t^{S} \dt+ \bm{Q}_{\sigma}^{S}\bcdot\dBt$ the corresponding source term) and $\rho_{\text{\tiny BQ}}(T,p,S)$ obtained by the equation of state.
Derivation details are provided in Appendix~\ref{sec:details_Boussinesq}.
In addition, a stochastic version of the so-called \emph{simple Boussinesq} equations are also provided in this appendix.
Let us emphasize that well-posedness of a  similar system under an extended stochastic hydrostatic assumption has been recently established in \citep{Debussche-Memin-Moneyron26}.

It is worth noting that in the small noise limit  (when $\sdBt\mapsto0$), the system consistently reduces to the usual advection-diffusion equation for temperature transport, while still accounting for radiative transfers and compressibility effects.
When noise is present, the system~\eqref{eq:boussinesq_EOS} explicitly captures interactions between turbulent small scales and the resolved forcing, particularly through the drift-work terms.
The quadratic variation terms ensure energetic consistency and are key to the stochastic energy balance.

These source and sink terms in the temperature evolution equation represent one of the principal outcomes of this study.
As discussed, these additional terms may be exploited to parameterise discrepancies relative to hydrostatic physics and the classical primitive equations.

\section{Numerical Applications}\label{sec:numerical}
We provide in this section a numerical application for the case of Boussinesq-hydrostatic approximation (Section~\ref{sec:Boussinesq}). 
Specifically, we are interested in evaluating the different contributions of each term of the stochastic temperature equation of (\ref{eq:boussinesq_EOS}) in a free convection large eddy simulation (LES).
The LES is performed under non-hydrostatic assumptions, and we aim at evaluating the ability of the stochatic modelling to account for a part of non-hydrostatic processes in convection events.
Following standard practices, this 3D LES is horizontally averaged and we evaluate the capabilities of our stochastic model to represent some aspects of the turbulent dynamics of the system for the time evolution of the horizontally averaged temperature vertical profile.

\subsection{Large Eddy Simulation}\label{subsec:num_exp}
The numerical simulation is performed with the Coastal and Regional Ocean Community model (CROCO ; \url{http://www.croco-ocean.org}), which is a new ocean model that is build upon the structure of the ROMS-AGRIF primitive equation solver \citep{shchepetkin2005,debreu2012}. 
The non-hydrostatic, non-Boussinesq \citep[NQB ; ][]{auclair2018} capabilities of CROCO are used here to produce a three-dimensional representation of convective processes in a stratified ocean\footnote{Note that non-Boussinesq dynamics is used at the ``fast'' time step of the split-explicit time stepping, but the model outputs we are analysing are associated with the dynamics at the ``slow'' time step which satisfies the Boussinesq 3D non-divergent condition, but remain non-hydrostatic.}.
The configuration is inspired by free convection studies \citep[\eg][]{souza2020}, where an horizontally uniform surface cooling is applied to a constantly stratified, horizontally uniform ocean.
We take $u_\text{\tiny ref}=\SI{1}{m/s}$, $L_\text{\tiny ref}=\SI{1}{m}$ and $\rho_\text{\tiny ref}=\SI{1024}{kg/m^3}$.
From this horizontally uniform and vertically constant stratified initial condition, the model is stepped forward in time on a 100$\times$100$\times$50 discretised grid with isotropic resolution of \SI{10}{m}, and exposed to a constant (in time and space) surface radiative flux of $Q_{rad} = \SI{500}{W m^{-2}}$. 
This leads to a cooling of the upper ocean layers, which ultimately become unstable through gravitational instabilities as a result of a negative buoyancy frequency $(N^2=- \frac{g}{\rho} \partial_zb)<0$, and thus undergoes convection.
The current settings is run with no Coriolis effect,
and is initialised with stochastic perturbations on ocean upper layers temperature decaying with depth to trigger the formation of convective plumes.
Momentum is advected with the 5$^{th}$ order advective scheme WENO5 in the three directions, and temperature is advected with with 5$^{th}$ order advective scheme WENO5 in the horizontal but with a centred 2$^{nd}$ order advective scheme in the vertical to avoid complexity in diagnosing sub-grid scale temperature vertical fluxes.
We also use the KPP \citep{large1994} vertical mixing closure scheme to mimic unresolved vertical fluxes near the surface.
Note that KPP in CROCO uses an enhanced vertical diffusion (EVD) approach in case of gravitationally unstable flow (\ie $N^2 < 0$),
with a dissipative coefficient set to \SI{0.1}{m^2s^{-1}}, which turns out to represent most of the situation in our free convection LES. The non-local capabilities of KPP are thus not used in these situations.
Initial stratification is set to $N^2=2\times10^{-6}$\,\SI{}{s^{-2}} and the equation of state is linear in temperature.
We run this experiment for \SI{3}{days} and output snapshots of the model state every \SI{10}{min}, including temperature vertical mixing associated with KPP.

\begin{figure}
    \centering
    \includegraphics[width = \textwidth]{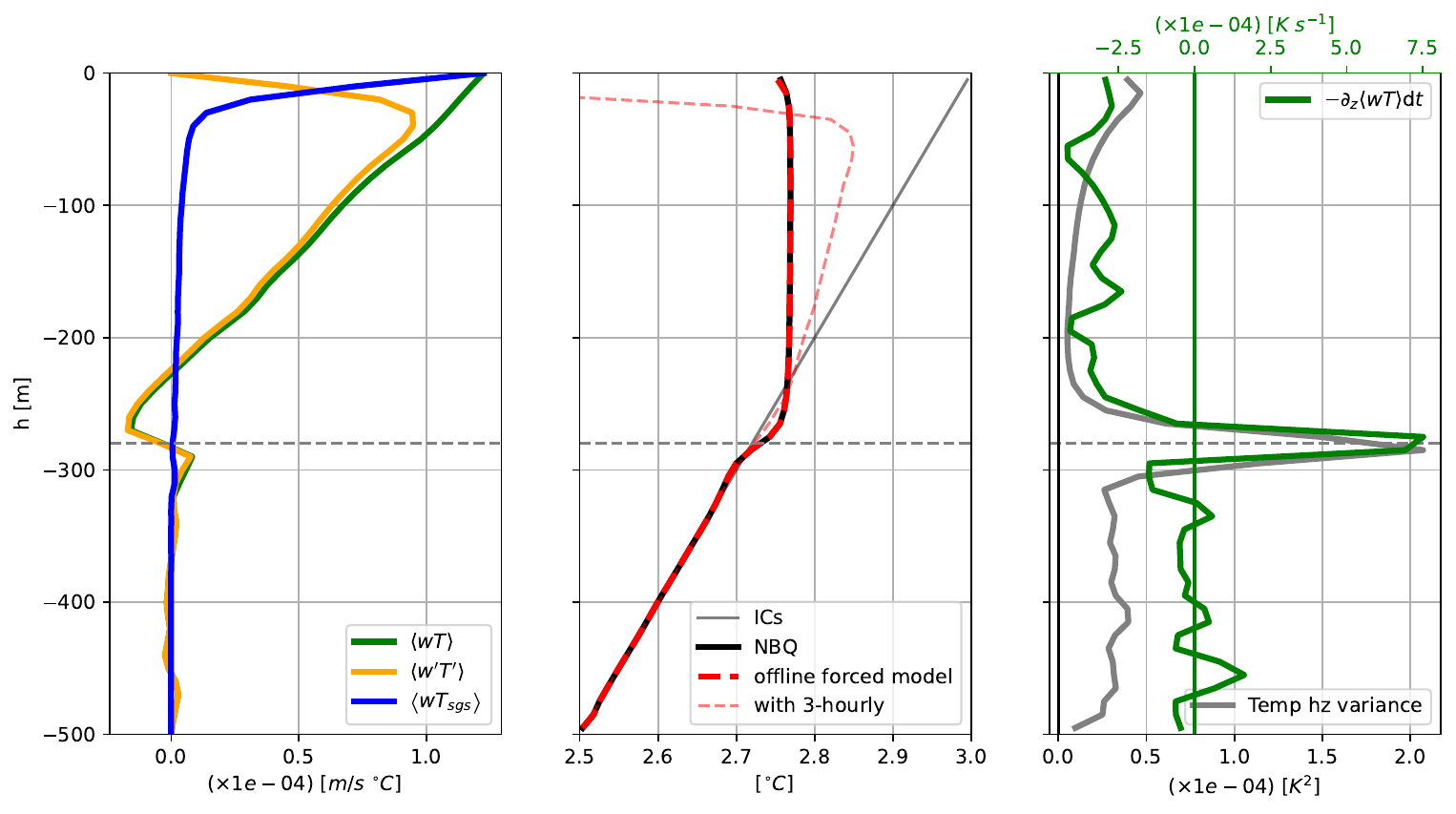}
    \caption{(Left) Horizontally averaged vertical temperature fluxes $\oxy{wT}$ after 60 hours of simulation (green), 
    decomposed into a resolved component associated with turbulent dynamics $\oxy{w'T'}$ (yellow),
    and a sub-grid scale component associated with KPP parametrisation $\oxy{w'T'_{sgs}}$ (blue).
    The contribution of mean components $\oxy{w}\oxy{T}$
    is several orders of magnitude weaker, as expected.
    (Center) Horizontally averaged temperature profile from the non-hydrostatic NBQ simulation
    (black) and from our \ti{offline forced model} (\ref{eq:zeroth_order_model}) based on 10-minutes model snapshots (thick dashed red) and based on 3 hourly averaged model outputs after 60 hours of integration (thin dashed red). Initial temperature profile is shown in gray.
    (Right) Divergence of vertical turbulent fluxes $-\pz \oxy{wT}$ (resolved and sub-grid scale; green), and horizontal variance of temperature (gray line) at time 60 hours. 
    Horizontal dashed line indicate the base of the mixed layer, computed as the maximum of stratification \citep{serazin2023}.}
    \label{fig:wT}
\end{figure}

In such a setting, the deterministic horizontally averaged temperature equation reduces to
\beq
    \pt  \oxy{T} = -\pz \oxy{wT},
    \label{eq:temp_eq}
\eeq
where sub-grid scale effects are lumped into vertical temperature fluxes, as explained below, and surface fluxes provide upper boundary conditions to these sub-grid scale effects. 
The averaging operator $\oxy{\bullet}$ which filters these sub-grid scales is formally an horizontal average, but in our double periodic and horizontally homogeneous-isotropic setting this is equivalent to an ensemble average. 
Left panel of Figure~\ref{fig:wT} shows the horizontally averaged vertical total temperature fluxes $\oxy{wT}$ after \SI{60}{h} of simulation (green), where upper ocean has been mixed within the nearly 300 upper meters. 
Near the surface, $\oxy{wT}$ tends toward $1.25\times10^{-4}$\,\SI{}{\meter.\second^{-1}\degreeCelsius}, \ie the prescribed surface radiative temperature fluxes of $\frac{Q_{rad}}{\rho_0C_p}$, then linearly decreases with depth. 
Near the bottom of the mixed layer, temperature vertical fluxes become negative before fading off. 
The vertical divergence of these fluxes (right panel, green line in figure~\ref{fig:wT}) controls the horizontally averaged temperature profile,
driving a cooling of most of the upper mixed layer, but a warming at the base of the mixed layer (centre panel, black line). 
In this free convection experiment, the warming at the base of the mixed layer is associated with penetrative convective events \cite[\eg][]{cushman1982},
and is usually not well captured by sub-grid scale convective parameterisations \cite[see discussion in, \eg][]{hourdin2002,suselj2019,giordani2020}.

The total vertical temperature fluxes is composed of a LES-resolved, turbulent contribution $\lb w'T'\rb$, with $\bullet'$ the residual from the horizontal averaging resolved by our 3D LES (left panel, orange line),
and a LES sub-grid scale contribution $\oxy{wT_{sgs}}$ (left panel, blue line),
computed following the dissipative down-gradient approach $\oxy{wT_{sgs}} = -\oxy{\nu_T \pz T}$ with $\nu_T$ the dissipative coefficient computed by the vertical mixing scheme (here KPP). 
As is usually done in models such as CROCO, the surface heat fluxes are provided as the upper boundary conditions for the dissipative coefficient $\nu_T$.
The resolved turbulent fluxes drive most of the dynamics within the mixed layer, reach a maximum at about \SI{30}{m} depth, then decrease toward zero at the surface to satisfy the non-penetrative boundary condition $w(z=\eta)=0$.
Near the surface, most of the vertical temperature fluxes are driven by the sub-grid scale fluxes, which host the prescribed surface cooling as upper boundary conditions.
Note that the LES sub-grid scale turbulent fluxes at the base of the mixed layer are weak, suggesting that, although our LES has relatively coarse vertical resolution (\ie \SI{10}{m}), it still provides reasonable estimates of turbulent (resolved) dynamics within this strongly stratified region. 

\begin{figure}
    \centering
    \includegraphics[width = \textwidth]{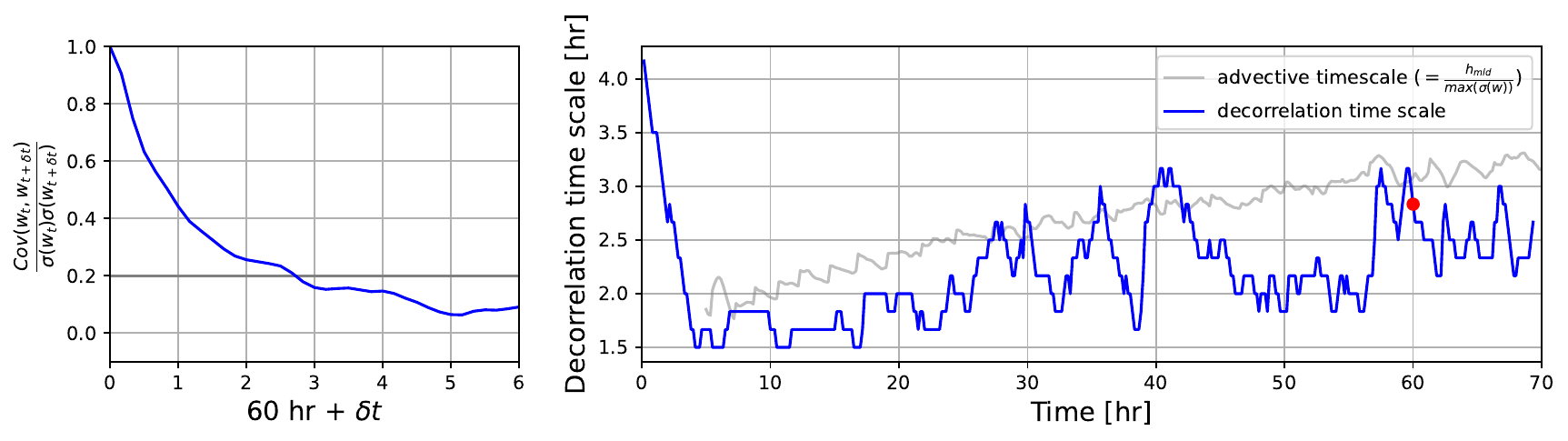}
    \caption{Estimation of the decorrelation time scale associated with convective plumes. (Left) evolution of the normalized 2 times covariance as a function of delay time $\delta t$ after \SI{60}{h} of simulation time, and (right) evolution of the decorrelation timescale, defined as the value of $\delta t$ needed for the normalized 2 time covariance to fall below $0.2$.
    Advective timescale computed as the ratio between mixed layer depth and maximum of vertical velocities standard deviation (gray line). Red dot indicate the time frame used throughout the analysis.}
    \label{fig:time_scale}
\end{figure}

Based on the LES snapshots sampled every \SI{10}{min}, we compute an offline estimate of the time evolution of the 1D vertical temperature profile with a simple forward Euler time stepping:
\begin{equation}
    \oxy{T}^{(n+1)}=\oxy{T}^{(n)}-\Delta t\lp \pz\oxy{wT}\rp^{(n)}.
    \label{eq:zeroth_order_model}
\end{equation}
Starting from initial conditions and stepping forward this \ti{offline forced model} for 3 days leads to a near perfect representation of  temperature profile time evolution (Figure~\ref{fig:wT}, centre panel, red dotted line). 
The fact that our \ti{offline forced model} for 1D vertical temperature profile is able to track so well the results of a 3D, fully non-linear model subject to (horizontal) averaging demonstrates the consistency of the diagnostics of the Reynolds stresses, and of the numerical time scheme equation~\eqref{eq:zeroth_order_model} associated with the time step $\Delta t$.
In  particular, it confirms that \textit{i)} the knowledge of the vertical derivative is enough due to homogeneity and isotropy in the horizontal direction, and \textit{ii)} the \SI{10}{min} model output is sufficiently high in frequency compared to typical timescales of convective plumes.
Indeed, typical timescales of the convective plume are of about 2 to 3 hours depending on the depth of the mixed layer (Figure \ref{fig:time_scale}), well longer than LES high frequency model outputs that we have.
In the next section, we evaluate the ability of our stochastic model to reproduce similar dynamics when time filtering is applied to the system.
We pay a particular attention to the warming at the base of the mixed layer associated with penetrative convective events, which are typically misrepresented by standard models.
This region is characterized by a positive contribution of the (negative) vertical divergence of vertical temperature fluxes (Figure~\ref{fig:wT}, right panel, green line), as noted earlier, and by strong horizontal variations in temperature associated with penetrative plumes.
Such plumes are usually described as non-local vertical dynamics driven by inertial processes such as entrainment and detrainment, and are therefore of an advective nature, but they remain highly localized in the horizontal plane.
When a convective plume penetrates below the mixed layer, it entrains water from the mixed layer into the cooler water beneath, inducing large horizontal temperature variations.
In contrast, the temperature within the mixed layer is relatively well mixed, so that horizontal temperature variations remain weak.
Taking the horizontal variance in a 3D dataset as a proxy for what a stochastic 1D vertical model should reproduce, we therefore expect such a model applied to temperature to be particularly skillful at the base of the mixed layer, where horizontal temperature fluctuations are large.
This will be demonstrated in the next section.

\subsection{Stochastic temperature equation}\label{subsec:1D_temp}
\begin{figure}
    \centering
    \includegraphics[width = 1.\textwidth]{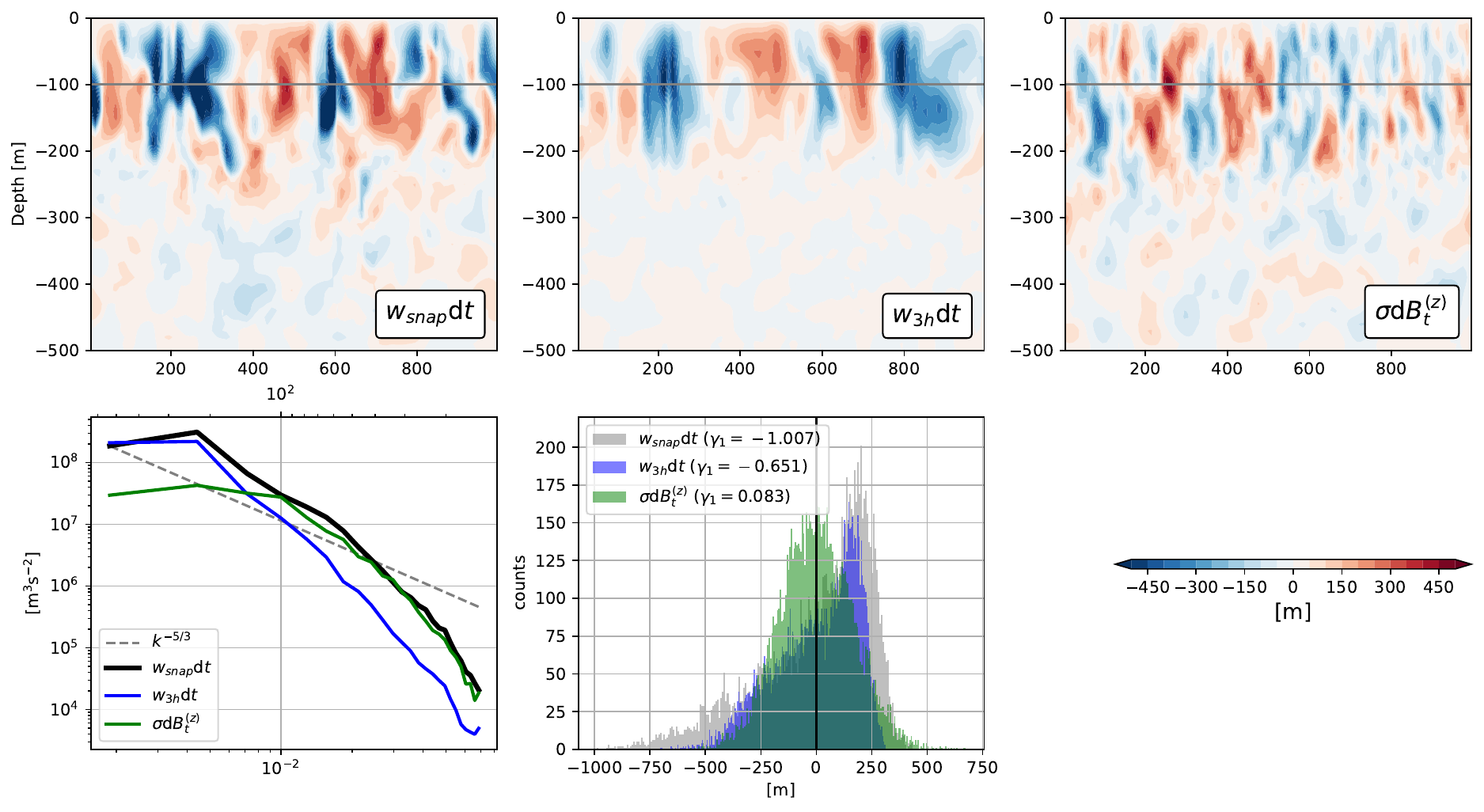}
    \caption{(Top) Vertical cross section of snapshot model output vertical displacement ($w_{snap}\mathrm{d}t$ top left), decomposed into a smooth-in-time, 3-hourly time-filtered displacement ($w_{3h}\mathrm{d}t$ top centre) and a stochastic reconstruction of the residual ($\sigma \mathrm{d}B_t^{(z)}$, top right). (Bottom) Spectra (left) and distribution (right) of these three quantities along an horizontal section at -100 m depth.}
    \label{fig:LU_illustr}
\end{figure}

Stochastic LU modelling inherently accounts for the coexistence of two components: a time-differentiable part and a martingale part.
The martingale component, which is time-decorrelated, represents a fast contribution (small-scale turbulence) relative to the slow, resolved component (the large-scale plume), the two being distinguished by their temporal behaviour.
These contributions are extracted from the LES through time filtering, where the filtered quantities represent the time-differentiable component, and the residuals are used to construct the martingale part as a Brownian motion.
Based on decorrelation timescale of convective plumes (Figure \ref{fig:time_scale}), we select a typical timescale of \SI{3}{h} to filter the equations.
An example of this time filtering, applied to vertical displacement, is shown in Figure~\ref{fig:LU_illustr}, where a model output snapshot is decomposed into a 3-hourly time filtered component (referred to as smooth in time), and a high frequency, time-uncorrelated noise component $(\sigma \dif B_t)^{(z)}$, computed as a stochastic reconstruction of the residual associated with that time filtering, \textit{i.e.} the vertical component of:
\beq
    \bsig(\bx,t) \dif \B_t = \sqrt{\tau} \sum_{k}\sqrt{\lambda}_k \bs{\phi}_{k}(\bx, t) \mathrm{d}\beta_t^k,
    \label{eq:modal_noise_num}
\eeq
where $\lambda_k$ and $\bs{\phi}_k$ denote the Fourier modal amplitudes and spatial modes, respectively, and $\mathrm{d}\beta_t^k$ are independent standard Brownian motions.
Note that $\bm{\varphi}_k=\sqrt{\tau}\bm{\phi}_k$ (see Eq.~\eqref{eq:modal_noise}) provide the appropriate scaling such that $\bm{\varphi}_k$ forms a basis of Brownian displacements, while $\bm{\phi}_k$ corresponds to the associated velocity modes.
In equation~\eqref{eq:modal_noise_num}, $\tau$ denotes the decorrelation timescale of the velocity anomalies, which is adjusted so that spectral estimates match (Figure~\ref{fig:LU_illustr}, bottom left panel).
At a depth of $-100$,m, \textit{i.e.}, in the middle of the mixed layer, this yields a decorrelation timescale of 10 minutes, corresponding to the model output frequency.
With such filtering, the smooth-in-time component of the transport dominates at scales larger than $100\,$m, while the Brownian component dominates at scales smaller than $100\,$m.
The two components also differ in their distributions: the smooth-in-time transport retains some asymmetry toward downward (\textit{i.e.}, negative) vertical transport (skewness $\gamma_1 = -0.651$, Figure~\ref{fig:LU_illustr}, bottom centre panel), although weaker than in the original field ($\gamma_1 = -1.0$), whereas the Brownian transport has a symmetric distribution, as expected from its definition.
As will be shown in the following, our stochastic model exhibits limited skill within the mixed layer (at least at leading order), but performs very well at the base of the mixed layer.
To ensure consistency between the noise definition and the forthcoming results, we adjust the free parameter $\tau$ to the characteristic timescales of this region, which are about three times smaller than those within the mixed layer (not shown).
At this depth (approximately $-220$,m), the model snapshot of vertical transport exhibits a more symmetric distribution ($\gamma_1 = -0.55$) and is largely dominated by the sub-filtered residual component; it is therefore primarily represented by the Brownian component of the transport in our stochastic model.

Associated with the transport noise, we define the corresponding 1-point 1-time variance tensor $\ba$ as:
\beq
    \ba \mathrm{d}t = \langle \bsig\dif \B_t, (\bsig\dif \B_t)^{\transp} \rangle_t = \tau \sum_{k}\lambda_k \bs{\phi}_{k}(\bx, t)\bs{\phi}_{k}^{\transp}(\bx, t)\mathrm{d}t,
\eeq
with these quantities defined, we are now in a position to evaluate the different terms contributing to the temperature equation~\eqref{eq:boussinesq_EOS}.
Before proceeding, we introduce a few additional simplifications associated with our idealised setup.

First, our LES assumes a linear, temperature driven, equation of state
\beq
    \rho_{\textit{\tiny BQ}}=\rho_0(1-\beta_T(T-T_0)),
    \label{eq:rho_bq2}
\eeq
with $\beta_T=2.048\times10^{-4}$\,\SI{}{kg.m^{-3} K^{-1}} the thermal expansion coefficient and $T_0$ a reference temperature. Hence, the stochastic transport of the Boussinesq density $\rho_{\textit{\tiny BQ}}$ can be directly expressed in terms of the stochastic transport of temperature:
\beq
   \mathbb{D}_t \rho_{\textit{\tiny BQ}} = \rho_0 g \beta_T~z \mathbb{D}_tT,
\eeq
where we have neglected the contribution of the free surface $\eta$, as it is weak in our configuration.
We interpret the transport of Boussinesq density as a contribution of temperature in the form of potential energy, which will be discussed further below.
Finally, the temperature equation of (\ref{eq:boussinesq_EOS}) includes a source term $\dot{Q}$, which represents 3D distributed heat sources.
In our numerical application, heat sources arise solely from radiative fluxes at the surface.
Following their implementation in the LES (see Section~\ref{subsec:num_exp}), we incorporate this contribution into the subgrid-scale fluxes.
Accordingly, the simplified temperature equation does not explicitly include radiative fluxes, although they contribute to the vertical divergence of subgrid-scale temperature fluxes in the upper layer.
To summarize, the temperature equation of (\ref{eq:boussinesq_EOS}) reduces to:
\beq
\begin{aligned}
\underbrace{\rho_0(\frac{1}{\gamma}+g\beta_Tz)}_{(M)} \mathrm{d}_t T= 
        &\underbrace{\rho_0(\frac{1}{\gamma}+g\beta_Tz)}_{(M)} 
        \left(
            - \underbrace{\bu\bcdot\bnabla T \mathrm{d}t}_{(A)}
            - \underbrace{\left(\frac{1}{2}\bnabla \bcdot \ba\right)\bcdot\bnabla T \mathrm{d}t}_{(B)}
            - \underbrace{\sdBt\Grad T}_{(C)}
            + \underbrace{\frac{1}{2}\bnabla\bcdot\ba\bnabla T \mathrm{d}t}_{(D)}
        \right)\\
        &+\underbrace{\left(\frac{1}{2}\bnabla \bcdot\ba \dt-\sdBt\right)\bcdot\Grad p_{2}}_{(D_t)}
         +\underbrace{\frac{1}{2} \bnabla\bcdot\ba\bcdot\Grad \dd p_t^\sigma}_{(D_{\sigma})}
        -A_T -A_u,
      \label{eq:temp_eq_sto_bq}
\end{aligned}
\eeq
where the stochastic transport operator (\ref{eq:stochastic_transport})  has been developed for a divergent-free noise and expressed in the form $\mathrm{d}_t T = RHS$.
A peculiarity of our diagnostic approach is that both the Brownian and the smooth-in-time component of the transport are divergent-free, since they are defined from the time-filtered model output.
However, this property does not hold for the Itô-Stokes drift contribution $-\frac{1}{2}\bnabla \bcdot \ba$.
This differs slightly from its definition in (\ref{eq:boussinesq_EOS}) and is a consequence of the diagnostic approach.
As a result, the terms $C$ (the second part of $D_t$) and $A$ can be written in flux form.
This is, however, not the case for the Itô-Stokes drift contribution ($B$, first the part of $D_t$ and $D_{\sigma}$), as its divergence is not guaranteed to vanish.

The LHS of (\ref{eq:temp_eq_sto_bq}) represents the changes between initial and final time (\ie $\dif_t T=T^f-T^i$) over a time interval  $\dif t$, here taken as 3 hours.
The multiplicative term (denoted $(M)$) expresses contributions in the form of internal energy ($\frac{1}{\gamma}$) and potential energy\footnote{Note that our non-dimensionalisation (\ref{eq:adim}) includes a term $\frac{u_{ref}^2}{c_p^{\phi}}$ in the expression of $\beta_T$.} ($g\beta_T z$) similarly to \cite{perrot2025} (see their appendix F3.2).
More precisely, this last term arises from the work of compression ($P_t$ in equation~\eqref{eq:non_conservative_T}), which, strictly speaking, transfers kinetic energy to internal energy \citep[\textit{e.g.}][]{tailleux2012}, but is expressed here in terms of potential energy through the hydrostatic balance, continuity and state equations.
Terms $(A)$-$(D)$ on the RHS are associated with stochastic transport, and also contribute to both internal and potential energy.
 The remaining terms on the RHS are source terms associated with: the work of the Itô-Stokes drift $(\bu^*-\bu)$ and stochastic transport $\bsig \dif \B_t$ acting on the gradient of the quasi-non-hydrostatic pressure $p_2$ (denoted $D_t$ in (\ref{eq:non_conservative_T})); the work of the Itô-Stokes drift $(\bu^*-\bu)$ on the stochastic pressure gradient $\Grad \dif p_t^{\sigma}$ (denoted $(D_{\sigma})$); and quadratic covariation terms associated with temperature and momentum ($A_u$ and $A_T$, respectively).
 Our aim is to evaluate the contributions of these different contributions from our LES, identify the leading order balance, and comment on their interpretation.
 Note that further simplifications could be made based on the geometry of our idealised setup (horizontal homogeneity, isotropy, doubly periodic boundary conditions).
 However, we choose to work with full 3D estimates of the different terms, and assess their relevance \textit{a posterior}.
 A complete description is provided in Appendix~\ref{sec:sto_Teq_details}, which we summarise below.

Compared with our previous \ti{offline forced model} (eq.~\eqref{eq:temp_eq}), in which all the Reynolds stresses was, by construction, properly captured by high frequency model outputs, the application of time filtering reduces the capabilities of such a simple model.
An illustration is provided in Figure~\ref{fig:wT}, where the offline reconstruction of the 1D vertical temperature profile based on 3-hourly filtered data\footnote{Note that, here, LES subgrid-scale vertical fluxes are computed at a 10-minute frequency and then averaged over a 3-hour window, rather than being diagnosed from 3-hourly averaged temperature and numerical subgrid-scale dissipation provided by KPP. While the latter would be more consistent, it leads to negative vertical temperature fluxes at the base of the mixed layer, thereby violating the down-gradient assumption inherent to KPP.} (thin dashed red line) significantly departs from the reference LES (thick black line).

We first observe an imbalance between the upper and mid-depth layer, which we interpret as a direct consequence of overly weak vertical temperature fluxes to homogenise these regions.
More importantly, the time-filtered estimates fail to capture the warming at the base of the mixed layer, and thus the associated penetrative convection. As we will show below, stochastic modelling can help to improve these estimates.

Although the covariance tensor of $\sdBt$ is diagnosed from the LES in this proof-of-concept study, it should be noted that modelling the statistics of the small-scale velocity field is more straightforward than modelling Reynolds fluxes of thermodynamic variables such as $\langle wT \rangle$.
Representing these fluxes using a purely diffusive closure, as in the standard Boussinesq approximation for the momentum equation, remains open to debate and is often supplemented by eddy mass-flux parameterisations.
A possible model-based strategy (not explored in this paper) would be to define the noise variance through a transport equation for TKE, similarly to \citet{li2025}.

\begin{figure}
    \centering
    \includegraphics[width = 1.\textwidth]{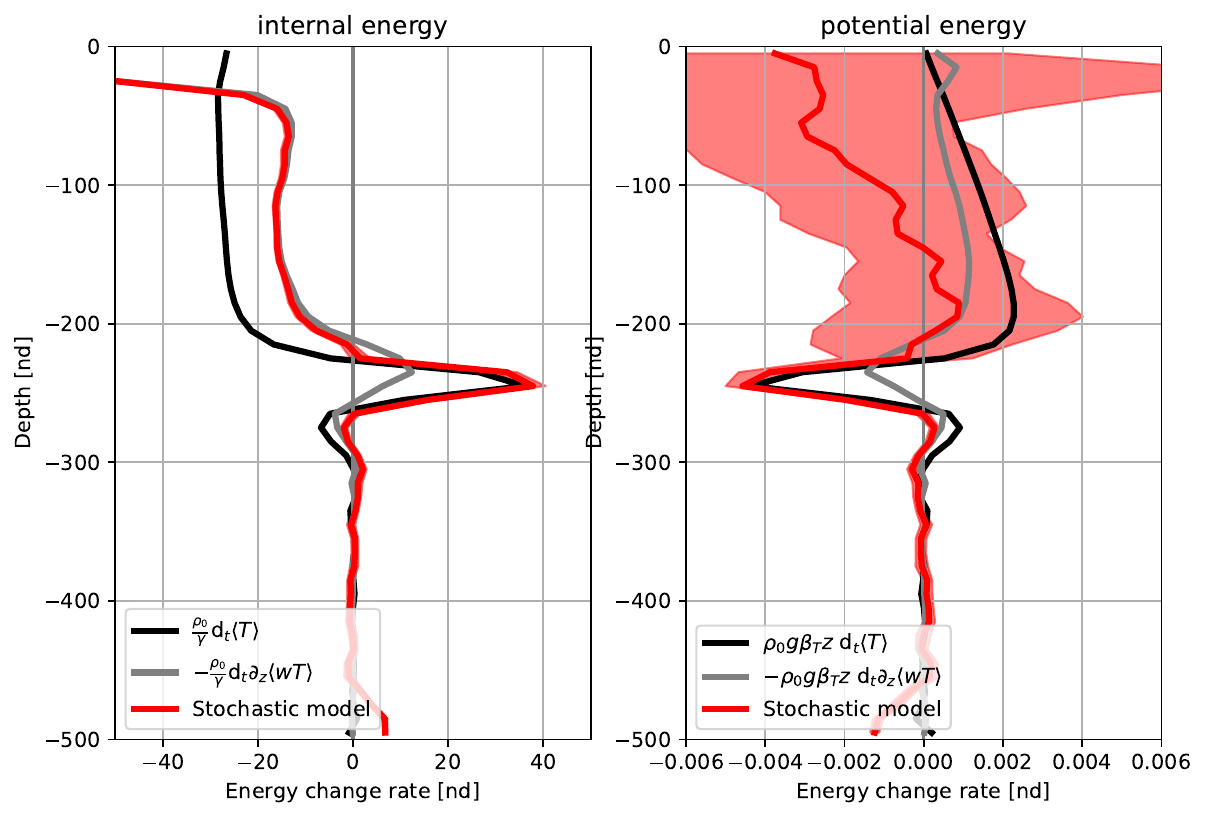}
    \caption{Temperature changes over the 3-hour window period in form of internal (left) and potential (right) energy. Sum of the RHS of our stochastic 1D model (\ref{eq:temp_eq_sto_bq}) (red) 
    is compared to the reference temperature changes (black). Contribution of vertical temperature fluxes associated with the 3-hour filtered vertical velocities and temperature is shown in gray. Contributions from the Brownian terms (\ie associated with $\mathrm{d}B_t$) are estimated based on 500 realisations and blue shading shows one standard deviation around the mean. Decorrelation time scale has been set to $\tau=\SI{175}{s}$ to match our stochastic 1D model with estimates of temperature changes at the base of the mixed layer. Note that vertically averaged internal and potential energy changes are of $-10.5$ and $+5.8\times10^{-4}$, respectively, in agreement with numbers reported on Figure \ref{fig:nrj_time_series}.}
    \label{fig:sum_dt_e_pe}
\end{figure}

Figure \ref{fig:sum_dt_e_pe} shows the horizontally averaged internal (left) and potential (right) energy changes associated with changes in temperature.
In each of these plots, the black line represents the change of energy between two, 3~hours distant snapshots started at time 40 hours.
These are associated with $\dif_tT$ in (\ref{eq:temp_eq_sto_bq}), multiplied by their respective (internal or potential) energy factor, and are our reference.
The red lines represent the reconstructions of these energy changes by adding all the terms on the RHS of equation (\ref{eq:temp_eq_sto_bq}).
The specific contribution of 3-hourly, time filtered fluxes $\bu\bcdot\bnabla T$ (A in (\ref{eq:temp_eq_sto_bq})), is shown in gray, and the contribution of Brownian terms (\ie $C$, $D_t$ and $D_{\sigma}$ in (\ref{eq:temp_eq_sto_bq})) is shown in light red shading as $\pm$1 standard deviation based on 100 realisations.

\begin{figure}
    \centering
    \includegraphics[width = \textwidth]{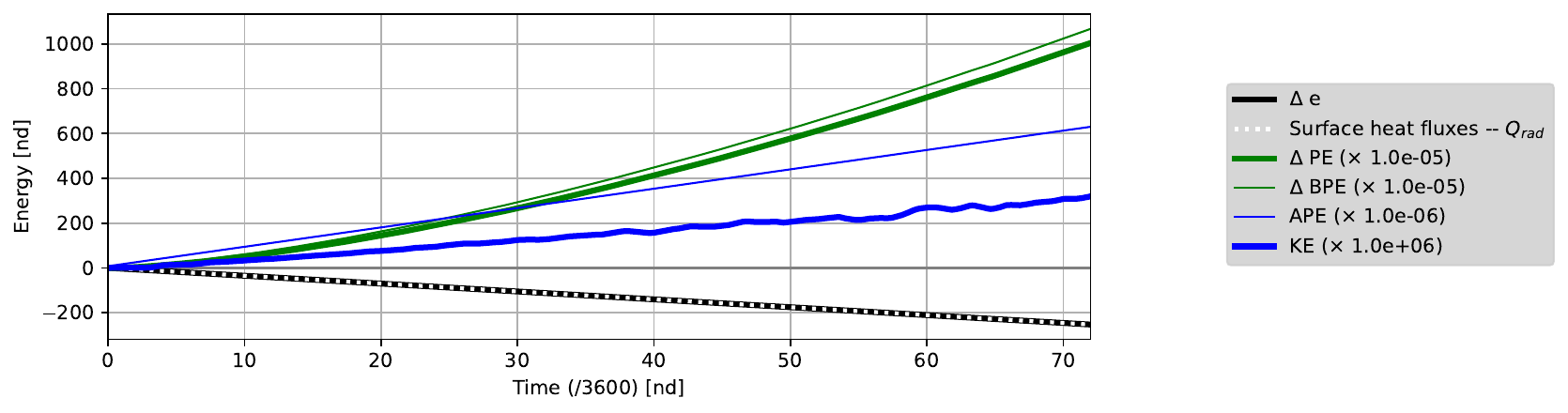}
    \caption{Time series of domain averaged internal energy ($\Delta e$), Potential Energy ($\Delta PE$) and Background Potential Energy ($\Delta BPE$) anomalies, Available Potential Energy ($APE$) and Kinetic Energy ($KE$), in response to net surface heat fluxes $Q_{rad}$. $\Delta e$, $\Delta PE$ and $\Delta BPE$ are anomalies from initial values. Note the different scale factors between $\Delta e$ and $Q$, $\Delta PE$ and $\Delta BPE$, and $APE$ and $KE$.}
    \label{fig:nrj_time_series}
\end{figure}

From the estimates shown in Figure \ref{fig:sum_dt_e_pe}, the leading-order balance is associated with the transport of temperature in the form of internal energy (left panel).
At this order, we find that the transport terms ($(A)$–$(D)$ in (\ref{eq:temp_eq_sto_bq})) dominate the RHS and are significantly larger than the other contributions (\ie terms $(D_t)$, $(D_{\sigma})$, $A_T$, and $A_u$).
This result is consistent with the Boussinesq approximation, in which dynamics and thermodynamics are effectively decoupled.
From an energetic perspective, this approximation is justified by the fact that internal energy constitutes a much larger reservoir than the other forms of energy. This is illustrated in Figure \ref{fig:nrj_time_series}, where changes in internal energy $\Delta e$ in response to radiative surface cooling $Q_{rad}$ are several orders of magnitude larger than the corresponding changes in potential and kinetic energy, as expected.

In (\ref{eq:temp_eq_sto_bq}), the stochastic transport of temperature (term $C$) vanishes on average and exhibits random variations at the base of the mixed layer (Figure \ref{fig:Tsto_bgt_internal}, upper right panel).
As a consequence, Itô-Stokes drift and stochastic diffusion ($B$ and $D$, respectively) contribute significantly in this region, in good agreement with the diagnosed vertical temperature fluxes.
This suggests that our stochastic model has strong potential to mimic turbulent fluxes at the base of the mixed layer.
This is a particularly encouraging result, as it relates to penetrative convection processes that are typically difficult to represent in subgrid-scale models.

These findings are consistent with the role of the non-local term in the $k\epsilon t$ model of \cite{legay2024b}.
By decomposing this term, they identified a contribution at the base of the mixed layer primarily driven by the maximum of temperature variance.
In our case, the temperature fluxes are reproduced remarkably well.
We attribute this to the fact that our stochastic model does not impose any \textit{a priori} form for the turbulence closures.
Instead, they are derived from physical invariants and from a single idealised decorrelation assumption for the unresolved processes at the grid scale.
In particular, the impact of horizontal variability at the base of the mixed layer is naturally accounted for.

However, temperature changes within the mixed layer are driven almost exclusively  by the 3-hour time-filtered vertical temperature fluxes, with little to no contribution from the other stochastic terms.
This results in a significant underestimation of temperature changes within the mixed layer.
Note that when filtering the model output, both the residual velocity and temperature are removed, and their covariations is found to account for approximately half of the total temperature fluxes (not shown).
While we reconstruct the stochastic small-scale velocity field, the associated fluctuating temperature field is not reconstructed. 
A potential way to reintroduce temperature anomalies would be to move from our current diagnostic approach to an effective stochastic model with an appropriate prognostic time-stepping, thereby allowing temperature anomalies to emerge through stochastic transport.
This is left for future, dedicated work.
Additional research direction could also focus on TKE-based model within the LU framework.
Indeed, in the $k\epsilon t$ model of \cite{legay2024b}, the non-local term -- responsible of the good agreement within the mixed layer -- is governed by the various closure terms involved in their formulation.
For prognostic modelling, the noise variance could be parameterised using a TKE-based model, thereby combining the strengths of both approaches.

Finally, our stochastic model shows significant differences from the reference near the surface, a discrepancy likely arising from an imbalance between the stochastic representation and the actual subgrid-scale fluxes diagnosed in the LES.
This primarily reflects the fact that the LES does not resolve such small-scale fluctuations, which therefore cannot be adequately captured by the stochastic terms.

Temperature also contributes to potential energy, which -- compared to internal energy -- has a magnitude order much closer to that of the kinetic energy reservoir (Figure \ref{fig:nrj_time_series}).
Changes in total potential energy are largely controlled by variations in background potential energy, as expected in our diabatically forced system.
Once this background component is removed, the remaining part -- namely the ``available'' potential energy that can interact with turbulence and thus with kinetic energy -- is of the same order of magnitude as the latter, \ie $10^{-6}$ in nondimensional units.
Therefore, analyzing vertical temperature fluxes in terms of potential energy, rather than internal energy, appears more appropriate for highlighting exchanges between energy reservoirs.

In our view, this also raises questions about the energetic consistency of procedures used in TKE-based approaches (see \cite{umlauf2005} for a review), in which ``\textit{stability functions}'' are inferred by solving a system with as many equations as unknowns, while turbulent vertical temperature fluxes are implicitly associated with the internal energy reservoir under the Boussinesq approximation (\ie assuming a decoupling between dynamics and thermodynamics).
Once again, these findings rely solely on diagnostics of terms derived from the stochastic system.
Here, such a connection is ensured through the incorporation of the work of compression $P_t$ which, under the Boussinesq approximation, reduces to a transport equation for the Boussinesq density $\mathbb{D}_t \rho_{\textit{\tiny BQ}}$ (see equation~\ref{eq:Pt_boussinesq}), in line with recent developments \citep{tailleux2024,perrot2025}. 

Estimates for our stochastic 1D model are shown in the right panel of Figure~\ref{fig:sum_dt_e_pe}.
As expected, changes in potential energy (black line) are now positive and increase linearly with depth due to their dependence on $z$, a behaviour common to all transport terms (\ie $(A)$, $(B)$, $(C)$, and $(D)$ in (\ref{eq:temp_eq_sto_bq})).
At the base of the mixed layer, penetrative convection processes lead to a local reduction in potential energy, and we recover a good representation from our stochastic model, again primarily driven by the Itô–Stokes drift and stochastic diffusion.
At this order, the quadratic covariation term $A_u$ as well as the drift contributions $D_{t}$ and $D_{\sigma}$ become significant.

It is interesting to note that both $A_u$ and $D_{\sigma}$ are associated with the stochastic pressure $\mathrm{d}p_t^{\sigma}$, which involves vertical integration of stochastic transport of vertical velocities, \ie $\bsig\mathrm{d}\B_t w$, most of which is driven by horizontal transport.
This term therefore encodes coupling between horizontal and vertical dynamics and, as a pressure-like contribution, introduces non-local (in the vertical) effects.
The quadratic covariation term $A_u$ is positive definite, so that $-A_u$ contributes negatively to the budget.
Starting from a zero bottom boundary condition, it decreases monotonically within the mixed layer and reaches a minimum at the surface.
The structure induced by this pressure-like, non-local contribution shares similarities with the vertical profile of potential energy changes within the mixed layer, although a clear offset between the two remains, the interpretation of which is not straightforward.
A possible improvement would be to introduce a random surface pressure boundary condition term related to wind forcing or surface waves as in \citep{li2025} -- instead of a zero boundary condition at the bottom resulting in an physically uncontrolled value at the surface.
The stochastic drift-work contribution $D_{\sigma}$ corresponds to the work of the stochastic pressure on the Itô–Stokes drift $(\bu^*-\bu)$ and exhibits a significant spread throughout the mixed layer.
This supports the fact that this coupling between kinetic and potential energy should be treated as an inherently random process.

\section{Conclusion}
\label{sec:conclusion}
This paper proposes a stochastic representation of the compressible Navier–-Stokes equations within the framework of location uncertainty.
The system is derived from the conservation laws of mass, species concentration, momentum, and total energy, each subject to stochastic transport.
The resulting equations retain a structure similar to their deterministic counterparts, while introducing additional physically meaningful terms.
Among these, we identify a work contribution arising from the alignment between the time-differentiable pressure gradient and the drift velocity.
This small-scale–induced work is analogous to the baropycnal work encountered in compressible large-eddy simulations.

These terms arise from a rigorous derivation grounded in conservation laws and stochastic calculus, rather than from phenomenological arguments.
This provides a clear justification for their form, role, and physical interpretation, as they primarily originate from statistical correlations between small-scale velocity fluctuations and forcing terms.
Importantly, these contributions are entirely determined by the covariance structure of the stochastic displacement and do not rely on a Boussinesq-type approximation for the Reynolds stresses associated with thermodynamic variables.

We have also shown that applying the Boussinesq approximation to the stochastic compressible system consistently recovers previously derived stochastic Boussinesq systems, while yielding an enhanced form of the temperature transport equation in which temperature fully behaves as an active tracer accounting for energetic exchanges.
From a modelling perspective, these systems form a nested hierarchy of increasing complexity, with the general stochastic compressible formulation naturally incorporating thermodynamic effects that are absent in simpler models.

We have assessed the stochastic contributions in a large-eddy simulation (LES) of oceanic convective plumes.
The results demonstrate the ability of the stochastic terms to qualitatively reproduce part of the missing Reynolds stresses in a horizontally averaged temperature equation.
In particular, these terms show strong skill in representing vertical temperature fluxes at the base of the mixed layer.
To our knowledge, no existing parameterisation captures these effects at a comparable level of closure.
We interpret this as a consequence of the fact that our stochastic modelling does not rely on explicit closure assumptions (such as Boussinesq-type diffusive or counter-gradient formulations), but instead only on a decorrelation hypothesis.
Both the Itô–Stokes drift and stochastic diffusion contribute to this process, while compressibility effects are found to be several orders of magnitude weaker.
Accordingly, the leading-order balance is consistent with the Boussinesq approximation.
Within the mixed layer, however, little to no contribution from stochastic terms is identified.
Further analyses are required to draw firm conclusions regarding the missing physics, but we nonetheless highlight several possible directions in this regard.

As a proof of concept, we have focused here on the 1D vertical temperature profile equation, \ie its first-order moment \citep{stull1988}.
Compared with TKE-based vertical mixing schemes, our model is able to capture key features at the base of the convective plume.
However, within the plume, the relatively low variability of the turbulent field leads to negligible contributions from the stochastic terms, whereas TKE-based models perform well.
This gap highlights an important avenue for improvement.

TKE-based approaches typically rely on prognostic equations for higher-order moments, usually second-order quantities such as $\oxy{w'T'}$, together with empirical closures. 
The success of the modelling strategy in \citet{legay2024b} within the mixed layer may be related to their representation of the vertical transport of small-scale structures along the plume. 
In a similar spirit, building on the framework proposed by \cite{Debussche-Memin25}, which provides prognostic equations for the noise characteristics, incorporating the transport of small scales through an evolution equation for the noise correlation tensor may offer a promising direction for improving model behaviour within the mixed layer.
A complementary alternative, in the spirit of \citet{li2025}, would be to introduce an additional random surface pressure forcing to better account for the effects of wind stress and wave forcing.
Moreover, defining mass fluxes through a Girsanov drift \citep{li2023} also appears to be a promising avenue.

Otherwise, while the leading-order balance satisfies the Boussinesq approximation, the quadratic variation term $A_u$ as well as the drift work $D_{\sigma}$ associated with the (non-hydrostatic) stochastic pressure prove capable of qualitatively capturing unresolved fluxes within the mixed layer. This pressure work is associated with energy transfers from mechanical to internal energy through compression, providing new insights into the energetics of convective events.
To our knowledge, such a contribution has not been reported elsewhere.

Finally, while we have focused here on the temperature equation, this represents only one part of the full system (\ref{eq:boussinesq_EOS}).
The work of \cite{jamet2022c} extends these results by demonstrating the benefits of including a direct stochastic non-hydrostatic pressure correction in the form of $p_2$ within a quasi-nonhydrostatic set of equations.
In particular, they show that the leading-order contribution of horizontal stochastic diffusion plays a key role in shaping the structure of convective plumes, bringing their stochastic quasi-nonhydrostatic model into closer agreement with reference LES simulations.
A noted limitation is the inhibition of penetrative convection, due to excessive damping of vertical velocities.
A possible perspective is that accounting for both dynamical and thermodynamical stochastic effects within such quasi-nonhydrostatic models could provide an accurate representation of convective events without the need for a numerically expensive non-hydrostatic solver.

Overall, this work provides a structured modelling framework and opens promising research directions for further exploring the potential of stochastic representations in the numerical simulation of oceanic and geophysical flows.

\clearpage

\acknowledgments
The authors acknowledge the support of the ERC EU project 856408-STUOD, the French National program LEFE (Les Enveloppes Fluides et l’Environnement) and warmly thank Valentin Resseguier, Long Li, Alexandre Legay and Bruno Deremble for very fruitful and stimulating discussions on this work. 


\datastatement
LES numerical details and post-processing tools used in Section~\ref{sec:numerical} are available at \url{https://doi.org/10.5281/zenodo.8059690}.

\appendix
\section{Quadratic (co-)variation}\label{sec:bracket}
In stochastic calculus, quadratic covariation of two real-valued processes $X$ and $Y$ plays a fundamental role. Quadratic variation is a bounded variation process defined as:
\begin{equation}\label{eq:defbracket}
\langle X, Y \rangle_t = \lim_{n\rightarrow 0} \sum_{i=1}^{p_n} (X_{i}^{n} - X_{i-1}^{n})(Y_{i}^{n} - Y_{i-1}^{n}),
\end{equation}
where $0=t_0^n<t_1^n<\cdots<t_{p_n}^n=t$ is a partition of the interval $[0,t]$ and this limit, if it exists, is defined in the sense of convergence in probability.

Assuming that $X$ and $Y$ are two real-valued continuous semimartingales, defined as $X_t = X_0 + A_t + M_t, Y_t = Y_0 + B_t + N_t$ with $M, N$ martingales and $A, B$ finite variation processes, then their quadratic covariation \eqref{eq:defbracket} exists, and is given by 
\begin{equation}\label{eq:semibracket}
\langle X, Y \rangle_t = \langle M, N \rangle_t.
\end{equation}
In particular, the quadratic variation of a standard Brownian motion $B$ (as a martingale) is given by $\langle B \rangle_t = t$, the quadratic variation of two bounded variation processes $f$, and $g$ (such as deterministic functions) can be shown to be zero ($\langle f,g\rangle_t=0$), as well as the covariation between a martingale and a bounded variation process ($\langle f,M\rangle_t=0$).

The quadratic (co-)variations play an important role in the It\^{o} calculus and its generalisation of the chain rule. In particular, they are involved in the It\^{o} integration by parts formula:
\begin{equation}\label{eq:IPP}
\dif (XY) = X \dif Y + Y \dif X + \dif \langle X, Y \rangle_t.
\end{equation}
The quadratic variation of the It\^{o} integrals of two adapted processes with respect to martingale, $M$ and $N$, respectively, is provided by the following important  formula:
\begin{equation}
    \biggl\langle \int_0^{\cdot} \Theta_s \dif M_s, \int_0^{\cdot} \Theta'_s \dif N_s \biggr\rangle_t = \int_0^t \Theta_s \Theta'_s \dif \langle M, N \rangle_s .
\end{equation}
This property is involved in the It\^{o} isometry, allowing to express the covariance of two It\^{o} integrals:
\begin{equation}\label{eq:isometry}
\Exp \left[ \left(\int_0^t f \dif M_s\right) \left(\int_0^t g \dif N_s\right) \right] = \Exp \left[ \int_0^t fg \,\dif  \langle M, N \rangle_s \right],
\end{equation}
where $f$ and $g$ are two adapted processes such that $\int_0^t f^2 \dif \langle M,M \rangle_s$ and $\int_0^t g^2 \dif \langle N,N \rangle_s$ are integrable.

\section{Stochastic Reynolds transport theorem with source terms}
\label{sec:strat-ito}
The aim of this section is to demonstrate the stochastic Reynolds transport theorem with source terms presented in~\eqref{eq:SRTT_transport}.
To that end, we follow the steps of \citet[Appendix C]{bauer2020a} then, we pass from the Stratonovich to the It\^o form.
Stratonovich and It\^o forms are equivalent {for regular enough processes.
The conversion rule between Stratonovich and It\^{o}--integrals is   provided by the following theorem \citep{kunitabook}:
\begin{theorem}
If $X$ and $Y$ are two real valued continuous semimartingales, the following Stratonovich integral is well defined:
\begin{equation}\label{eq:ItoStra}
X_t\circ\dif Y_t = X_t\dif Y_t + \frac{1}{2}\dif\big\langle X, Y \big\rangle_t.
\end{equation}
\end{theorem}
The bracket in the above expression stands for the martingale covariation. The definition of the quadratic covariation process is recalled in Appendix~\ref{sec:bracket}.
Stratonovich integral has the advantage to be associated to a classical Leibniz product rule, while It\^o calculus involves quadratic variation terms -- see Appendix~\ref{sec:bracket}.
Formal computations are much more easier in Stratonovich setting.
However, opposite to It\^o integral, Stratonovich integral is not anymore a martingale.
As a consequence, the analytical specification of expectation or correlations can only be done in the It\^o representation.

Using the above conversion rule, it can be shown that the flow \eqref{eq:stdisplacement} reads in Stratonovich form \citep{bauer2020a} as
\begin{equation}
\dif \bm{X}_t = \left(\bm{u}-\frac{1}{2}\Div\mathsfbi{a}+\frac{1}{2}\boldsymbol{\sigma}_t(\Div\boldsymbol{\sigma}_t) \right)\dif t + \strat{\boldsymbol{\sigma}_t}{\dBt}.
\end{equation}
We define the characteristic function $\phi(x,t)$ transported by the flow, such that
\begin{equation}
    \phi\left(\bm{X}_t(\bm{x}_0)\right)=g(\bm{x}_0),
\end{equation}
within a compact spatial support $\mathcal{V}(t)$ of non-zero values {that} does not include points on the domain boundary.
We can then write
\begin{equation}
        \dd\int_{\mathcal{V}(t)} (q\phi)(\bm{x},t)\dx
         = \dd\int_{\Omega} (q\phi)(\bm{x},t)\dx
         =
         \int_{\Omega} \strat{\dd_t}{q}\, \phi + q\,\strat{\dd_t}{\phi} \dx.
\end{equation}
Since $\phi$ is transported, we have
\begin{equation}
    \begin{split}
       & \dd\phi(\bm{X}_t,t)=\strat{\dd_t}{\phi}+\Grad{\phi}\bcdot\dd \bm{X}_t=0,\\
       &\strat{\dd_t}{\phi}+\left(u-\frac{1}{2}\Div\mathsfbi{a}+\frac{1}{2}\boldsymbol{\sigma}_t(\Div\boldsymbol{\sigma}_t)\right)\bcdot\Grad{\phi}\,\dif t+\strat{(\Grad{\phi}{\,\bcdot\,}\boldsymbol{\sigma}_t)}{\dBt}=0.
    \end{split}
\end{equation}
We have then
\begin{multline}
        \dd\int_{\mathcal{V}(t)} (q\phi)(\bm{x},t)\dx
         \\=
         \int_{\Omega} \strat{\dd_t}{q}\, \phi - q\left(\left(u-\frac{1}{2}\Div\mathsfbi{a}+\frac{1}{2}\boldsymbol{\sigma}_t(\Div\boldsymbol{\sigma}_t)\right)\bcdot\Grad\phi\,\dif t+\strat{(\Grad{\phi}{\,\bcdot\,}\boldsymbol{\sigma}_t)}{\dBt}\right) \dx
         \\=
         \int_{\Omega} \left[ \strat{\dd_t}{q}
         +\Div\left( q\left(\bigl(u-\frac{1}{2}\Div\mathsfbi{a}+\frac{1}{2}\boldsymbol{\sigma}_t(\Div\boldsymbol{\sigma}_t)\bigr)\dt+\strat{\boldsymbol{\sigma}_t}{\dBt}\right)\right)\right]\phi \dx
        .
\end{multline}
We add now a force and obtain the SRTT in Stratonovich form:
\begin{equation}
    \strat{\dd_t}{q}
    +\Div\left( q\left(\left(u-\frac{1}{2}\Div\mathsfbi{a}+\frac{1}{2}\boldsymbol{\sigma}_t(\Div\boldsymbol{\sigma}_t)\right)\dt+\strat{\boldsymbol{\sigma}_t}{\dBt}\right)\right)
    =Q_t\dt+\strat{\bm{Q}_\sigma}{\dBt}.
\end{equation}
Let us now write this expression in It\^o form:
\begin{equation}
        \ddt{}{x_i}\left(q\strat{\boldsymbol{\sigma}_{t}^{ij}}{\dBt^j}\right)
        =
        \ddt{}{x_i}\left(q\boldsymbol{\sigma}_{t}^{ij}\dBt^j\right)
        +\frac{1}{2}\underbrace{\covariation{\dd_t\left(\ddt{}{x_i}\left(q{\boldsymbol{\sigma}_{s}^{ij}}\delete{(\bm{X}_s)}\right)\right)}{\dBs^j}}_J
    .
\end{equation}
Since $\boldsymbol{\sigma}_t$ is time differentiable {in the Eulerian grid}, we have $$\covariation{\dd_t\boldsymbol{\sigma}_{s}^{ij}}{\dBs^j}=0.$$
Then,
\begin{equation}
    \begin{split}
        J=&\covariation{\left(\ddt{}{x_i}\left(\dd_t q{\boldsymbol{\sigma}_{s}^{ij}}\delete{(\bm{X}_s)}\right)\right)}{\dBs^j}
        \\=&
        \covariation{\left(\ddt{}{x_i}\left(\left(\bm{Q}_\sigma^k\bcdot\dBs^k-\ddt{}{x_l}\left( q\boldsymbol{\sigma}_{s}^{lm}\dBs^m\right) \right){\boldsymbol{\sigma}_{s}^{ij}}\delete{(\bm{X}_s)}\right)\right)}{\dBs^j}
        \\=&
        \ddt{}{x_i}\left(\bm{Q}_\sigma^j\boldsymbol{\sigma}_{t}^{ij}\right)\dt - \ddt{}{x_i}\left( \ddt{}{x_l}\left(q\boldsymbol{\sigma}_{t}^{lj}\right)\boldsymbol{\sigma}_{t}^{ij} \right)\dt 
        \\=&
        \ddt{}{x_i}\left(\bm{Q}_\sigma^j\boldsymbol{\sigma}_{t}^{ij}\right)\dt
         - \ddt{}{x_i}\left( \ddt{q}{x_l}\boldsymbol{\sigma}_{t}^{lj}\boldsymbol{\sigma}_{t}^{ij} \right)\dt 
         - \ddt{}{x_i}\left( q \ddt{\boldsymbol{\sigma}_{t}^{lj}}{x_l}\boldsymbol{\sigma}_{t}^{ij} \right)\dt 
        \\=&
        \Div(\boldsymbol{\sigma}_t\bm{Q}_\sigma)\dt
        -\Div(\mathsfbi{a}\Grad q)\dt
        - \Div(q\boldsymbol{\sigma}_t(\Div\boldsymbol{\sigma}_t))\dt.
    \end{split}
\end{equation}

In addition, we make {the} hypothesis {that} $\dd \bm{Q}_\sigma$ is time-differentiable {in the Lagrangian frame}, such that we have
\begin{equation}
    \begin{split}
        {\dd\int_{\mathcal{V}(t)} \bm{Q}^j_\sigma \dx}
        &= \int_{\mathcal{V}(t)}\dd_t \bm{Q}_\sigma^j + \Div(\bm{Q}^j_\sigma(\udrift\dt+\sdBt))+\Diffst{\bm{Q}^j_\sigma}\dt \dx
        \\&= \int_{\mathcal{V}(t)} F\dt\dx.
    \end{split}
\end{equation}
We can then write
\begin{equation}
    \begin{split}
        {\int_{\mathcal{V}(t)}}\strat{\bm{Q}_\sigma}{\dBt}\,{\dx}
        =&
        {\int_{\mathcal{V}(t)}}\bm{Q}_\sigma\dBt
        +\frac{1}{2}\covariation{\dd_t \bm{Q}_\sigma^i}{\dBs^i}\,{\dx}
       \\=&
        {\int_{\mathcal{V}(t)}}\bm{Q}_\sigma\dBt
        -\frac{1}{2}\covariation{\Div( \bm{Q}_\sigma^i\sdBs)}{\dBs^i}\,{\dx}
       \\=&
        {\int_{\mathcal{V}(t)}}\bm{Q}_\sigma\dBt
        -\frac{1}{2}\covariation{\ddt{}{x_j}( \bm{Q}_\sigma^i\boldsymbol{\sigma}_{s}^{jk}\dBs^k)}{\dBs^i}\,{\dx}
       \\=&
        {\int_{\mathcal{V}(t)}}\bm{Q}_\sigma\dBt
        -\frac{1}{2}\ddt{}{x_j}( \bm{Q}_\sigma^i \boldsymbol{\sigma}_{t}^{ji})\dt\,{\dx}
       \\=&
        {\int_{\mathcal{V}(t)}}\bm{Q}_\sigma\dBt
        -\frac{1}{2}\Div( \boldsymbol{\sigma}_t\bm{Q}_\sigma)\dt\,{\dx},
    \end{split}
\end{equation}
where the correlation operator product $(\boldsymbol{\sigma}_t\bm{Q}_\sigma)$ should be understood in terms of matrix kernel product.
Assembling everything {and dropping the space integral}, we obtain
\begin{multline}
        \strat{\dd_t}{q}
        +\Div\left( q\left(\left(\bm{u}-\frac{1}{2}\Div\mathsfbi{a}+\frac{1}{2}\boldsymbol{\sigma}_t(\Div\boldsymbol{\sigma}_t)\right)\dt+\strat{\boldsymbol{\sigma}_t}{\dBt}\right)\right)
        - Q_t\dt
        -\strat{\bm{Q}_\sigma}{\dBt}
        \\=
         \dd_t q
        +\Div\left( q\left(\left(\bm{u}-\frac{1}{2}\Div\mathsfbi{a}+\frac{1}{2}\boldsymbol{\sigma}_t(\Div\boldsymbol{\sigma}_t)\right)\dt+\sdBt\right)\right)
        \\+
        \frac{1}{2}\left[
            \Div(\boldsymbol{\sigma}_t\bm{Q}_\sigma)\dt
           -\Div(\mathsfbi{a}\Grad q)\dt
           - \Div(q\boldsymbol{\sigma}_t(\Div\boldsymbol{\sigma}_t))\dt
        \right]
        \\
        - Q_t\dt
        - \bm{Q}_\sigma\dBt
        +\frac{1}{2}\Div( \boldsymbol{\sigma}_t\bm{Q}_\sigma)\dt.
\end{multline}
After simplification, we obtain
\begin{equation}
        \dd_t q
       +\Div\left( q\left(\left(\bm{u}-\frac{1}{2}\Div\mathsfbi{a}\right)\dt+\sdBt\right)\right)
       +\Div(\boldsymbol{\sigma}_t\bm{Q}_\sigma)\dt
       = \Diffst{q}\dt
       +
       Q_t\dt
       +
       \bm{Q}_\sigma\dBt
       ,
    \label{eq:SRTT_transport_Strat}
\end{equation}
which is exactly equation~\eqref{eq:SRTT_transport}.
It can be shown after some algebraic manipulations, that this expression is consistent with the implicit form provided by \citet[Appendix D]{resseguier2017e}.

\section{Calculation rules}
\label{sec:calc}
\subsection{Distributivity of the stochastic transport operator}
\label{sec:distr}
The distributivity of the stochastic transport operator is detailed in this section for the case where it is balanced by a random right-hand side forcing term.
If the evolutions of two variables $f$ and $g$ are given by
\begin{equation}
    \begin{split}
       & \mathbb{D}_tf = F_t\,\mathrm{d}t + \bm{F}_\sigma\bcdot\dBt \\
       & \mathbb{D}_tg = G_t\,\mathrm{d}t + \bm{G}_\sigma\bcdot\dBt
      ,
   \end{split}
\end{equation}
then the evolution of the product $fg$ satisfies
\begin{equation}
   \mathbb{D}_t(fg) = f\,\mathbb{D}_tg + g\,\mathbb{D}_tf + \bm{F}_{\sigma} \bcdot \bm{G}_{\sigma}\, \mathrm{d}t - \sadv{(\boldsymbol{\sigma}_t\bm{F}_\sigma)}{g}\dt- \sadv{(\boldsymbol{\sigma}_t\bm{G}_\sigma)}{f}\dt,
   \label{eq:Dfg}
\end{equation}
or less formally:
\begin{multline}
       \mathbb{D}_t(fg)
       =
       f\,\mathbb{D}_tg
       + g\,\mathbb{D}_tf
         + \covariation{ \bm{F}_\sigma\bcdot\dBs }{ \bm{G}_\sigma\bcdot\dBs }
       \\
         - \covariation{ \bm{F}_\sigma\bcdot\dBs }{ \Adv{\sdBs}{g} }
         - \covariation{ \bm{G}_\sigma\bcdot\dBs }{ \Adv{\sdBs}{f} },
   \label{eq:Dfg_nonformal}
\end{multline}
where the It\^o integration by part formula \eqref{eq:IPP} has been used.
This relation proves useful for transforming the conservative form of the Navier--Stokes equations into non-conservative form.

\begin{proof}
\begin{multline}
    \mathbb{D}_t(fg) =
     \dd_t(fg) + \Adv{\udrift}{(fg)}\dt + \Adv{\sdBt}{(fg)} -\Diffst{(fg)}\dt  \\ 
     = f \dd_t g + g \dd_t f
    + \dd_t\langle f,g \rangle
    + f \Adv{\udrift}{g}\dt+ g \Adv{\udrift}{f}\dt
    \\ 
    + f \Adv{\sdBt}{g}+ g \Adv{\sdBt}{f}
    -\frac{1}{2}\Div{(f\mathsfbi{a}\Grad{g} +g\mathsfbi{a}\Grad{f})} \\
    = f \dd_t g + g \dd_t f
    + \covariation{ - \Adv{\sdBs}{f} + \bm{F}_\sigma\bcdot\dBs }{ - \Adv{\sdBs}{g}+\bm{G}_\sigma\bcdot\dBs }
    \\ 
    + f \Adv{\udrift}{g}\dt
    + g \Adv{\udrift}{f}\dt
    + f \Adv{\sdBt}{g}+ g \Adv{\sdBt}{f}
    \\
    -\frac{1}{2}\Big( f\Div{(\mathsfbi{a}\Grad{g})} + {\left((\mathsfbi{a}\Grad{g})\bcdot\nabla\right)f} + g\Div{(\mathsfbi{a}\Grad{f})} + {\left((\mathsfbi{a}\Grad{f})\bcdot\nabla\right)g} \Big) \dt
   .
   \label{eq:proof1}
\end{multline}
Developing only the covariation term:
\begin{multline}
        \covariation{ - \Adv{\sdBs}{f} + \bm{F}_\sigma\bcdot\dBs }{ - \Adv{\sdBs}{g}+\bm{G}_\sigma\bcdot\dBs }
         =
         {\Adv{(\mathsfbi{a}\Grad{f})}{g}} \dt
         \\
         + \covariation{ \bm{F}_\sigma\bcdot\dBs }{ \bm{G}_\sigma\bcdot\dBs }
         - \covariation{ \bm{F}_\sigma\bcdot\dBs }{ \Adv{\sdBs}{g} }
         \\
         - \covariation{ \bm{G}_\sigma\bcdot\dBs }{ \Adv{\sdBs}{f} }
        ,
   \label{eq:proof2}
\end{multline}
which reads
\begin{equation}
         {\Adv{\mathsfbi{a}\Grad{f}}{g}} \dt
         + \bm{F}_\sigma\bcdot\bm{G}_\sigma \dt
         - \sadv{(\boldsymbol{\sigma}_t\bm{F}_\sigma)}{g} \dt
         - \sadv{(\boldsymbol{\sigma}_t\bm{G}_\sigma)}{f}\dt.
   \label{eq:proof3}
\end{equation}
Substituting~\eqref{eq:proof3} into~\eqref{eq:proof1}, we obtain equation~\eqref{eq:Dfg}.
\end{proof}

\subsection{Work of martingale forces}
\label{sec:work_random}
For the sake of clarity, we begin by detailing the calculation rules in the context of point mechanics, in order to define the work of martingale forces.

We consider a force whose impulse is given by a martingale $\dd \bm{F}_\sigma$.
We define its elementary work in a weak sense for any differentiable test function $\phi(t)$ satisfying $\phi(0)=\phi(T)=0$:
\begin{equation}
    \int_0^T\phi(t)\dd W_\sigma = \int_0^T \phi(t) \left(\ddt{}{t}\int_0^t\dd \bm{F}_\sigma\right)\bcdot \dd \bm{X}.
    \label{eq:weakwork}
\end{equation}
The expression $\ddt{}{t}\int_0^t\dd \bm{F}_\sigma$ is written in a formal sense, since $\dd \bm{F}_\sigma$ is a martingale and cannot be differentiated in time.
The work defined in equation~\eqref{eq:weakwork} can be decomposed into two contributions: $\dd W_{\sigma,u}$ associated with the displacement $\bm{u}\,\dd t$, and $\dd W_{\sigma,\sigma}$ associated with the displacement $\sdBt$.
We treat these two terms separately.
For the first component, we have:
\begin{equation}
    \begin{split}
        \int_0^T\phi(t)\dd W_{\sigma,u} &= \int_0^T \phi(t) \left(\ddt{}{t}\int_0^t\dd \bm{F}_\sigma\right)\bcdot \bm{u}(x,t)\dt.
        \\&=-\int_0^T \left(\int_0^t\dd \bm{F}_\sigma\right)\bcdot\left(\phi(t) \bm{u}(x,t)\right)'\dt.
    \end{split}
\end{equation}
The final expression is well-defined and provides a rigorous way to interpret this contribution to the work.

To go further, we note that $\int_0^t\dd \bm{F}_\sigma$ is homogeneous to a Brownian.
Let us now consider the following identity, valid for any time-differentiable function $\bm{\psi}(x,t)$.
Expanding the differential of the product, we have:
\begin{equation}
    \begin{split}
        \int_0^T \dd\left(\bm{\psi}(x,t)\bcdot\int_0^t\dd \bm{F}_\sigma\right)
        &= \int_0^T \left(\int_0^t\dd \bm{F}_\sigma\right)\bcdot\dd\bm{\psi}(x,t)
           + \int_0^T\bm{\psi}(x,t) \dd \bm{F}_\sigma
        \\&
          = \int_0^T \left(\int_0^t\dd \bm{F}_\sigma\right)\bcdot\bm{\psi}'(x,t)\dt
           + \int_0^T\bm{\psi}(x,t) \dd \bm{F}_\sigma.
    \end{split}
\end{equation}
By taking $\bm{\psi}(x,t)=\phi(t)\bm{u}(x,t)$, we obtain
\begin{equation}
    \begin{split}
        \int_0^T\phi(t)\dd W_{\sigma,u} &=\int_0^T\phi(t)\bm{u}(x,t) \dd \bm{F}_\sigma-\int_0^T \dd\left(\phi(t)\bm{u}(x,t)\bcdot\int_0^t\dd \bm{F}_\sigma\right) 
        \\&
        =\int_0^T\phi(t)\bm{u}(x,t) \dd \bm{F}_\sigma - \phi(T)\bm{u}(x,T)\bcdot \bm{F}_\sigma(x,T) 
        \\&
        =\int_0^T\phi(t)\bm{u}(x,t) \dd \bm{F}_\sigma,
    \end{split}
\end{equation}
where the last equality holds because $\phi(T)=0$ by assumption.
From this, we identify the pointwise expression for the work contribution:
\begin{equation}
    \dd W_{\sigma,u} =\bm{u}(x,t) \dd \bm{F}_\sigma.
\end{equation}

The second term $\dd W_{\sigma,\sigma}$ is not well defined, even in weak form.
Informally, it should balance with the kinetic energy associated with the displacement  $\sdBt$, which operates at the same scale.
This highly irregular term that is also not well defined (possibly infinite), has not been considered in the definition of total energy.
Discarding this balance is similar and consistent with the derivation of the momentum in \cite{memin2014}, where the acceleration associated with $\sdBt$ being highly irregular is assumed to be in balance with some forces components of the same nature (\ie of the order of the (weak) acceleration of white noise).

As a consequence, in our model, there is no work contribution from the random forces associated with the Brownian-motion-induced displacement of the control surface.

\section{Displacement of a transported control surface}
\label{sec:displacement_control_surface}
Let us apply the SRTT \eqref{eq:SRTT_transport} to a characteristic function ($q=1$ in $\Omega(t)$, $q=0$ outside) transported by the flow -- meaning $\mathbb{D}_t q=0$.
Using the divergence theorem, we obtain the expression for the volume variation associated with a control surface transported by the stochastic flow.
\begin{equation}
    \begin{split}
       \dd V(t) = {\dd}\int_{\Omega(t)} 1 \,\dx
        &= \int_{\Omega(t)} \Div(\udrift\dt+\sdBt) \, \dx
       \\&=
         \int_{\partial \Omega(t)} \underbrace{(\udrift\dt+\sdBt)}_{\dd \bm{X}_{d,t}}\bcdot\bm{n}\,\dd S.
    \end{split}
\end{equation}
Hence, the normal displacement of the control surface is given by $\dd \bm{X}_{d,t}\bcdot{\bm{n}}$, which involves the modified advection velocity.
As a result, the modified drift velocity must be used in  the definitions of elementary work terms involving  surface integrals.
This drift naturally incorporates the contribution of spatially inhomogeneous noise.

\section{Detailed derivation of the non-conservative stochastic Navier--Stokes equations}
\label{sec:details_non_conservative}
\subsection{Continuity}
Mass conservation follows from applying the stochastic Reynolds transport theorem (SRTT) to the density, \ie by setting $q=\rho$ and assuming the absence of any mass source:
\begin{equation}
    \dd_t \rho
    + \Div\Bigl(\bigl((\bm{u}-\frac{1}{2}\Div\mathsfbi{a})\dt+\sdBt\bigr)\rho\Bigr)
    =
    \Diffst{\rho}\dt.
    \label{eq:continuity}
\end{equation}
Here, the stochastic diffusion term has been moved to the right-hand side to emphasize its role as a sink term, compensating for the source induced by the transport via $\sdBt$ through quadratic variation effects.
As previously mentioned, one can verify using the It\^{o} integration by parts formula that mass is conserved path-wise (\ie for each individual realisation), under appropriate boundary conditions.

Similarly, the conservation of the mass concentration of the species $i$ can be written as:
\begin{multline}
    \label{eq:continuity_frac}
    \dd_t (\rho Y_i)
    + \Div\Bigl(\bigl((\bm{u}-\frac{1}{2}\Div\mathsfbi{a})\dt+\sdBt\bigr)\rho Y_i\Bigr)
    + \Div\bigl(\boldsymbol{\sigma}_t(\rho\bm{Q}_{\sigma}^{Y_i})\bigr)\dt
    =
    \\
    \Diffst{\rho Y_i}\dt
    + \frac{1}{ReSc_i}\Div\bigl(\Grad (\rho Y_i)\bigr)\dt
    + \rho Q_t^{Y_i} \dt
    + \rho \bm{Q}_{\sigma}^{Y_i}\bcdot\dBt.
\end{multline}

Combining equations~\eqref{eq:continuity} and~\eqref{eq:continuity_frac} and applying the stochastic distributivity rule \eqref{eq:Dfg} (see Appendix~\ref{sec:calc}\ref{sec:distr}), yields the stochastic advection-diffusion-reaction equation:
\begin{multline}
    \mathbb{D}_tY_i
    +\sum_k\covariation{\left(\sdBs\right)^k}{\ddt{}{x_k}\bm{Q}_\sigma^{Y_i}\bcdot\dBs}
    =
    \frac{1}{ReSc_i}\Div(\Grad Y_i)\dt
    + Q_{Y_i} \dt
    + \bm{Q}_{\sigma}^{Y_i}\bcdot\dBt.
    \label{eq:transport_Y}
\end{multline}
To obtain this non-conservative form, the same steps as those outlined in Appendix~\ref{sec:momentum} for the momentum equation are followed.
We have thus obtained expressions~\eqref{eq:non_conservative_rho} and~\eqref{eq:non_conservative_Y}.

\subsection{Momentum}
\label{sec:momentum}
The SRTT is now applied to the momentum component $\rho u_i$, balanced by forces, with $u_i\in\{u,v,w\}$:
\begin{multline}
    \dd_t(\rho u_i)
    + \Div\left(\left((\bm{u}-\frac{1}{2}\Div\mathsfbi{a})\dt+\sdBt\right)\rho u_i\right)
    + \Div(\boldsymbol{\sigma}_t\bm{F}_\sigma^{\rho u_i})\dt
    =
    -\ddt{p}{x_i} \dt
     - \ddt{\dd p_t^\sigma}{x_i}-\rho g \delta_{i3}\dt
    \\
    + \Rem\ddt{\tau_{ij}(\bm{u})}{x_j}\dt +
     \Rem\ddt{\tau_{ij}(\sdBt)}{x_j}
    +\Diffst{(\rho u_i)}\dt
    ,
    \label{momentum}
\end{multline}
For the sake of generality, we consider the molecular viscosity stress tensor
\begin{equation}
    \mathsfbi{\tau}(\bm{u})=\mu\left(\nabla \bm{u}+\left(\nabla \bm{u}\right)^T\right)+\left( \mu_b - \frac{2}{3}\mu\right)\Div\bm{u}\,\mathbb{I}, 
\end{equation}
where $\mu_b$ denotes the bulk viscosity.
Similarly to the pressure decomposition, the viscous stress has a finite variation friction contribution due to $\bm{u}\dt$ and a martingale contribution due to $\sdBt$.

Using the distributivity rule~\eqref{eq:Dfg_nonformal}, with $\bm{F}^{\rho}_\sigma\bcdot\dBt= - \rho \Div(\sdBt)$, we obtain:
\begin{multline}
     \rho \dd_t u_i + u_i\dd_t \rho
     -\covariation{ \rho \Div(\sdBs) }{ \bm{F}_{\sigma}^{u_i}\bcdot{\dBs} }
    \\
     + \covariation{ \rho\Div(\sdBs) }{ \Adv{\sdBs}{u_i} }
     - \covariation{ \bm{F}_\sigma^{u_i}\bcdot{\dBs} }{ \Adv{\sdBs}{\rho}}
    \\
     + \rho\Adv{(\bm{u}-\frac{1}{2}\Div\mathsfbi{a})\dt+\sdBt}{u_i}
     +
     u_i\Div\left(\rho\left((\bm{u}-\frac{1}{2}\Div\mathsfbi{a})\dt+\sdBt\right)\right)
     + \Div(\boldsymbol{\sigma}\bm{F}_\sigma^{\rho u_i})\dt
    \\
     -\rho\Diffst{(u_i)}\dt
     -u_i\Diffst{(\rho)}\dt
    =
     - \ddt{p}{x_i} \dt - \ddt{\dd p_t^\sigma}{x_i}
     -\rho g \delta_{i3}\dt
     \\
     + \Rem\Div(\tau_i(\bm{u}))\dt + \Rem\Div(\tau_i(\sdBt))
    ,
    \label{eq:momentum0}
\end{multline}
which simplifies using the continuity equation~\eqref{eq:continuity} to
\begin{multline}
        \rho \dd_t u_i
        +
        \rho\Adv{(\bm{u}-\frac{1}{2}\Div\mathsfbi{a})\dt+\sdBt}{u_i}
        -\covariation{ \Div(\rho \sdBs) }{ \bm{F}_{\sigma}^{u_i}\bcdot{\dBs} }
       \\
        + \covariation{ \rho\Div(\sdBs) }{ \Adv{\sdBs}{u_i} }
        + \Div(\boldsymbol{\sigma}\bm{F}_\sigma^{\rho u_i})\dt
        -\rho\Diffst{(u_i)}\dt
       \\=
        - \ddt{p}{x_i} \dt - \ddt{\dd p_t^\sigma}{x_i}
        -\rho g \delta_{i3}\dt
        + \Rem\Div(\tau_i(\bm{u}))\dt + \Rem\Div(\tau_i(\sdBt))
       .
    \label{eq:momentum1}
\end{multline}
We now, detail the covariation terms in \eqref{eq:momentum1}:
\begin{equation}
    -\covariation{ \Div(\rho\sdBs) }{ \bm{F}_{\sigma}^{u_i}\bcdot{\dBs}}
    =
    -\sum_k\covariation{ \ddt{}{x_k}(\rho\sdBs)^k }{ \bm{F}_{\sigma}^{u_i}\bcdot{\dBs}}
    ,
\end{equation}
or equivalently
\begin{multline}
        \covariation{ \rho\Div(\sdBs) }{ \Adv{\sdBs}{u_i} }
       =\\\sum_{jklm}
        \covariation{ \rho\ddt{}{x_j}\int_\Omega \check{\boldsymbol{\sigma}}^{jk}(\bm{x},\bm{x}')\dBs^k(\bm{x}') \dx'}{ \int_\Omega \check{\boldsymbol{\sigma}}^{lm}(\bm{x},\bm{x}'')\dBs^m(\bm{x}'')\dx '' \ddt{u_i}{x_l} }
        \\=\sum_{j,k,l}
        \rho \int_\Omega \ddt{\boldsymbol{\sigma}^{jk}(\bm{x},\bm{x}')}{x_j} \boldsymbol{\sigma}^{lk}(\bm{x},\bm{x}') \dd \bm{x}' \ddt{u_i}{x_l} \dt
        =
        {\rho \boldsymbol{\sigma}_t(\Div\boldsymbol{\sigma}_t)\bcdot\Grad u_i\dt}
        ,
\end{multline}
and noting from \eqref{eq:momentum0} that $\bm{F}^{u_i}_\sigma= \rho^{-1} \bm{F}_\sigma^{\rho u_i}$,
\begin{equation}
    \begin{split}
        \Div(\boldsymbol{\sigma}\bm{F}_\sigma^{\rho u_i})\dt
        =&
           \sum_{k} \ddt{}{x_k}\covariation{\left(\sdBs\right)^k}{\rho\bm{F}^{u_i}\bcdot\dBs}
        \\=&
           \sum_k\covariation{ \ddt{}{x_k}(\rho\sdBs)^k }{ \bm{F}_{\sigma}^{u_i}\bcdot{\dBs}}
        \\& +
           \sum_{k}\covariation{\rho\left(\sdBs\right)^k}{\ddt{}{x_k}\bm{F}^{u_i}\bcdot\dBs}
         .
    \end{split}
\end{equation}
We finally obtain
\begin{multline}
        \rho\,\mathbb{D}_tu_i
       +\rho\sum_{k}\covariation{\left(\sdBs\right)^k}{\ddt{}{x_k}\bm{F}^{u_i}\bcdot\dBs}
       \\=
        -\ddt{p}{x_i} \dt - \ddt{ \dd p_t^\sigma}{x_i}
        -\rho g \delta_{i3}\dt
        + \Rem\Div(\tau_i(\bm{u}))\dt + \Rem\Div(\tau_i(\sdBt)),
\end{multline}
with $\bm{F}_\sigma^{u_i}\bcdot\dBs = \frac{1}{\rho}\left(-\ddt{\dd p_s^\sigma}{x_i} +\Rem\ddt{\tau_{ij}(\sdBs)}{x_j}\right)$ the martingale part of the right-hand side.
This corresponds to the expression presented in equation~\eqref{eq:non_conservative_u}.

\subsection{Energy}
As in the deterministic framework \citep{woodsbook,andersonbook,landaubook}, we now consider the conservation of the total energy and derive a corresponding transport equation for the temperature.

\subsubsection{General formulation}
The work of forces and heat fluxes acting on a transported control volume induce variations in the specific total energy $E$, such that
\begin{multline}
        \dd_t (\rho E)
        + \Div\left(\left((\bm{u}-\frac{1}{2}\Div\mathsfbi{a})\dt+\sdBt\right)\rho E\right)
        + \Div(\boldsymbol{\sigma}_t\bm{F}^{\rho E}_\sigma)\dt
        \\
        =
        \Diffst{(\rho E)}\dt
        +\dd W- \Div(\dd \bm{q})
        ,
    \label{eq:energy_conservation}
\end{multline}
The martingale part of these RHS terms is written $\bm{F}_\sigma^{\rho E}\bcdot\dBt$, where $\bm{F}^{\rho E}_\sigma$ denotes, as before, a correlation operator.

Similarly to Appendix~\ref{sec:momentum}, and using the distributivity rule~\eqref{eq:Dfg_nonformal} together with the continuity equation~\eqref{eq:continuity}, we obtain:
\begin{equation}
   \rho \mathbb{D}_t(E)
     + \rho\sum_k\covariation{ (\sdBs)^k }{ \ddt{}{x_k}\left(\bm{F}_\sigma^{E}\bcdot\dBs\right)}
    = \dd W- \Div(\dd \bm{q}),
   \label{eq:energy_transport}
\end{equation}
with $\bm{F}_\sigma^{E}=\frac{1}{\rho}\bm{F}_\sigma^{\rho E}$.

\subsubsection{Transport equation of temperature}
\label{sec:transport_T}
By decomposing the total energy into its internal, kinetic, and potential components, and by expressing the kinetic energy contribution using the momentum equation~\eqref{eq:non_conservative_u} along with the transport product distributivity rule~\eqref{eq:Dfg_nonformal} (see steps~\ref{eq:expansion_total_energy} to~\ref{eq:substitution_KE}), we obtain -- after a series of rearrangements detailed below (steps~\ref{eq:rearrangements_T} to~\ref{eq:final_T}) -- the transport equation for the temperature.
We start by the SRTT applied on the total energy transformed in non-conservative form \eqref{eq:energy_transport}:
\begin{equation*}
   \rho \mathbb{D}_t(E)
     + \rho\sum_k\covariation{ (\sdBs)^k }{ \ddt{}{x_k}\left(\bm{F}_\sigma^{E}\bcdot\dBs\right)}
    = \dd W- \Div(\dd \bm{q}).
\end{equation*}
Transport of total energy can be expanded as follows
\begin{multline}
        \rho\mathbb{D}_tE =
           \frac{\rho}{\gamma} \mathbb{D}_tT
         + \frac{\rho}{2}\mathbb{D}_t\|\bm{u}\|^2
         + \rho \mathbb{D}_t (gz)
         \\=
         \frac{\rho}{\gamma} \mathbb{D}_tT
          + \sum_i\rho u_i\mathbb{D}_tu_i
         +\frac{\rho}{2}\sum_i\covariation{ \bm{F}_\sigma^{u_i}\bcdot\dBs }{ \bm{F}_\sigma^{u_i}\bcdot\dBs }
         \\
         -\rho\sum_i\covariation{ \bm{F}_\sigma^{u_i}\bcdot\dBs }{ \Adv{\sdBs}{u_i} }
            + \rho g(w^\star \dt + \left(\sdBt\right)_z)
            -\frac{\rho g}{2}\Div \mat{a}_{\bullet z} \dt
         ,
    \label{eq:expansion_total_energy}
\end{multline}
where the distributivity rule of the transport operator~\eqref{eq:Dfg_nonformal} has been used.
It can be noted that the potential energy is a stochastic variable, of the same nature than the internal and kinetic energies.
This can be viewed by considering the position vertical coordinate as an integral of the vertical velocity, which is a stochastic variable.
Thus, it is valid to apply the stochastic transport operator to the potential energy.

From the momentum equation~\eqref{eq:non_conservative_u} we have
\begin{multline}
    \sum_i \rho u_i\mathbb{D}_tu_i=\sum_i\Big[
     - \rho\sum_k \covariation{(\sdBs)^k}{u_i\ddt{}{x_k}\left( \bm{F}_\sigma^{u_i}\bcdot{\dBs}\right)}
    \\
    -u_i\ddt{p}{x_i} \dt -u_i \ddt{\dd p_t^\sigma}{x_i} + \Rem u_i\Div(\tau_i(\bm{u}))\dt + \Rem u_i \Div(\tau_i(\sdBt))\Big]
    -\rho w g\dt
    .
    \label{eq:expansion_kinetic_energy}
\end{multline}
By substituting~\eqref{eq:expansion_total_energy} and~\eqref{eq:expansion_kinetic_energy} into~\eqref{eq:energy_transport}, we obtain
\begin{multline}
    \frac{\rho}{\gamma} \mathbb{D}_tT
    +\sum_i\Big[
       -\rho\sum_k\covariation{ (\sdBs)^k }{ u_i\ddt{}{x_k}\left( \bm{F}_\sigma^{u_i}\bcdot{\dBs}\right) } 
       - u_i\ddt{p}{x_i} \dt
       - u_i \ddt{ \dd p_t^\sigma }{x_i}
       \\
       + \Rem u_i\Div\tau_i(\bm{u})\dt
       + \Rem u_i \Div\tau_i(\sdBt)
    \Big]
    +\frac{\rho}{2}\sum_i\covariation{ \bm{F}_\sigma^{u_i}\bcdot\dBs }{ \bm{F}_\sigma^{u_i}\bcdot\dBs }
    \\
    -\rho\sum_i\covariation{ \bm{F}_\sigma^{u_i}\bcdot\dBs }{ \Adv{\sdBs}{u_i} }
    + \rho g((w^\star-w) \dt + \left(\sdBt\right)_z)
    -\frac{\rho g}{2}\Div \mat{a}_{\bullet z} \dt
    \\
    + \rho\sum_k\covariation{(\sdBs)^k}{ \ddt{}{x_k}\left( \bm{F}^{E}_\sigma\bcdot\dBs\right) }
    =
    - \Div(p(\udrift\dd t +\sdBt))
    - \Div{\left(\udrift\dd p_t^\sigma\right)}
    \\
    + \Rem \Div \left(\tau(\bm{u})(\udrift\dt+\sdBt)\right)
    + \Rem \Div\left(\tau(\sdBt)\udrift\right)
    +\sum_i \frac{1}{Re Sc_i} \Div \left((\rho Y_i c_{v,i}T) \Grad Y_i\right) \dd t
    \\
    +\RePrm \Div(\Grad T)\dt
    + \dot{Q}\dt
     .
    \label{eq:substitution_KE}
\end{multline}
This expression can be rearranged as follows
\begin{multline}
    \frac{\rho}{\gamma} \mathbb{D}_tT
    +\sum_i\Big[
       {\frac{\rho}{2}\covariation{ \bm{F}_\sigma^{u_i}\bcdot\dBs }{ \bm{F}_\sigma^{u_i}\bcdot\dBs }}
       -\rho\covariation{ \bm{F}_\sigma^{u_i}\bcdot\dBs }{ \Adv{\sdBs}{u_i} }
    \Big]
    \\
    + \rho\sum_k\covariation{(\sdBs)^k }{ \ddt{}{x_k}\left(\bm{F}^{E}_\sigma\bcdot\dBs\right)-\sum_iu_i\ddt{}{x_k}\left( \bm{F}_\sigma^{u_i}\bcdot\dBs\right)}
    = -p\Div(\udrift\dd t +\sdBt)
    \\
    -\dd p_t^\sigma\Div(\udrift)
    \left((\udrift-\bm{u})\dt+\sdBt\right)\bcdot\left( - \Grad p + \rho\bm{g}\right)
    -\left(\udrift-\bm{u}\right)\bcdot\Grad \dd p_t^\sigma
    +\frac{\rho g}{2}\Div \mat{a}_{\bullet z} \dt
    \\
    +\Rem\tau(\bm{u}):\Grad\left(\udrift \dt +\sdBt\right)
    +\Rem\tau(\sdBt):\Grad\udrift
    +\Rem\left((\udrift-\bm{u})\dt+\sdBt\right)\bcdot\left(\Div\tau(\bm{u})\right)
    \\
    +\Rem\left(\udrift-\bm{u}\right)\bcdot\left(\Div\tau(\sdBt)\right)
    +\sum_i \frac{1}{Re Sc_i} \Div \left((\rho Y_i c_{v,i}T) \Grad Y_i\right) \dd t
    +\frac{1}{RePr}\Div(\Grad T)\dt
    + \dot{Q}\dt
    .
    \label{eq:rearrangements_T}
\end{multline}
Since a transport equation of temperature is sought, it is convenient to express quadratic co-variation terms in function of the martingale RHS of the transport of temperature -- noted $\bm{F}_\sigma^T$ following our notation convention -- by remarking that
\begin{multline}
        \sum_k\covariation{(\sdBs)^k }{ \ddt{}{x_k}\left(\bm{F}^{E}_\sigma\bcdot\dBs\right)-\sum_iu_i\ddt{}{x_k}\left(\bm{F}_\sigma^{u_i}\bcdot\dBs\right)}
        \\=
        \sum_k\covariation{(\sdBs)^k }{ \ddt{}{x_k}\left(\bm{F}^{E}_\sigma\bcdot\dBs\right)-\sum_i\ddt{}{x_k}\left( u_i\bm{F}_\sigma^{u_i}\bcdot\dBs\right)}
        \\
        +\sum_{i,k}\covariation{(\sdBs)^k }{ \bm{F}_\sigma^{u_i}\bcdot\dBs\ddt{u_i}{x_k}}
        \\=
        \sum_k\dd_t\Bigg\langle\int_0^{\bcdot}(\sdBs)^k , \int_0^{\bcdot} \ddt{}{x_k}\underbrace{\left(\bm{F}^{E}_\sigma\bcdot\dBs-\sum_iu_i\bm{F}_\sigma^{u_i}\bcdot\dBs\right)}_{\frac{1}{\gamma}\bm{F}^T_\sigma\bcdot\dBs - \bm{g}\bcdot\sdBs}\Bigg\rangle_t
        \\
        +\sum_i\covariation{\Adv{\sdBs}{u_i}}{ \bm{F}_\sigma^{u_i}\bcdot\dBs}
        .
\end{multline}
We recall that 
\begin{equation}
    \begin{split}
       \bm{F}_\sigma^{u_i}\bcdot\dBt = & -\frac{1}{\rho}\ddt{\dd p_t^\sigma}{x_i} +\frac{1}{\rho Re}\Div(\tau_i(\sdBt))
       \\
       \bm{F}_\sigma^{\rho u_i}\bcdot\dBt = & -\ddt{\dd p_t^\sigma}{x_i} +\Rem\Div(\tau_i(\sdBt))
       \\
       \bm{F}_\sigma^{\rho E}\bcdot\dBt = & -\Div(p\sdBt) - \Div(\udrift\dd p_t^\sigma)
       + \Rem\Div(\boldsymbol{\tau}(\bm{u})\sdBt)
       \\&
       + \Rem\Div(\boldsymbol{\tau}(\sdBt)\udrift)
       \\
       \bm{F}_\sigma^{E}\bcdot\dBt = & \frac{1}{\rho}\bm{F}_\sigma^{\rho E}\bcdot\dBt
       \\
       \frac{\rho}{\gamma} \bm{F}^T_\sigma\bcdot\dBt
       = &
       -p\Div\sdBt
       -\dd p_t^\sigma\Div\udrift
       +\Rem\mathsfbi{\tau}(\bm{u}):\Grad\sdBt
       \\&
       +\Rem\mathsfbi{\tau}(\sdBt):\Grad\udrift
       +\sdBt\bcdot\left(-\Grad p + \rho \bm{g} + \Rem\Div\mathsfbi{\tau}(\bm{u})\delete{+\rho\bm{g}}\right)
       \\&
       +\left(\udrift-\bm{u}\right)\bcdot\left(-\Grad \dd p_t^\sigma + \Rem\Div\mathsfbi{\tau}(\sdBt)\right)
       .
    \end{split}
\end{equation}
Observing $\bm{F}^T_\sigma$, defined by the martingale part of the right-hand-side of equation~\eqref{eq:rearrangements_T}, it can be remarked that it is natural to obtain
\begin{equation}
    \frac{\rho}{\gamma}\bm{F}^T_\sigma\bcdot\sdBt=\left(\bm{F}^{\rho E}_\sigma\bcdot\dBt-\sum_iu_i\rho\bm{F}_\sigma^{u_i}\bcdot\dBt\right)+\rho\bm{g}\bcdot\sdBt,
\end{equation}
since the transport equation of internal energy (then temperature) is obtained from the transport equation of total energy to which the contribution of kinetic and potential energy are expressed and subtracted from the LHS and RHS.
As a consequence, the respective contributions of total energy minus kinetic and potential energy appear in the martingale terms of the temperature transport equation.
Moreover, by remarking that
\begin{equation}
    -\rho\covariation{\left(\sdBs\right)_z}{\ddt{}{z} \bm{g}\bcdot\sdBs}=\frac{\rho g}{2}\Div \mat{a}_{\bullet z} \dt,
\end{equation}
the system reduces to
\begin{multline}
    \frac{\rho}{\gamma} \mathbb{D}_tT
    + \frac{\rho}{\gamma}\sum_k \covariation{(\sdBs)^k }{ \ddt{}{x_k}\left(\bm{F}^T_\sigma\bcdot\dBs\right)}
    + \frac{\rho}{2}\sum_i
    \covariation{ \bm{F}_\sigma^{u_i}\bcdot\dBs }{ \bm{F}_\sigma^{u_i}\bcdot\dBs }
    \\
    = -p\Div(\udrift\dd t +\sdBt)
    -\dd p_t^\sigma\Div(\udrift)
    + \left((\udrift-\bm{u})\dt+\sdBt\right)\bcdot\left(-\Grad p + \rho\bm{g} + \Rem\Div\tau(\bm{u})\right)
    \\
    +\left(\udrift-\bm{u}\right)\bcdot\left(-\Grad \dd p_t^\sigma + \Rem\Div\tau(\sdBt)\right)
    +\Rem\tau(\bm{u}):\Grad\left(\udrift \dt +\sdBt\right)
    +\Rem\tau(\sdBt):\Grad\udrift
    \\
    +\sum_i \frac{1}{Re Sc_i} \Div \left((\rho Y_i c_{v,i}T) \Grad Y_i\right) \dd t
    +\frac{1}{RePr}\Div(\Grad T)\dt
    + \dot{Q}\dt,
   \label{eq:final_T}
\end{multline}
which is the expression presented in equation~\eqref{eq:non_conservative_T}.

\subsection{Equation of state}
To close the system, it is necessary to specify an equation of state.
For generality, we adopt the formal representation
\begin{equation}
    p=f(\rho,T,Y_i).
\end{equation}
As in the deterministic framework, since we have evolution equations for both density and temperature, the pressure can be evaluated  explicitly from the equation of state.
This, however, comes at the cost of a Courant-Friedrichs-Lewy (CFL) condition that is constrained by the acoustic (sound) speed.
The martingale pressure component can be identified by differentiating the equation of state using It\^{o} calculus.
Since the equation of state is deterministic -- \ie the functional relation $p=f(\rho,T,Y_i)$ does not depend explicitly on random events -- we can apply the It\^{o} formula to obtain an explicit stochastic evolution equation for the pressure:
\begin{multline}
    \dd_t p = \ddt{f}{\rho}\dd_t\rho + \ddt{f}{T}\dd_tT
    + \sum_i\ddt{f}{Y_i}\dd_tY_i
    + \frac{1}{2}\dddt{f}{\rho}\dd_t\langle \rho,\rho\rangle_t
    + \frac{1}{2}\dddt{f}{T} \dd_t\langle T, T\rangle_t
    + \sum_{i,j}\frac{1}{2}\ddt{^2f}{Y_i\partial Y_j} \dd_t\langle Y_i, Y_j\rangle_t
    \\
    + \ddt{^2f}{\rho\partial T}\dd_t\langle \rho,T\rangle_t
    + \sum_i\ddt{^2f}{\rho\partial Y_i}\dd_t\langle \rho,Y_i\rangle_t
    + \sum_i\ddt{^2f}{T\partial Y_i}\dd_t\langle T,Y_i\rangle_{t}
    =
    \ddt{\widetilde{p}}{t}\dt + \frac{\dd p_t^\sigma}{{\tau}},
    \label{eq:diff_EOS}
\end{multline}
where $\widetilde{p}$ denotes the finite variation (time-differentiable) part of the pressure, which includes, in particular, all the covariation (quadratic variation) terms.
The second term, $\dd p_t^\sigma{/\tau}$, corresponds to the martingale contribution, with $\tau$ denoting a characteristic decorrelation time.
This decorrelation time represents the typical time over which the martingale pressure remains coherent and effectively transfers momentum.
It is assumed to be the same as the decorrelation time introduced in previous works \citep[\eg][]{kadriharouna2017,bauer2020a} in the context of relating the variance tensor to the variance of velocity fluctuations, via the relation: (\ie $\mathsfbi{a} = \tau\, \mathbb{E} (\bm{u}' {\bm{u}'}^T)$).
Accordingly, the term $\dd p^\sigma$, identified from~\eqref{eq:diff_EOS}, represents the martingale pressure component that appears in the stochastic momentum equation.

If we assume that the pressure arises from a homentropic process -- that is, of acoustic nature with the same entropy level between fluid particles -- we can write
\begin{equation}
    \mathbb{D}_t p = \ddt{p}{\rho}\Big\vert_s \mathbb{D}_t\rho = c^2 \mathbb{D}_t\rho,
    \label{eq:homentropic_Lagrangian}
\end{equation}
with $c$ denoting the speed of sound and $s$ the entropy.
This expression is written in the Lagrangian frame since the transported fluid parcel is submitted to isentropic transformations.
We neglect here quadratic variations between right-hand side of the stochastic transport of $\rho$ (and then $p$) and $\sdBt$; approximation being exact for divergence-free noise.
This allows to amalgamate the stochastic transport operator $\mathbb{D}_t$ as the material derivative \citep[see][for further details]{resseguier2017e}.
We neglect as well $\partial (c^2) / \partial \rho$ arising in the Itô formulae when differentiating the equation of state.
Substituting~\eqref{eq:homentropic_Lagrangian} from the continuity equation~\eqref{eq:continuity}, we identify
\begin{equation}
    \dd p_t^\sigma = - {\tau}\Big(\Adv{\sdBt}{p} + \rho c^2\Div(\sdBt)\Big).
   \label{eq:dpt_ac}
\end{equation}
This modelling choice considers the random pressure to be of acoustic nature, and could be useful in some flow regimes.
In the results section, the stochastic pressure will physically represent non-hydrostatic effects.
A combination of both processes could be considered to relax the Boussinesq approximation, as it is proposed in \citet{tissotSTUOD2023}.

\section{Detailed derivation of the Boussinesq approximation}
\label{sec:details_Boussinesq}
\subsection{Density}
Substituting the expansion of density~\eqref{eq:density_expansion} in the continuity equation~\eqref{eq:non_conservative_rho} and neglecting terms $\mathcal{O}(\epsilon)$, we obtain:
\begin{equation}
    \Div\udrift=0
    \quad
    ;
    \quad
    \Div\sdBt=0
    \quad
    ;
    \quad
    \mathbb{D}_t\left(\rho_1+\rho_2\right)=0.
    \label{eq:continuity_BQ}
\end{equation}
Both the drift velocity and the noise are divergence-free, and the density perturbations are stochastically transported by the flow.
Since $\Div\sdBt=0$, the transport operator $\mathbb{D}_t(\bcdot)$ can be directly used in the SRTT \eqref{eq:SRTT_transport} formulation.

The terms of order $\epsilon$ of equation~\eqref{eq:continuity_BQ} can be expressed in terms of buoyancy, $b=-\epsilon\,g\rho_2/\rho_0$, leading to:
\begin{equation}
    \frac{\rho_0}{g}\mathbb{D}_tb=\left(w^* \dt+(\sdBt)_z\right)\ddt{\rho_1}{z}-\frac{1}{2}\Div\left(\mathsfbi{a}_{\bullet z}\ddt{\rho_1}{z}\right)\dt
    ,
    \label{eq:simpleBoussinesq}
\end{equation}
with
\begin{equation}
    \mathsfbi{a}=
    \begin{pmatrix}
        \mathsfbi{a}_{HH^T} & \mathsfbi{a}_{Hz}\\
        \mathsfbi{a}_{zH^T}             & a_{zz}
    \end{pmatrix}
    ,
\end{equation}
and $\mathsfbi{a}_{zH^T}=\mathsfbi{a}_{Hz}^T${, for $H=(x\ y)^T$}.

\subsection{Thermodynamic effects}
\label{sec:Boussinesq_thermo}
Equation~\eqref{eq:simpleBoussinesq} is part of the stochastic analogue of what is commonly referred to as the \emph{simple Boussinesq} equations.
This system typically includes, in addition, the so-called traditional approximation of the Coriolis term, which neglects the horizontal component of the Earth's rotation.
In the ocean, thermodynamic effects can play a significant role.
To account for these, we follow the approach of \cite{vallisbook} by coupling the buoyancy and energy equations.
In particular, the impact of the heat sources $\dot{Q}\dt$ on the fluid dynamics is described by the following set of equations.

Assuming a linear equation of state for sea water around the reference state ($\rho_0$, $T_0$, $S_0$, $p_0$), the density can be expressed as:
\begin{equation}
    \rho=\rho_0\Big(1-\beta_T(T-T_0)
    +\beta_S(S-S_0)
    +\beta_p (p-p_0)\Big),
    \label{eq:linearEOS}
\end{equation}
where $S$ is the salinity, $\beta_T=-1/\rho_0\ddt{\rho}{T}$, $\beta_S=1/\rho_0\ddt{\rho}{S}$ and $\beta_p=1/\rho_0c^2$ are the thermal expansion, haline contraction, and compressibility coefficients, respectively, derived from the Taylor expansion.

The transport equation~\eqref{eq:non_conservative_Y} applied to salinity, written for $Y_i=S$, becomes
\begin{equation}
    \mathbb{D}_tS
    =
    \molec{\frac{1}{ReSc}\Delta S \dt }
    Q_{S} \dt
    +
    \bm{Q}_{S}\bcdot\dBt
    ,
    \label{eq:transport_S}
\end{equation}
where the source terms model physical processes such as evaporation and precipitation at the sea surface, dilution from freshwater inputs, or salinity changes near regions of iceberg formation and melting.
Notably, ocean-atmosphere exchanges can occur on timescales that are fast relative to oceanic dynamics, which justifies the inclusion of a martingale source term $\bm{Q}_{S}\bcdot\dBt$.
It can be remarked that the covariation terms are $\mathcal{O}(\epsilon)$, and are thus neglected in this framework.

We apply the stochastic transport operator to eq.~\eqref{eq:linearEOS} and  obtain
\begin{equation}
    \begin{split}
        &\mathbb{D}_t\rho=
         -\rho_0\beta_T\mathbb{D}_tT
         +\rho_0\beta_S\mathbb{D}_tS
         +\frac{1}{c^2}\mathbb{D}_t p,
        \\&
        \mathbb{D}_t\left(\rho-\frac{1}{c^2}p\right)=-\gamma \beta_T \dd Q_T
    \molec{+ \rho_0\beta_S\left(\frac{1}{ReSc}\Delta S \dt + Q_{S} \dt\right),}
    + \rho_0\beta_S\dd Q_{S},
    \end{split}
\end{equation}
where $\dd Q_T$ collects all work, source, and quadratic variation terms, as detailed in equation~\eqref{eq:non_conservative_T}, and $\dd Q_S = Q_{S} \dt+\bm{Q}_{S}\bcdot\dBt$.
We introduce the potential buoyancy $b_\phi$, which directly captures the effects of work and heat source:
\begin{equation}
    b_\phi\triangleq-\frac{g}{\rho_0}\left(\epsilon( \rho_1 + \rho_2) + \frac{\rho_0gz}{c^2}\right)=b_{N}+b-g\frac{z}{H_p},
\end{equation}
where $b_N=-\epsilon\,g\rho_1/\rho_0$ is associated with the background stratification, and  $H_p=c^2/g$ is a pressure scale height.
The evolution of $b_\phi$ then reads:
\begin{equation}
    \mathbb{D}_t b_\phi
    =\frac{\gamma g\beta_T}{\rho_0}\dd Q_T
     \molec{- g\beta_S\left(\frac{1}{ReSc}\Delta S \dt + Q_{S} \dt\right),}
     - g\beta_S \dd Q_{S}.
\end{equation}
In the absence of viscosity and under divergence-free velocity, the thermal source term simplifies to:
\begin{equation}
    \dd Q_T=\rho_0{\left((w^\star-w)\dt+\left(\sdBt\right)_z\right)b -\left(\udrift-\bm{u}\right)\bcdot\Grad \dd p_t^\sigma}
           + \dot{Q}\dt
           - A_u - A_T
           .
\end{equation}

By defining the buoyancy frequency as
\begin{equation}
    N^2(z)\triangleq\ddt{}{z}\left(-\epsilon g\frac{\rho_1}{\rho_0}-g\frac{z}{H_p}\right)=-\epsilon\frac{g}{\rho_0}\ddt{\rho_1}{z}-\frac{g^2}{c^2},
\end{equation}
the effects of stratification and thermodynamics can be explicitly incorporated into the buoyancy equation.
The full buoyancy evolution then reads
\begin{equation}
    \mathbb{D}_tb+\left(w^\star \dt+(\sdBt)_z\right)N^2-\frac{1}{2}\Div\left(\mathsfbi{a}_{\bullet z}N^2\right)\dt
    =\frac{\gamma g\beta_T}{\rho_0}\dd Q_T
    \molec{- g\beta_S\left(\frac{1}{ReSc}\Delta S \dt + Q_{S} \dt\right).}
    - g\beta_S \dd Q_{S}.
    \label{eq:buoyancy_full}
\end{equation}

\subsection{Momentum}
Concerning the momentum equation, we neglect the viscous terms.
Within this framework, the gravitational acceleration $g$ is assumed to be of order $\mathcal{O}(1/\epsilon)$.
We recall that the pressure field is decomposed as follows:
\begin{equation}
    p=\underbrace{p_0(z)}_{\mathcal{O}(\frac{1}{\epsilon})}+ p_1 + p_2 + \mathcal{O}(\epsilon),
\end{equation}
where $p_0$ and $p_1$ satisfy hydrostatic balance:
\begin{equation}
    \ddt{p_0}{z}=- g\rho_0
    \quad \text{and} \quad
    \ddt{p_1}{z}=-\epsilon g\rho_1.
\end{equation}
The momentum equation \eqref{eq:non_conservative_u} then becomes
\begin{multline}
    \Big(\rho_0+\epsilon(\rho_1+\rho_2)\Big)\,\mathbb{D}_tu_i
    -\rho\sum_k\covariation{(\sdBs)^k}{\ddt{}{x_k}\left(\frac{1}{\rho}\ddt{\dd p_s^\sigma}{x_i}\right)}
    =
    -\ddt{p_2}{x_i} \dt - \ddt{ \dd p_t^\sigma}{x_i}
    \\-\epsilon\rho_2g\delta_{i3}\dt.
\end{multline}
As in Section~\ref{sec:Boussinesq_thermo}, the covariation terms are neglected, which yields
 \begin{equation}
    \begin{split}
        \mathbb{D}_tu_i
       =
        -\frac{1}{\rho_0}\ddt{p_2}{x_i} \dt - \frac{1}{\rho_0}\ddt{ \dd p_t^\sigma}{x_i} \underbrace{-\epsilon\frac{\rho_2}{\rho_0}g}_{b}\delta_{i3}\dt.
        \label{eq:momentum_Boussinesq}
    \end{split}
\end{equation}
The pressure components $(p_2,\dd p_t^\sigma)$ are in quasi-hydrostatic balance, as detailed in~\eqref{eq:dpt_boussinesq}.

\subsection{Summary}
By collecting the equations~\eqref{eq:buoyancy_full}, \eqref{eq:momentum_Boussinesq}, \eqref{eq:dpt_boussinesq}, we obtain the following stochastic Boussinesq system with thermodynamic forcing
\begin{equation}
    \left\{
    \begin{aligned}
       &
       \mathbb{D}_tu_i
       =
       -\frac{1}{\rho_0}\ddt{p_2}{x_i} \dt - \frac{1}{\rho_0}\ddt{ \dd p_t^\sigma}{x_i}
       \quad\text{for}\quad i=\{u,v\}\\&
        w=\frac{1}{2}(\Div\mathsfbi{a})_z-\int_{-H}^z\left(\ddt{u^\star}{x}+\ddt{v^\star}{y}\right)\,\dd z
       \\&
       \Div(\sdBt)=0
       \\&
       p_2=-\rho_0\int_z^\eta \left( \Adv{\frac{1}{2}\Div\mathsfbi{a}}{w}+\Diffst{w} + b \right)\,\dd z 
       \\&
       \dd p_t^\sigma = \rho_0\int_z^\eta \Adv{\sdBt}{w}\,\dd z
       \\&
       \mathbb{D}_tb+\left(w^\star \dt+(\sdBt)_z\right)N^2-\frac{1}{2}\Div\left(a_{\bullet z}N^2\right)\dt
        =\frac{\gamma g\beta_T}{\rho_0}\dd Q_T
        - g\beta_S \dd Q_{S}
       .
    \end{aligned}
    \right.
    \label{eq:boussinesq}
\end{equation}

Note that if the rigid-lid assumption is relaxed, an additional evolution equation \citep[of stochastic shallow water type;][]{brecht2021} for the free surface $\eta$ should be introduced in system~\eqref{eq:boussinesq}.
In this case, the surface pressure is associated to the divergence of the barotropic component.
In system~\eqref{eq:boussinesq}, the pressure is obtained through a relaxed hydrostatic balance, and the vertical velocity is diagnosed kinematically from the divergence-free condition on the drift velocity.

Neglecting thermodynamic effects and assuming a strong hydrostatic balance (\ie weak to moderate noise regime), we recover the simple Boussinesq system presented in \cite{resseguier2017e} -- albeit here without the traditional Coriolis correction.
Of course, including the Coriolis force could be done straightforwardly without major difficulties.

In some applications, a more accurate evaluation of the buoyancy is required, and can be obtained through an equation of state $\rho_{\text{\tiny BQ}}(T,p,S)$ (\eg eq.~\eqref{eq:linearEOS}), combined with transport equations for temperature and salinity.
Under the assumptions introduced earlier, we consider the temperature transport \eqref{eq:non_conservative_T} while neglecting molecular diffusion terms.
Following \citet{tailleux2012}, the work of compression cannot be ignored under the Boussinesq approximation, because velocity divergence, which scales as $\mathcal{O}(\epsilon)$, is multiplied by the pressure, which is in hydrostatic balance and scales as $\mathcal{O}(1/\epsilon)$.
This leads to the following expression for the pressure work term:
\begin{equation}
    P_t = -p\Div\left(\udrift \dt +\sdBt\right)
    = \left(\rho_0 g (\eta-z)\right)\left(-\frac{1}{\rho_0}\mathbb{D}_t\rho_{\textit{\tiny BQ}}\right)
    =- g (\eta-z)\mathbb{D}_t\rho_{\textit{\tiny BQ}}
    .
    \label{eq:Pt_boussinesq}
\end{equation}
Here, buoyancy is diagnosed through the equation of state.
The full system can then be written as:
\begin{equation}
    \left\{
    \begin{aligned}
       &
       \mathbb{D}_tu_i
       =
       -\frac{1}{\rho_0}\ddt{p_2}{x_i} \dt - \frac{1}{\rho_0}\ddt{ \dd p_t^\sigma}{x_i}
       \quad\text{for}\quad i=\{u,v\}\\&
        w=\frac{1}{2}(\Div\mathsfbi{a})_z-\int_{-H}^z\left(\ddt{u^\star}{x}+\ddt{v^\star}{y}\right)\,\dd z
       \\&
       \Div(\sdBt)=0
       \\&
       p_2=-\rho_0\int_z^\eta \left( \Adv{\frac{1}{2}\Div\mathsfbi{a}}{w}+\Diffst{w} + b \right)\,\dd z
       \\&
       \dd p_t^\sigma = \rho_0\int_z^\eta \Adv{\sdBt}{w}\,\dd z
       \\&
       \frac{\rho_0}{\gamma} \mathbb{D}_tT=
       - g (\eta-z)\mathbb{D}_t\rho_{\textit{\tiny BQ}}
       +{\left(\frac{1}{2}\Div\mathsfbi{a}\dt-\sdBt\right)\bcdot\Grad p_2}
       +{\frac{1}{2}\Div\mathsfbi{a}\bcdot\Grad \dd p_t^\sigma}
       - A_T
       - A_u 
       + \dot{Q}\dt
       \\&
       \mathbb{D}_tS
       =
       Q_t^{S} \dt+ \bm{Q}_{\sigma}^{S}\bcdot\dBt
       \\&
       b=g\left(1+\epsilon\frac{\rho_1}{\rho_0}-\frac{\rho_{\text{\tiny BQ}}}{\rho_0}\right)
       .
    \end{aligned}
    \right.
    \label{eq:boussinesq_EOS_app}
\end{equation}
This corresponds to the system~\eqref{eq:boussinesq_EOS} considered in the numerical application.

\section{Stochastic temperature equation in free convection LES}\label{sec:sto_Teq_details}
We here provide some additional guidelines for the numerical application of Section~\ref{sec:numerical}.
First, our LES assumes a linear, temperature driven equation of state:
\beq
    \rho_{\textit{\tiny BQ}}=\rho_0(1-\beta_T(T-T_0)),
    \label{eq:rho_bq}
\eeq
with $\beta_T=2.048\times10^{-4}$\,\SI{}{kg.m^{-3} K^{-1}} the thermal expansion coefficient and $T_0$ a reference temperature.
Using (\ref{eq:rho_bq}) in (\ref{eq:boussinesq_EOS}) simplifies the stochastic temperature equation to:
\beq
\begin{aligned}
\rho_0\left(\frac{1}{\gamma}-g\beta_T(\eta-z)\right) \mathbb{D}_tT=
      {\left(\frac{1}{2}\bnabla \bcdot\ba \dt-\sdBt\right)\bcdot\Grad p_{2}}
      +{\frac{1}{2} \bnabla\bcdot\ba\bcdot\Grad \dd p_t^\sigma}
        -A_T -A_u.
      \label{eq:temp_eq2}
\end{aligned}
\eeq
Note that surface radiative fluxes are been lumped into sub-grid scale vertical temperature fluxes $\lb wT_{sgs}\rb$ as upper boundary conditions within $\mathbb{D}_tT$, in agreement with current practices in general circulation models (GCMs ; See Section~\ref{sec:numerical}\ref{subsec:num_exp}).

Next, the transport noise is defined through its Mercer representation 
\beq
    \bsig(\bx, t) \dif \B_t = \sqrt{\tau} \sum_{n} \sqrt{\lambda_n}(t) \bs{\phi}_n(\bx,t)\dif \beta_t^{n},
    \label{eq:noise_spectral}
\eeq
with its associated variance as:
\beq
    \ba(\bx,t)\mathrm{d}t = \tau \sum_{n}\lambda_n(t) \bs{\phi}_n(\bx,t) \bs{\phi}_n^\transp(\bx,t)\mathrm{d}t,
    \label{eq:var_spectral}
\eeq
with $\bs{\phi_n}$ and $\lambda_n$ the eigenfunctions basis and corresponding eigenvalues, respectively, with the dimension of a velocity.
In our doubly periodic lateral boundary conditions, Fourier modes is a suitable basis for a spectral decomposition \citep[\eg][]{berkooz1993},
leading to several simplifications as we now show.
Let us decompose the three-dimensional noise $\sdBt=(\sigma \dif B_t^{(x)}, \sigma \dif B_t^{(y)}, \sigma \dif B_t^{(z)})$ as horizontal Fourier modes, such that:
\beq
    \sdBt(\bs{x}_H,z,t) = \sum_{m,n} \hat{\bs{b}}_{m,n}(z,t)  \bs{\phi}_{m,n}(\bs{x}_H) \dif \B_t^{(m,n)},
\eeq
with $\bs{\phi}_{m,n} = e^{2\pi i \bs{k}^{(m,n)} \cdot \bs{x}_H}$ a complex horizontal plane wave, 
with $\bs{x}_H=(x,y)^T$ the horizontal unit vector and $\bs{k}=(k,l)^T$ the horizontal wavevector.
The Brownians $\dif \B_t^{(m,n)}$ are constrained to be organised in dependent complex-conjugated pairs to ensure $\sdBt$ to be real-valued.
The summation indices
$(m,n)$ are associated with $(k,l)$ such that $\bs{k}^{(m,n)}\cdot \bs{x}_H=(k^mx, l^ny)$.
In addition, $\hat{\bs{b}}_{m,n}=(\hat{a}_{m,n}, \hat{b}_{m,n}, \hat{c}_{m,n})^T$ are the complex Fourier modes associated with three-dimensional noise (\ie $\hat{a}$ is associated with the $x$ component of noise, $\hat{b}$ with $y$ component, and $\hat{c}$ with $z$ component).
Following \cite{Li-Memin-Tissot-2023}, we first define the two-point covariance matrix
$Q(\bs{x},\bs{x}',t)=Cov \lp \bsig \dif \B_t(\bs{x}), \bsig \dif \B_t(\bs{x}') \rp$ as:
\beq
\begin{aligned}
    Q(\bs{x},\bs{x}',t) 
    &= \Exp \left[ \bsig \dif \B_t (\bsig \dif \B_t)^{\dagger} \right] \\
    &= \Exp \left[ \lp \sum_{m,n} \hat{\bs{b}}_{m,n}(z,t)  \bs{\phi}_{m,n}(\bs{x}_H) \dif \B_t^{(m,n)} \rp \lp \sum_{p,q} {\hat{\bs{b}}_{p,q}(z',t)}^{\dagger}  \bs{\phi}_{p,q}^{\dagger}(\bx'_H) \overline{\dif \B_t}^{(p,q)} \rp \right] \\
    &= \sum_{m,n} \sum_{p,q} \Exp \left[ \dif \underbrace{\lb \B^{(m,n)}, \overline{\B}^{(p,q)}  \rb_t}_{\delta_{(m,n)(p,q)}\dt}   \right] 
    \hat{\bs{b}}_{m,n}(z,t)  \hat{\bs{b}}_{p,q}^{\dagger}(z',t)  \bs{\phi}_{m,n}(\bs{x}_H)  \bs{\phi}_{p,q}^{\dagger}(\bx'_H)   \\
    &= \sum_{m,n} \hat{\bs{b}}_{m,n}(z,t)  \hat{\bs{b}}_{m,n}^{\dagger}(z',t)  \bs{\phi}_{m,n}(\bs{x}_H)  \bs{\phi}^{\dagger}_{m,n}(\bx'_H) \dif t
    \label{eq:qxx_local}
\end{aligned}
\eeq
with $\cdot^{\dagger}$ the transpose-conjugate operation,
$\cdot\transp$ the transpose operation and $\overline{\cdot}$ the complex conjugate.
With this,  the diagonal of the two-point covariance matrix reads:
\beq
\begin{aligned}
       Q(\bs{x}_H,z,t,\bs{x}_H,z,t) 
       &= \sum_{m,n} \hat{\bs{b}}_{m,n}(z,t) \hat{\bs{b}}_{m,n}^{\dagger} (z,t) |e^{2\pi i \bs{k}^{(m,n)} \cdot (\bs{x}_H-\bs{x}_H)}|^2 \dif t \\
       &=
       \sum_{m,n}
       \abs{
       \begin{bmatrix}
        \hat{a}_{m,n}^2 & \hat{b}_{m,n}\hat{a}_{m,n} & \hat{c}_{m,n}\hat{a}_{m,n}\\
        \hat{a}_{m,n}\hat{b}_{m,n} & \hat{b}_{m,n}^2 & \hat{c}_{m,n}\hat{b}_{m,n} \\
        \hat{a}_{m,n}\hat{c}_{m,n} & \hat{b}_{m,n}\hat{c}_{m,n} & \hat{c}_{m,n}^2
       \end{bmatrix}}(z,t) \dif t,
       \label{eq:a_fourier}
\end{aligned}
\eeq
\ie the noise variance $\ba(\bs{x}_H,z,t)=Q(\bs{x}_H,z,t,\bs{x}_H,z,t)/\dt$ is horizontally homogeneous,
and the vertical and temporal structure of $\ba$
are set by the one-point one-time covariance matrix of the horizontal Fourier coefficients of the three-dimensional noise. 
In addition, through horizontal isotropy of the system, off-diagonal elements are identically zero and $\hat{a}_{m,n}^2=\hat{b}_{m,n}^2$.
This last conditions are not perfectly met numerically, where we rather observe small residual in the off-diagonal elements of $\ba$ (Figure~\ref{fig:a_coef_fourier_physic}) and small differences between $\hat{a}_{m,n}^2$ and $\hat{b}_{m,n}^2$.
We have verified that computing the 1-point covariance as the ensemble mean of stochastic velocities, \ie $\left< \bsig \dif \B_t (\bsig \dif \B_t)^{\mgt{\dagger}} \right>$, convergences toward these estimates (top and bottom right panels). 
In order to insure the definition of a divergent-free noise, we express it on the original model grid (\ie Arakawa C type), where $x$, $y$ and $z$ components of the noise are estimated on east, north and upper faces of a tracer cell.
Then, the noise is constructed based on a randomly selected time frame of residual velocities within the filtering window, and this operation is repeated several (100 in our case) times in order to obtain an ensemble estimate.
This procedure insures a divergent-free noise definition at machine precision.
\begin{figure}
    \centering
    \includegraphics[width = \textwidth]{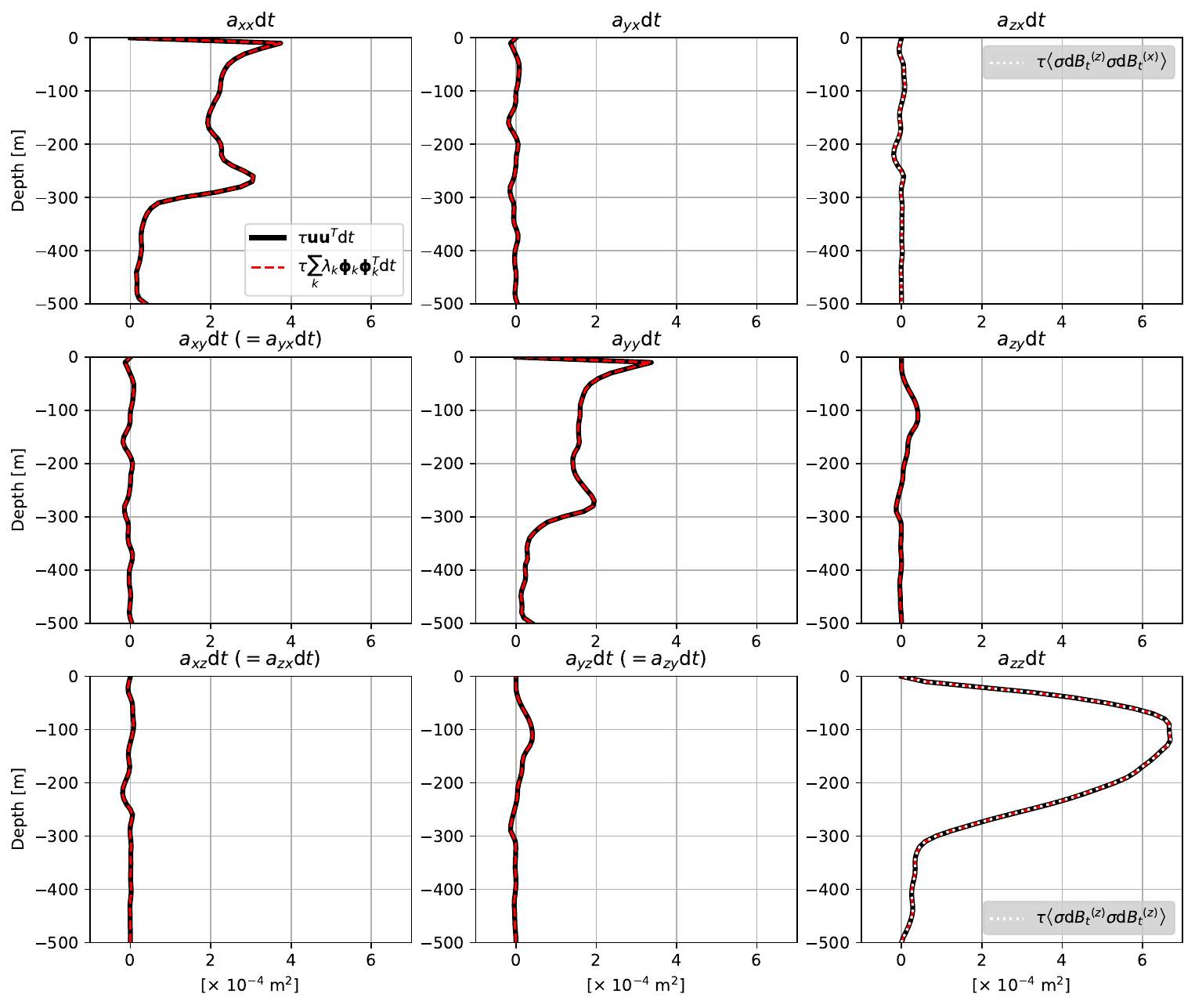}
    \caption{Estimates of the horizontally averaged 1-time, 1-point covariance of the noise $\ba\mathrm{d}t$ from non-Boussinesq, non-hydrostatic simulations of free convection (details in Section~~\ref{subsec:num_exp}). Stochastic estimates are provided in dashed red, and estimates of 1-time, 1-point covariance of velocity $\tau\bu\bu^{\transp}\mathrm{d}t$ are provided in black for comparison, with $\tau=$\SI{1}{s}. Top right and bottom right panels also show the estimate of $a_{zx}$ and $a_{zz}$ in white dashed line, computed as the ensemble mean of the covariance between vertical and zonal stochastic velocities, \ie 
    $\left< (\sigma \dif B_t)^{(z)} (\sigma \dif B_t)^{(x)} \right>$ and $\left< (\sigma \dif B_t)^{(z)} (\sigma \dif B_t)^{(z)} \right>$, respectively, where $\lb \cdot \rb$ stands for the ensemble mean taken over 100 realisations. } 
    \label{fig:a_coef_fourier_physic}
\end{figure}

In this study we consider the horizontally averaged temperature equation
\beq
    \dif_t \oxy{T}=\dif_t \lp \frac{1}{A}\int_{xy} T\,\dd x \dd y \rp,
\eeq
with $A=L_xL_y$ the area of the domain, which leads to several simplifications.
Owing to the divergent-free definition of the noise $\Div \sdBt=0$ and the resolved flow $\Div \bu =0$, double periodic lateral boundary conditions insures zero horizontal flux at the side of the domain. Thus, we have: 
\beq
    \oxy{\sdBt}= 0, 
\eeq
\beq
    \oxy{\sdBt \bcdot \bnabla q}= \pz \oxy{ (\sigma \dif B_t^{(z)} q) }, 
    \label{eq:div_sto_up}
\eeq
and 
\beq
    \oxy{\bu \bcdot \bnabla q}=\pz \oxy{wq}.
\eeq

Secondly, the Itô-Stokes drift reduces to:
\beq
    \oxy{\Adv{\Big(\frac{1}{2}\Div\mathsfbi{a}\Big)}{q}}
    = \oxy{\Big( \frac{1}{2} 
    \begin{bmatrix}
        \px a_{xx} \\
        \py a_{yy} \\
        \pz a_{zz} 
    \end{bmatrix}
    \cdot \nabla \Big) q }
    = 
    \frac{1}{2}\lp \pz a_{zz}\rp \pz \oxy{q},
    \label{eq:ustar_homo}
\eeq
where the first equality results from isotropic conditions and the second equality to horizontally homogeneous conditions of the variance tensor $\ba$. Similarly, the diffusion term reduces to:
\beq
    \oxy{\Diffst{q}}
    =\oxy{\frac{1}{2} \bnabla \bcdot
    \begin{bmatrix}
        a_{xx} \px q\\
        a_{yy} \py q\\
        a_{zz} \pz q
    \end{bmatrix}
    }
    =\frac{1}{2}\pz \lp a_{zz}\pz \oxy{q} \rp.
\eeq

Similar simplifications also arise in other terms of (\ref{eq:temp_eq2}), and we finally obtain an equation of the form :
\begin{multline}
    \rho_0\lp\frac{1}{\gamma}-g\beta_T(\eta-z)\rp \dif_t \oxy{T} =
    \rho_0\lp\frac{1}{\gamma}-g\beta_T(\eta-z)\rp \Bigg(
    - \underbrace{\pz \oxy{wT} \dt}_{(A)}
    + \underbrace{\frac{1}{2}\lp \pz a_{zz} \rp \pz \oxy{T} \dt}_{(B)}
   \\
    - \underbrace{\pz \oxy{(\sigma \dif B_t^{(z)} T) }}_{(C)}
    + \underbrace{\frac{1}{2}\pz \lp a_{zz}\pz \oxy{T} \rp \dt}_{(D)}
    \Bigg)
    +\underbrace{\frac{1}{2} \pz a_{zz} \lp \pz \oxy{p_{qnh}} \rp \dt
    -\pz \oxy{ (\sigma \dif B_t^{(z)} p_{qnh}) }
    }_{D_t}
    \\
    + \underbrace{\frac{1}{2}\lp\pz a_{zz}\rp\pz \oxy{\dd p_t^\sigma}}_{ D_{\sigma}}
    - \oxy{A_T} 
    - \oxy{A_u}.
    \label{eq:Teq_hz_avg}
\end{multline}
Further, the first term of $D_t$ involves the vertical gradient of $p_{qnh}$, with $p_{qnh}$ the quasi-non-hydrostatic pressure defined as the two first terms in (\ref{eq:boussinesq_EOS}), which equals to:
\beq
    \pz{p_{qnh}} = \rho_0 \left( \frac{1}{2}(\pz a_{zz})\pz w
    +\frac{1}{2} \pz (a_{zz}\pz w)\right),
    \label{eq:dp2dz_2}
\eeq
Upon horizontally averaging, this reduces to zero\footnote{We note the contribution of pressure at the surface $\eta$ (typically atmospheric pressure loading) has been ignored here.}.
The evaluation of second contribution of $D_t$, however, requires the full 3D structure of $p_{qnh}$, and we have used (\ref{eq:div_sto_up}) to express it as the divergence of vertical pressure flux.
Similarly, the martingal component of quasi-nonhydrostatic pressure reduces to:
\beq
    \pz \oxy{\dd p_t^\sigma} = -\rho_0 \oxy{\sigma \dif B_t^{(z)} \pz w}.
    \label{eq:dz_dptsigma}
\eeq
Thus, the final stochastic transport equation for horizontally averaged temperature
profile reads:
\begin{multline}
    \rho_0\lp\frac{1}{\gamma}-g\beta_T(\eta-z)\rp \dif_t \oxy{T} =
    \rho_0\lp\frac{1}{\gamma}-g\beta_T(\eta-z)\rp \Bigg(
    - \underbrace{\pz \oxy{wT} \dt}_{(A)}
    + \underbrace{\frac{1}{2}\lp \pz a_{zz} \rp \pz \oxy{T} \dt}_{(B)}
    \\
    - \underbrace{ \pz(\oxy{ \sigma \dif B_t^{(z)} T) }}_{(C)}
    + \underbrace{\frac{1}{2}\pz \lp a_{zz}\pz \oxy{T} \rp \dt}_{(D)}
    \Bigg)
   - \underbrace{\pz \oxy{ (\sigma \dif B_t^{(z)} p_{qnh}) }}_{D_t} 
   - \underbrace{\frac{\rho_0}{2}\lp\pz a_{zz}\rp \lb\sigma \dif B_t^{(z)}\pz w\rb }_{D_{\sigma}}
   \\
   - \oxy{A_T}
   - \oxy{A_u},
   \label{eq:Teq_hz_avg2}
\end{multline}
with quadratic covariations terms $A_u$ and $A_T$ further developed in what follows.
We have evaluated the respective contributions of each terms of (\ref{eq:Teq_hz_avg2}), the results of which are detailed in Figure~\ref{fig:Tsto_bgt_internal} and Figure~\ref{fig:Tsto_bgt_potential} for internal and potential energy contribution, respectively,
and are discussed in Section~\ref{sec:numerical}\ref{subsec:1D_temp}.
\begin{figure}
    \centering
    \includegraphics[width = \textwidth]{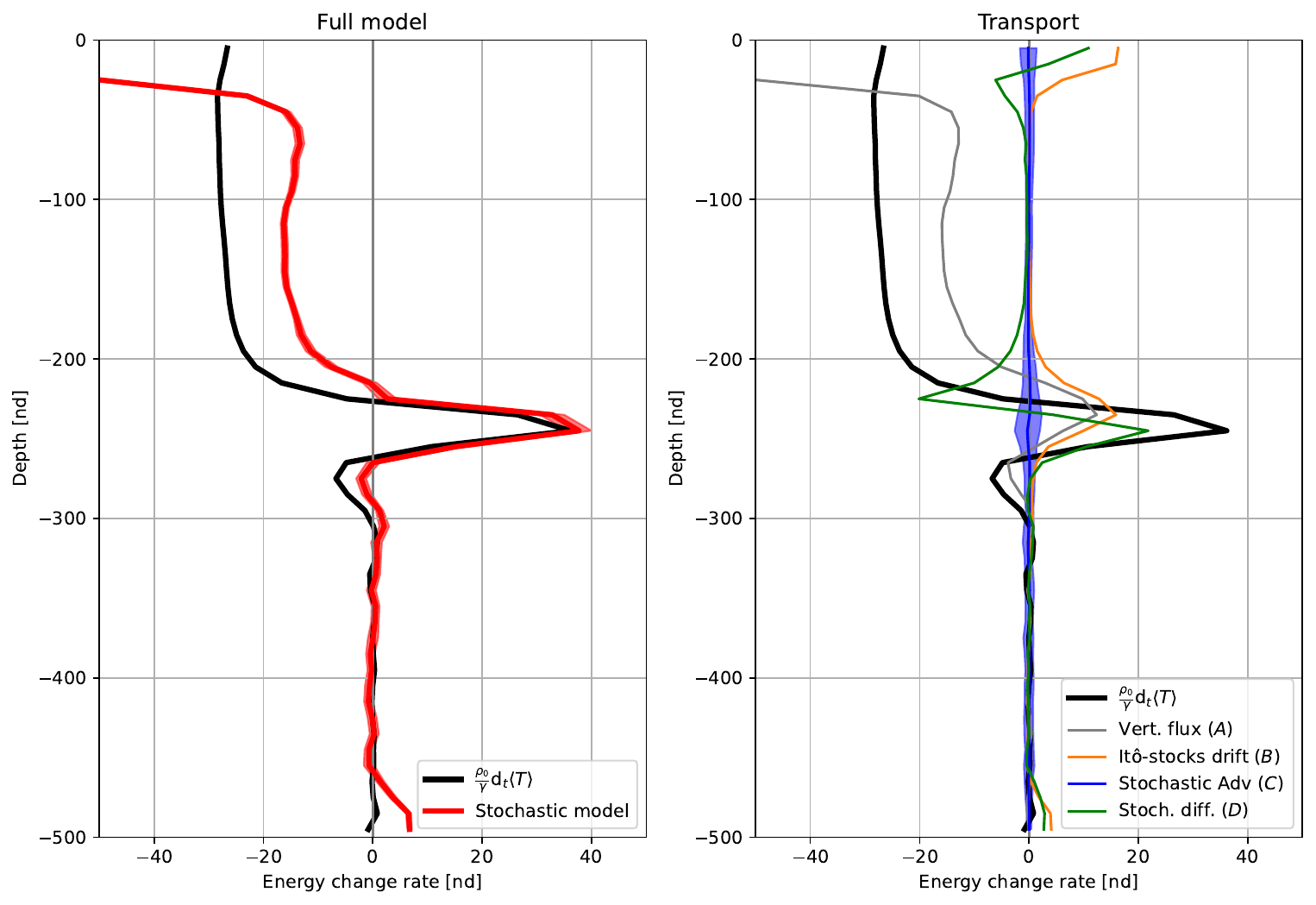}
    \caption{Contribution of the different terms of the  stochastic temperature equation (\ref{eq:temp_eq2}) for internal energy, with the full model (top left panel) and the transport terms (top right panel), quadratic co-variation terms (bottom left panel), and drift works (bottom right panels). Temperature changes over the 3-hour period (thick black) is provided for reference.} 
    \label{fig:Tsto_bgt_internal}
\end{figure}

\begin{figure}
    \centering
    \includegraphics[width = \textwidth]{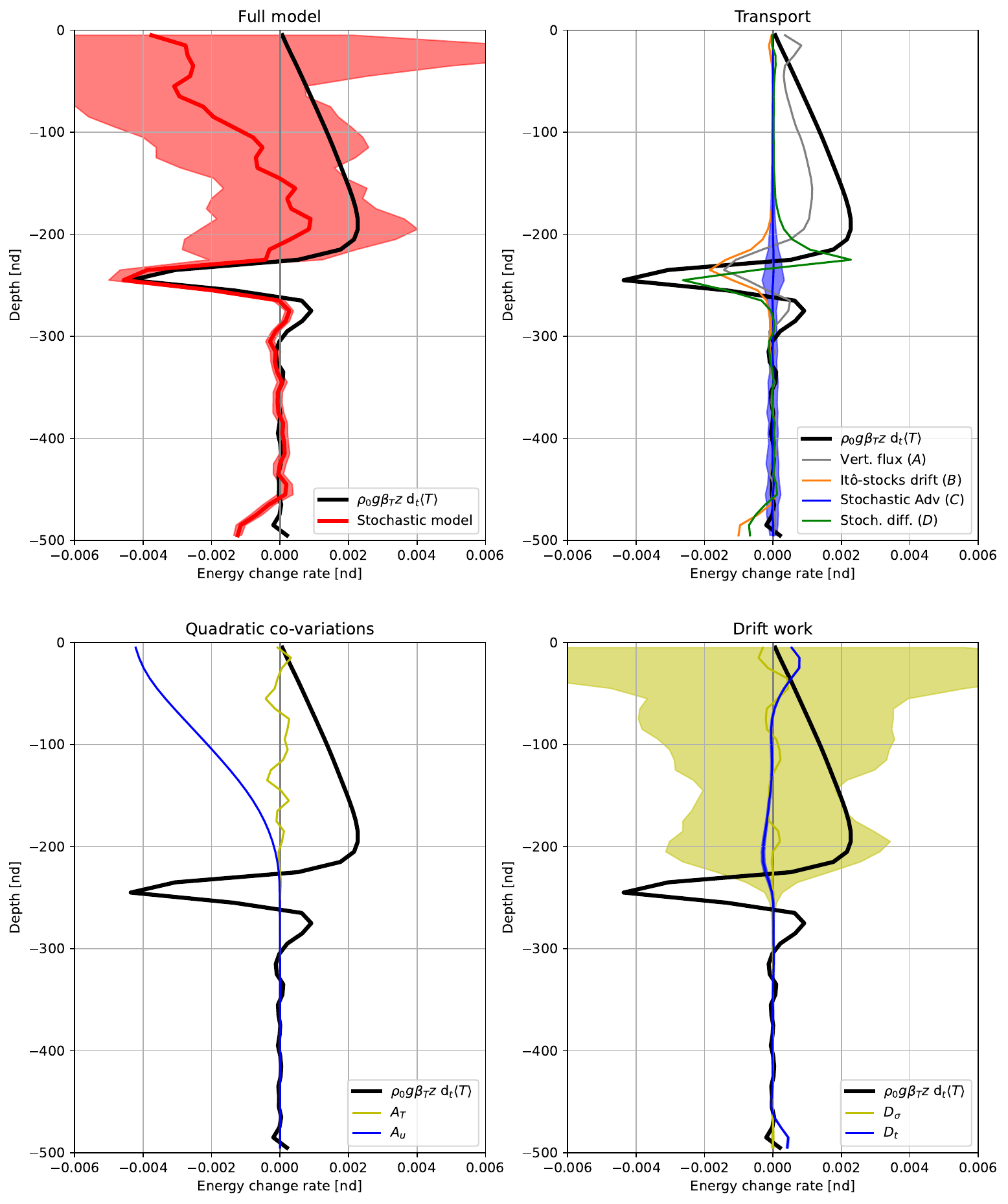}
    \caption{Same as Figure \ref{fig:Tsto_bgt_internal} but for potential energy contribution. In addition, contribution of quadratic co-variation terms (bottom left panel), and drift works (bottom right panels).} 
    \label{fig:Tsto_bgt_potential}
\end{figure}

\subsection{Quadratic covariation terms $A_u$, $A_T$}\label{subsec:quadr_covar}
Starting with the quadratic variation $A_u$ induced by the gradient of the stochastic pressure in the momentum equation, 
adapted from (\ref{eq:non_conservative_T}) for a Boussinesq system horizontally averaged with doubly periodic boundary conditions, we have (ignoring viscous contributions):
\beq
    A_u = \frac{\rho_0}{2} \sum_i \oxy{\dif_t \lb \int_0^{\cdot}\bs{F}_{\sigma}^{u_i} \bcdot \dif \B_s , \int_0^{\cdot} \bs{F}_{\sigma}^{u_i} \bcdot \dif \B_s \rb_t},
\eeq
with 
\beq
    \bs{F}_{\sigma}^{u_i} \bcdot \dif \B_s = -\frac{1}{\rho_0} \partial_{x_i} \dif p_s^{\sigma},
\eeq
and 
\beq
    \dif p_s^{\sigma} = \rho_0 \int_z^{\eta}  \bnabla \bcdot \lp \bsig\dif \B_s w \rp~\dd z.
\eeq

Considering the co-variation term $A_T$ in our doubly periodic boundary conditions setting, 
only the co-variation between the vertical component displacement $\bsig_s \dif \B_s$ with the vertical gradient of the martingale part of the RHS of the temperature stochastic transport equation $\bs{F}_{\sigma}^T \bcdot \dif \B_s$ will play a role in the horizontally averaged temperature equation. It thus reduces to:
\beq
    A_T = \frac{\rho_0}{\gamma} 
    \oxy{ \dif_t \lb
    \int_0^{\cdot} \lp \sigma_s \dif B_s \rp^{(z)},
    \int_0^{\cdot}\partial_z \lp \bs{F}_{\sigma}^T \bcdot \dif \B_t
    \rp \rb_t},
    \label{eq:A_T}
\eeq
with the martingale part of the RHS of the temperature stochastic transport equation defined as:
\beq
    \frac{\rho_0}{\gamma} \bs{F}_{\sigma}^T \bcdot \dif \B_s = 
    - \bsig_t \dif \B_t \bcdot \lp \nabla p_{qnh} \rp
     -(\bu^{*}-\bu) \bcdot \nabla \dif p_t^{\sigma}
     -\dif p_t^{\sigma} \nabla \bcdot \bu^*.
     \label{eq:FTdBs}
\eeq 
It appears from (\ref{eq:FTdBs}) that the full, 3D structure of $\bs{F}_{\sigma}^T \bcdot \dif \B_s$ is to be considered for the evaluation of $A_T$.

\bibliographystyle{ametsocV6}
\bibliography{biblio}

@article{Debussche-Memin25,
	author = {Arnaud Debussche and Etienne M\'{e}min},
	journal = {Physica D: Nonlinear Phenomena},
	title = {Variational principles for fully coupled stochastic fluid dynamics across scales},
	year = {2025},
	volume = {481},
	pages = {134777}
	}

@article{Debussche-Memin-Moneyron26,
	author = {Arnaud Debussche and Etienne M\'{e}min and Antoine Moneyron},
	journal = {Rendiconti Lincei – Matematica e Applicazioni},
	title = {Stochastic interpretations of the oceanic primitive equations with relaxed hydrostatic assumptions},
	year = {2026},
	volume = {in press}
	}

@article{shyy1997,
title = {Compressibility effects in modeling complex turbulent flows},
journal = {Progress in Aerospace Sciences},
volume = {33},
number = {9},
pages = {587-645},
year = {1997},
issn = {0376-0421},
author = {Shyy, W. and Krishnamurty, V. S.},
}

@article{Cotter-et-al-2020,
	author = {Cotter, C. and Crisan, D. and Holm, D. and Pan, W. and Shevchenko, I.},
	date = {2020/06/01},
	id = {Cotter2020},
	isbn = {1572-9613},
	journal = {Journal of Statistical Physics},
	number = {5},
	pages = {1186--1221},
	title = {Data Assimilation for a Quasi-Geostrophic Model with Circulation-Preserving Stochastic Transport Noise},
	volume = {179},
	year = {2020}}

@article{Gugole-Franzke-2019,
	author = {Gugole, F. and Franzke, C.},
	journal = {Mathematics of Climate and Weather Forecasting},
	month = {2023-08-03},
	number = {1},
	pages = {45--64},
	title = {Numerical Development and Evaluation of an Energy Conserving Conceptual Stochastic Climate Model},
	volume = {5},
	year = {2019}}

@unpublished{memin2023linear,
	author = {M{\'e}min, E. and Li, L. and Lahaye, N. and Tissot, G. and Chapron, B.},
	note = {ArXiv preprint arXiv:2304.10183},
	title = {Linear waves solutions of a stochastic shallow water model},
	year = {2023}}

@inproceedings{Dufee-Memin-Crisan-2023,
	address = {Cham},
	author = {Duf{\'e}e, B. and M{\'e}min, E. and Crisan, D.},
	booktitle = {Stochastic Transport in Upper Ocean Dynamics},
	date = {2023//},
	id = {10.1007/978-3-031-18988-3{\_}4},
	isbn = {978-3-031-18988-3},
	pages = {43--56},
	publisher = {Springer International Publishing},
	title = {Observation-Based Noise Calibration: An Efficient Dynamics for the Ensemble {K}alman Filter},
	year = {2023}}

@inproceedings{Li-Memin-Tissot-2023,
	address = {Cham},
	author = {Li, L. and M{\'e}min, E. and Tissot, G.},
	booktitle = {Stochastic Transport in Upper Ocean Dynamics},
	date = {2023//},
	id = {10.1007/978-3-031-18988-3{\_}11},
	isbn = {978-3-031-18988-3},
	pages = {179--193},
	publisher = {Springer International Publishing},
	title = {Stochastic Parameterization with Dynamic Mode Decomposition},
	year = {2023}}

@article{Tissot-et-al-2023,
	author = {Tissot, G. and Cavalieri, A. V. G. and M\'emin, E.},
	issue = {3},
	journal = {Phys. Rev. Fluids},
	month = {Mar},
	numpages = {21},
	pages = {033904},
	publisher = {American Physical Society},
	title = {Input-output analysis of the stochastic Navier-Stokes equations: Application to turbulent channel flow},
	volume = {8},
	year = {2023}}

@article{Flandoli:2022sb,
	author = {Flandoli, F. and Galeati, L. and Luo, D.},
	date = {2022/01/31},
	journal = {Philosophical Transactions of the Royal Society A: Mathematical, Physical and Engineering Sciences},
	journal1 = {Philosophical Transactions of the Royal Society A: Mathematical, Physical and Engineering Sciences},
	journal2 = {Philosophical Transactions of the Royal Society A: Mathematical, Physical and Engineering Sciences},
	month = {2023/07/20},
	number = {2219},
	pages = {20210096},
	publisher = {Royal Society},
	title = {Eddy heat exchange at the boundary under white noise turbulence},
	volume = {380},
	year = {2022},
	year1 = {2022}}

@article{Holm2015,
	author = {Holm, D.},
	journal = {Proceedings of the Royal Society A: Mathematical, Physical and Engineering Sciences},
	number = {2176},
	pages = {20140963},
	publisher = {The Royal Society Publishing},
	title = {Variational principles for stochastic fluid dynamics},
	volume = {471},
	year = {2015}}

@article{Mikulevicius-Rosovskii-2005,
	author = {R. Mikulevicius and B. L. Rozovskii},
	date = {2005/1/1},
	journal = {The Annals of Probability},
	journal1 = {The Annals of Probability},
	journal2 = {The Annals of Probability},
	month = {1},
	number = {1},
	pages = {137--176},
	title = {Global {L}$_{2}$-solutions of stochastic {N}avier--{S}tokes equations},
	volume = {33},
	year = {2005}}

@article{breit-et-al-2021,
author = {Breit, Dominic and Feireisl, Eduard and Hofmanov\'{a}, Martina and Zatorska, Ewelina},
title = {Compressible {N}avier--{S}tokes System with Transport Noise},
journal = {SIAM Journal on Mathematical Analysis},
volume = {54},
number = {4},
pages = {4465-4494},
year = {2022}
}

@ARTICLE{boadi2025,
       author = {Boadi, R. and Breit, D. and Moyo, T. C.},
        title = "{Compressible {E}uler equations with transport noise}",
      journal = {arXiv e-prints},
         year = 2025,
        month = nov,
          eid = {arXiv:2511.20131},
        pages = {arXiv:2511.20131},
          doi = {10.48550/arXiv.2511.20131},
archivePrefix = {arXiv},
       eprint = {2511.20131},
 primaryClass = {math.AP},
       adsurl = {https://ui.adsabs.harvard.edu/abs/2025arXiv251120131B}
}

@article{street-crisan2021,
	author = {O. Street and D. Crisan},
	journal = {Proceedings of the Royal Society A},
	number = {2247},
	pages = {20200957},
	publisher = {The Royal Society Publishing},
	title = {Semi-martingale driven variational principles},
	volume = {477},
	year = {2021}}

@article{Lang2023,
	author = {Lang, O. and Crisan, D. and M{\'e}min, E.},
	date = {2023/02/20},
	id = {Lang2023},
	isbn = {1422-6952},
	journal = {Journal of Mathematical Fluid Mechanics},
	number = {2},
	pages = {29},
	title = {Analytical Properties for a Stochastic Rotating Shallow Water Model Under Location Uncertainty},
	volume = {25},
	year = {2023}}

@article{Goodair-et-al2022,
  title={Existence and uniqueness of maximal solutions to {SPDEs} with applications to viscous fluid equations},
  author={Goodair, Daniel and Crisan, Dan and Lang, Oana},
  journal={Stochastics and Partial Differential Equations: Analysis and Computations},
  volume={12},
  number={2},
  pages={1201--1264},
  year={2024},
  publisher={Springer}
}

@article{Agresti-et-al-2022,
	author = {A. Agresti and M. Hieber and A. Hussein and M. Saal},
	journal = {Stochastics and Partial Differential Equations: Analysis and Computations},
	title = {The stochastic primitive equations with transport noise and turbulent pressure},
        volume = 12,
        pages = {51--133},
	year = {2022}}

@article{Flandoli-et-al-10,
	author = {Flandoli, F. and Gubinelli, M. and Priola, E.},
	date = {2010/04/01},
	id = {Flandoli2010},
	isbn = {1432-1297},
	journal = {Inventiones mathematicae},
	number = {1},
	pages = {1--53},
	title = {Well-posedness of the transport equation by stochastic perturbation},
	volume = {180},
	year = {2010}}

@article{Brzezniak-Slavik-2021,
	author = {Z. Brze{{\'z}}niak and J. Slav{\'\i}k},
	issn = {0022-0396},
	journal = {Journal of Differential Equations},
	pages = {617-676},
	title = {Well-posedness of the {3D} stochastic primitive equations with multiplicative and transport noise},
	volume = {296},
	year = {2021}}

@article{Crisan-Flandoli-Holm-2019,
	author = {Crisan, D. and Flandoli, F. and Holm, D.},
	journal = {Journal of Nonlinear Science},
	number = {3},
	pages = {813--870},
	publisher = {Springer},
	title = {Solution properties of a 3{D} stochastic {E}uler fluid equation},
	volume = {29},
	year = {2019}}

@article{Flandoli-Russo2023,
	archiveprefix = {arXiv},
	author = {F. Flandoli and F. Russo},
	eprint = {2305.19293},
	journal = {ArXiv},
	primaryclass = {math.PR},
	title = {Reduced dissipation effect in stochastic transport by {G}aussian noise with regularity greater than 1/2},
	volume = {2305.19293},
	year = {2023}}

@article{Flandoli-Galeati-Luo-2021,
	author = {F. Flandoli and L. Galeati and D. Luo},
	eprint = {https://doi.org/10.1080/03605302.2021.1893748},
	journal = {Communications in Partial Differential Equations},
	number = {9},
	pages = {1757-1788},
	publisher = {Taylor & Francis},
	title = {Delayed blow-up by transport noise},
	volume = {46},
	year = {2021}}

@article{Flandoli-Luo-2021,
	author = {Flandoli, F. and Luo, D.},
	date = {2021/06/01},
	id = {Flandoli2021},
	isbn = {1432-2064},
	journal = {Probability Theory and Related Fields},
	number = {1},
	pages = {309--363},
	title = {High mode transport noise improves vorticity blow-up control in 3{D} {N}avier--{S}tokes equations},
	volume = {180},
	year = {2021}}

@misc{Debussche-Hofmanova2023,
      title={Rough analysis of two scale systems}, 
      author={Debussche, A. and Hofmanov\'a, M.},
      year={2024},
      eprint={2306.15781},
      archivePrefix={arXiv},
      primaryClass={math.PR},
      url={https://arxiv.org/abs/2306.15781}, 
      publisher={arXiv}
}

@article{Debussche-Pappalattera2023,
   title={Second order perturbation theory of two-scale systems in fluid dynamics},
   ISSN={1435-9863},
   url={http://dx.doi.org/10.4171/JEMS/1501},
   DOI={10.4171/jems/1501},
   journal={Journal of the European Mathematical Society},
   publisher={European Mathematical Society - EMS - Publishing House GmbH},
   author={Debussche, Arnaud and Pappalettera, Umberto},
   year={2024},
   month=jul }

@book{DaPrato,
	author = {G. Da Prato and J. Zabczyk},
	publisher = {Cambridge University Press},
	title = {Stochastic equations in infinite dimensions},
	year = {1992}}

@article{Debusshe-Hug-Memin-2023,
	author = {Debussche, A. and Hug, B. and M{\'e}min, E.},
	date = {2023/01/17},
	id = {Debussche2023},
	isbn = {1422-6952},
	journal = {Journal of Mathematical Fluid Mechanics},
	number = {1},
	pages = {19},
	title = {A Consistent Stochastic Large-Scale Representation of the {N}avier--{S}tokes Equations},
	volume = {25},
	year = {2023}}

@article{tissotJFM2021,
	author = {Tissot, G. and Cavalieri, A. V. G. and M{\'e}min, E.},
	journal = {Journal of Fluid Mechanics},
	pages = {A51},
	publisher = {Cambridge University Press},
	title = {Stochastic linear modes in a turbulent channel flow},
	volume = {912},
	year = {2021}}

@InProceedings{tissotSTUOD2023,
    author="Tissot, G. and M{\'e}min, E. and Jamet, Q.",
    editor="Chapron, Bertrand and Crisan, Dan and Holm, Darryl and M{\'e}min, Etienne and Radomska, Anna",
    title="Stochastic Compressible {N}avier--{S}tokes Equations Under Location Uncertainty",
    booktitle="Stochastic Transport in Upper Ocean Dynamics II",
    year="2024",
    publisher="Springer Nature Switzerland",
    address="Cham",
    pages="293--319",
    isbn="978-3-031-40094-0"
}

@book{andersonbook,
	author = {Anderson, J. D. and Wendt, J.},
	publisher = {Springer},
	title = {Computational fluid dynamics},
	volume = {206},
	year = {1995}}

@article{bauer2020a,
	author = {Bauer, W. and Chandramouli, P. and Chapron, B. and Li, L. and M{\'e}min, E.},
	journal = {{Journal of Physical Oceanography}},
	month = Feb,
	publisher = {{American Meteorological Society}},
	title = {{Deciphering the role of small-scale inhomogeneity on geophysical flow structuration: a stochastic approach}},
        volume = "50",
        number = "4",
        pages=   {983--1003},
	year = {2020}}

@article{bauer2020b,
	author = {Bauer, W. and Chandramouli, P. and Li, L. and M{\'e}min, E.},
	hal_id = {hal-02666147},
	hal_version = {v1},
	journal = {{Ocean Modelling}},
	pages = {1-50},
	publisher = {{Elsevier}},
	title = {{Stochastic representation of mesoscale eddy effects in coarse-resolution barotropic models}},
	volume = {151},
	year = {2020}}

@article{brecht2021,
	author = {Brecht, R. and Li, L. and Bauer, W. and M{\'e}min, E.},
	journal = {{Journal of Advances in Modeling Earth Systems}},
	number = {12},
	pages = {1-28},
	publisher = {{American Geophysical Union}},
	title = {{Rotating shallow water flow under location uncertainty with a structure-preserving discretization}},
	volume = {13},
	year = {2021}}

@article{chandramouli2018,
	author = {Chandramouli, P. and Heitz, D. and Laizet, S. and M\'emin, E.},
	issn = {0045--7930},
	journal = {Computers \& Fluids},
	pages = {170--189},
	title = {Coarse large-eddy simulations in a transitional wake flow with flow models under location uncertainty},
	volume = {168},
	year = {2018}}

@book{chassaingbook,
  title={Variable density fluid turbulence},
  author={Chassaing, P. and Antonia, R. A. and Anselmet, F. and Joly, L. and Sarkar, S.},
  volume={69},
  year={2013},
  publisher={Springer Science \& Business Media}
}

@article{chorin1967,
	author = {Chorin, A. J.},
	issn = {0021-9991},
	journal = {Journal of Computational Physics},
	number = {1},
	pages = {12-26},
	title = {A numerical method for solving incompressible viscous flow problems},
	volume = {2},
	year = {1967}}

@article{dewar2015,
	author = {Dewar, W. K. and Schoonover, J. and McDougall, T. J. and Young, W. R.},
	journal = {Journal of Physical Oceanography},
	number = {1},
	pages = {149--156},
	publisher = {American Meteorological Society},
	title = {Semicompressible ocean dynamics},
	volume = {45},
	year = {2015}}

@article{eden2015,
	author = {Eden, C.},
	journal = {Journal of Physical Oceanography},
	number = {3},
	pages = {630--637},
	publisher = {American Meteorological Society},
	title = {Revisiting the energetics of the ocean in {B}oussinesq approximation},
	volume = {45},
	year = {2015}}

@article{graham2013,
	address = {Boston MA, USA},
	author = {Graham, F. S. and McDougall, T. J.},
	journal = {Journal of Physical Oceanography},
	number = {5},
	pages = {838 - 862},
	publisher = {American Meteorological Society},
	title = {Quantifying the Nonconservative Production of Conservative Temperature, Potential Temperature, and Entropy},
	volume = {43},
	year = {2013}}

@article{kadriharouna2017,
	author = {Kadri Harouna, S. and M{\'e}min, E.},
	hal_id = {hal-01394780},
	hal_version = {v2},
	journal = {{Computers and Fluids}},
	month = Aug,
	pages = {456-469},
	publisher = {{Elsevier}},
	title = {{Stochastic representation of the Reynolds transport theorem: revisiting large-scale modeling}},
	volume = {156},
	year = {2017}}

@book{kunitabook,
	author = {Kunita, H.},
	publisher = {Cambridge university press},
	title = {Stochastic flows and stochastic differential equations},
	volume = {24},
	year = {1997}}

@book{landaubook,
	author = {Landau, L. D. and Lifshitz, E. M.},
	publisher = {Elsevier},
	title = {Fluid Mechanics, Course of Theoretical Physics},
	volume = {6},
	year = {2013}}

@article{li2025, 
      author = {Li, L. and Mémin, E. and Chapron, B.},
      title = "A generalized stochastic formulation of the Ekman-Stokes model with statistical analyses",
      journal = "Journal of Physical Oceanography",
      year = "2025",
      publisher = "American Meteorological Society",
      address = "Boston MA, USA",
      doi = "10.1175/JPO-D-24-0116.1"
}

@article{mcdougall2021,
	author = {McDougall, T. J. and Barker, P. M. and Holmes, R. M. and Pawlowicz, R. and Griffies, S. M. and Durack, P. J.},
	journal = {Geoscientific Model Development},
	number = {10},
	pages = {6445--6466},
	title = {The interpretation of temperature and salinity variables in numerical ocean model output and the calculation of heat fluxes and heat content},
	volume = {14},
	year = {2021}}

@article{mcdougall2003,
	address = {Boston MA, USA},
	author = {McDougall, T. J.},
	journal = {Journal of Physical Oceanography},
	number = {5},
	pages = {945 - 963},
	publisher = {American Meteorological Society},
	title = {Potential Enthalpy: A Conservative Oceanic Variable for Evaluating Heat Content and Heat Fluxes},
	volume = {33},
	year = {2003}}

@article{pinier2019,
	author = {Pinier, B. and M{\'e}min, E. and Laizet, S. and Lewandowski, R.},
	hal_id = {hal-01947662},
	hal_version = {v3},
	journal = {{Physical Review E }},
	publisher = {{American Physical Society (APS)}},
	title = {{A stochastic flow approach to model the mean velocity profile of wall-bounded flows}},
	year = {2019},
        volume = {99},
        issue = {6},
        pages = {063101},
        numpages = {11},
        month = {Jun}}

@article{resseguier2021,
	author = {Resseguier, V. and Li, LL and Jouan, G. and D{\'e}rian, P. and M{\'e}min, E. and Chapron, B.},
	hal_id = {hal-02558016},
	hal_version = {v1},
	journal = {{Archives of Computational Methods in Engineering}},
	month = Jan,
	number = {1},
	pages = {215-261},
	publisher = {{Springer Verlag}},
	title = {{New trends in ensemble forecast strategy: uncertainty quantification for coarse-grid computational fluid dynamics}},
	volume = {28},
	year = {2021}}

@article{resseguier2017e,
	author = {Resseguier, V. and M{\'e}min, E. and Chapron, B.},
	journal = {{Geophysical and Astrophysical Fluid Dynamics}},
	month = Apr,
	number = {3},
	pages = {149-227},
	publisher = {{Taylor \& Francis}},
	title = {{Geophysical flows under location uncertainty, Part I, II, \& III}},
	volume = {111},
	year = {2017}}

@article{tailleux2015,
	author = {Tailleux, R.},
	issn = {1463-5003},
	journal = {Ocean Modelling},
	pages = {26-37},
	title = {Observational and energetics constraints on the non-conservation of potential/Conservative Temperature and implications for ocean modelling},
	volume = {88},
	year = {2015}}

@article{towne2018,
	author = {Towne, A. and Schmidt, O. T and Colonius, T.},
	journal = {Journal of Fluid Mechanics},
	pages = {821--867},
	publisher = {Cambridge University Press},
	title = {Spectral proper orthogonal decomposition and its relationship to dynamic mode decomposition and resolvent analysis},
	volume = {847},
	year = {2018}}

@article{tucciarone2023,
	author = {Tucciarone, F. L. and M{\'e}min, E. and Li, L.},
	journal = {Stochastic Transport in Upper Ocean Dynamics},
	pages = {287},
	title = {Primitive Equations Under Location Uncertainty: Analytical Description and Model Development},
	year = {2023}}

@book{vallisbook,
	author = {Vallis, G. K.},
	publisher = {Cambridge University Press},
	title = {Atmospheric and oceanic fluid dynamics},
	year = {2017}}

@book{woodsbook,
	author = {Woods, L. C.},
	publisher = {Oxford University Press},
	title = {The thermodynamics of fluid systems},
	year = {1975}}

@article{dubois2023,
    title={Acoustic and gravity waves in the ocean: a new derivation of a linear model from the compressible {E}uler equation},
    volume={970},
    journal={Journal of Fluid Mechanics},
    author={Dubois, J. and Imperiale, S. and Mangeney, A. and Bouchut, F. and Sainte-Marie, J.},
    year={2023},
    pages={A28}
}

@article{mortimer1980,
    author = {Mortimer, R. G.  and Eyring, H. },
    title = {{Elementary transition state theory of the Soret and Dufour effects}},
    journal = {Proceedings of the National Academy of Sciences},
    volume = {77},
    number = {4},
    pages = {1728-1731},
    year = {1980},
}

@article{young2010,
      author = "Young, W. R.",
      title = "Dynamic Enthalpy, Conservative Temperature, and the Seawater {B}oussinesq Approximation",
      journal = "Journal of Physical Oceanography",
      year = "2010",
      publisher = "American Meteorological Society",
      address = "Boston MA, USA",
      volume = "40",
      number = "2",
      pages=      "394 - 400",
}

@String{gafd    = {Geophys.\ Astrophys.\ Fluid Dynamics}}

@String{GAFD    = {Geophys.\ Astrophys.\ Fluid Dynamics}}

@String{jmr     = {J.\ Mar.\ Res.}}

@String{JMR     = {J.\ Mar.\ Res.}}

@String{jas     = {J.\ Atmos.\ Sci.}}

@String{JAS     = {J.\ Atmos.\ Sci.}}

@String{james   = {Journal of Advances in Modeling Earth Systems}}

@String{JAMES   = {Journal of Advances in Modeling Earth Systems}}

@String{nature  = {Nature}}

@STRING{om      = {Ocean~Model.}}

@STRING{OM      = {Ocean~Model.}}

@String{po      = {Prog.\ Oceanogr.}}

@String{PO      = {Prog.\ Oceanogr.}}

@String{rg      = {Rev.\ Geophys.}}

@String{RG      = {Rev.\ Geophys.}}

@String{science = {Science}}

@String{Science = {Science}}

@article{aluie2013,
  title={Scale decomposition in compressible turbulence},
  author={Aluie, Hussein},
  journal={Physica D: Nonlinear Phenomena},
  volume={247},
  number={1},
  pages={54--65},
  year={2013},
  publisher={Elsevier}
}

@article{auclair2018,
  title={{{A non-hydrostatic non-Boussinesq algorithm for free-surface ocean modelling}}},
  author={Auclair, Francis and Bordois, L and Dossmann, Yvan and Duhaut, T and Paci, A and Ulses, C and Nguyen, C},
  journal=om,
  volume={132},
  pages={12--29},
  year={2018},
  publisher={Elsevier}
}

@article{berkooz1993,
  title={The proper orthogonal decomposition in the analysis of turbulent flows},
  author={Berkooz, Gal and Holmes, Philip and Lumley, John L},
  journal={Annual review of fluid mechanics},
  volume={25},
  number={1},
  pages={539--575},
  year={1993},
  publisher={Annual Reviews 4139 El Camino Way, PO Box 10139, Palo Alto, CA 94303-0139, USA}
}

@article{chapron2018,
  title={{{Large-scale flows under location uncertainty: a consistent stochastic framework}}},
  author={Chapron, Bertrand and D{\'e}rian, Pierre and M{\'e}min, Etienne and Resseguier, Valentin},
  journal={Quarterly Journal of the Royal Meteorological Society},
  volume={144},
  number={710},
  pages={251--260},
  year={2018},
  publisher={Wiley Online Library}
}

@article{cushman1982,
  title={Penetrative convection in the upper ocean due to surface cooling},
  author={Cushman-Roisin, Benoit},
  journal=gafd,
  volume={19},
  number={1-2},
  pages={61--91},
  year={1982},
  publisher={Taylor \& Francis}
}

@article{debreu2012,
  title={{{Two-way nesting in split-explicit ocean models: Algorithms, implementation and validation}}},
  author={Debreu, Laurent and Marchesiello, Patrick and Penven, Pierrick and Cambon, Gildas},
  journal=om,
  volume={49},
  pages={1--21},
  year={2012},
  publisher={Elsevier}
}

@article{giordani2020,
  title={An Eddy-Diffusivity Mass-Flux Parameterization for Modeling Oceanic Convection},
  author={Giordani, Herv{\'e} and Bourdall{\'e}-Badie, Romain and Madec, Gurvan},
  journal=james,
  volume={12},
  number={9},
  pages={e2020MS002078},
  year={2020},
  publisher={Wiley Online Library}
}

@article{heuze2020,
  title={Antarctic bottom water and North Atlantic deep water in CMIP6 models},
  author={Heuz{\'e}, C{\'e}line},
  journal={Ocean Science Discussions},
  volume={2020},
  pages={1--38},
  year={2020},
  publisher={G{\"o}ttingen, Germany}
}

@article{hourdin2002,
  title={Parameterization of the dry convective boundary layer based on a mass flux representation of thermals},
  author={Hourdin, Fr{\'e}d{\'e}ric and Couvreux, Fleur and Menut, Laurent},
  journal=jas,
  volume={59},
  number={6},
  pages={1105--1123},
  year={2002},
  publisher={American Meteorological Society}
}

@article{large1994,
  title={{{Oceanic vertical mixing: A review and a model with a nonlocal boundary layer parameterization}}},
  author={Large, William G and McWilliams, James C and Doney, Scott C},
  journal=RG,
  volume={32},
  number={4},
  pages={363--403},
  year={1994},
  publisher={Wiley Online Library}
}

@book{madec2017,
    author={Madec, Gurvan and Bourdall{\'e}-Badie, Romain and Bouttier, Pierre-Antoine and Bricaud, Clement and Bruciaferri, Diego and Calvert, Daley and Chanut, J{\'e}r{\^o}me and Clementi, Emanuela and Coward, Andrew and Delrosso, Damiano and others},
   title={NEMO ocean engine},
    year      = "2017",
    publisher = "Institut Pierre-Simon Laplace (IPSL), France"
}

@article{mellor1982,
  title={Development of a turbulence closure model for geophysical fluid problems},
  author={Mellor, George L and Yamada, Tetsuji},
  journal={Reviews of Geophysics},
  volume={20},
  number={4},
  pages={851--875},
  year={1982},
  publisher={Wiley Online Library}
}

@article{memin2014,
  title={Fluid flow dynamics under location uncertainty},
  author={M{\'e}min, Etienne},
  journal=gafd,
  volume={108},
  number={2},
  pages={119--146},
  year={2014},
  publisher={Taylor \& Francis}
}

@article{sander1998,
  title={Dynamical equations and turbulent closures in geophysics},
  author={Sander, Johannes},
  journal={Continuum Mechanics and Thermodynamics},
  volume={10},
  pages={1--28},
  year={1998},
  publisher={Springer}
}

@article{serazin2023,
  title={A seasonal climatology of the upper ocean pycnocline},
  author={S{\'e}razin, Guillaume and Tr{\'e}guier, Anne Marie and de Boyer Mont{\'e}gut, Cl{\'e}ment},
  journal={Frontiers in Marine Science},
  volume={10},
  pages={1120112},
  year={2023},
  publisher={Frontiers Media SA}
}

@article{shchepetkin2005,
  title={{{The regional oceanic modeling system (ROMS): a split-explicit, free-surface, topography-following-coordinate oceanic model}}},
  author={Shchepetkin, Alexander F and McWilliams, James C},
  journal={Ocean modelling},
  volume={9},
  number={4},
  pages={347--404},
  year={2005},
  publisher={Elsevier}
}

@article{souza2020,
  title={{{Uncertainty Quantification of Ocean Parameterizations: Application to the K-Profile-Parameterization for Penetrative Convection}}},
  author={Souza, Andre Nogueira and Wagner, GL and Ramadhan, Ali and Allen, B and Churavy, V and Schloss, James and Campin, J and Hill, Chris and Edelman, Alan and Marshall, John and others},
  journal=james,
  volume={12},
  number={12},
  pages={e2020MS002108},
  year={2020},
  publisher={Wiley Online Library}
}

@article{suselj2019,
  title={A unified eddy-diffusivity/mass-flux approach for modeling atmospheric convection},
  author={Suselj, Kay and Kurowski, Marcin J and Teixeira, Joao},
  journal=jas,
  volume={76},
  number={8},
  pages={2505--2537},
  year={2019}
}

@article{tailleux2012,
  title={Thermodynamics/dynamics coupling in weakly compressible turbulent stratified fluids},
  author={Tailleux, R{\'e}mi},
  journal={International Scholarly Research Notices},
  volume={2012},
  year={2012},
  publisher={Hindawi}
}

@article{tailleux2024,
  title={{{A simple and transparent method for improving the energetics and thermodynamics of seawater approximations: Static energy asymptotics (SEA)}}},
  author={Tailleux, R{\'e}mi and Dubos, Thomas},
  journal=om,
  volume={188},
  pages={102339},
  year={2024},
  publisher={Elsevier}
}

@article{umlauf2003,
  title={A generic length-scale equation for geophysical turbulence models},
  author={Umlauf, Lars and Burchard, Hans},
  journal=jmr,
  volume={61},
  number={2},
  pages={235--265},
  year={2003},
  publisher={Sears Foundation for Marine Research}
}

@article{umlauf2005,
  title={Second-order turbulence closure models for geophysical boundary layers. A review of recent work},
  author={Umlauf, Lars and Burchard, Hans},
  journal={Continental Shelf Research},
  volume={25},
  number={7-8},
  pages={795--827},
  year={2005},
  publisher={Elsevier}
}

@article{jamet2022c,
  title={{{Toward a Stochastic Parameterization for Oceanic Deep Convection}}},
  author={Jamet, Quentin and M{\'e}min, Etienne and Dumas, Franck and Li, Long and Garreau, Pierre},
  journal={Stochastic Transport in Upper Ocean Dynamics Annual Workshop},
  pages={143--157},
  year={2022},
  organization={Springer Nature Switzerland Cham}
}

@article{legay2024b,
author = {Legay, Alexandre and Deremble, Bruno and Burchard, Hans},
title = {Derivation and Implementation of a Non-Local Term to Improve the Oceanic Convection Representation Within the $k-\epsilon$ Parameterization},
journal = {Journal of Advances in Modeling Earth Systems},
volume = {17},
number = {1},
pages = {e2024MS004243},
year = {2025}
}

@article{li2023,
  title={Stochastic Data-Driven Parameterization of Unresolved Eddy Effects in a Baroclinic Quasi-Geostrophic Model},
  author={Li, L. and Deremble, B. and Lahaye, N. and M{\'e}min, E.},
  journal={Journal of Advances in Modeling Earth Systems},
  volume={15},
  number={2},
  pages={e2022MS003297},
  year={2023},
}

@article{perrot2025,
  title={{{Energetically Consistent Eddy-Diffusivity Mass-Flux Convective Schemes: Theory and Models}}},
  author={Perrot, Manolis and Lemari{\'e}, Florian and Dubos, Thomas},
  journal=james,
  pages={2024MS004273},
  year={2025},
  publisher={Wiley Online Library}
}

@book{stull1988,
  title={{{An introduction to boundary layer meteorology}}},
  author={Stull, Roland B},
  volume={13},
  year={1988},
  publisher={Springer Science \& Business Media}
}

\end{document}